\documentclass[sigconf, nonacm]{acmart}

\newcommand\vldbpagestyle{empty} 
\usepackage{amsmath,amsfonts}
\usepackage{graphicx}
\usepackage{textcomp}
\usepackage{xcolor}
\def\BibTeX{{\rm B\kern-.05em{\sc i\kern-.025em b}\kern-.08em
		T\kern-.1667em\lower.7ex\hbox{E}\kern-.125emX}}
\usepackage{appendix}
\usepackage{etoolbox}
\usepackage{multirow}
\usepackage{tabularx}

\newcolumntype{C}{>{\centering\arraybackslash}X}
\usepackage{booktabs}
\usepackage{lipsum}
\usepackage[aboveskip=0pt]{subcaption}
\usepackage{amsthm}
\usepackage{pgfplots}
\pgfplotsset{compat=1.16}
\newtheorem{lem}{Lemma}
\newtheorem{theorem}{Theorem}
\newtheorem{corollary}{Corollary}[theorem]
\usepackage{algorithmic,float}
\usepackage[ruled,linesnumbered,noend]{algorithm2e}
\let\oldnl\nl
\newcommand{\nonl}{\renewcommand{\nl}{\let\nl\oldnl}}
\usepackage{etoolbox}
\makeatletter
\patchcmd\algocf@Vline{\vrule}{\vrule \kern-0.4pt}{}{}
\patchcmd\algocf@Vsline{\vrule}{\vrule \kern-0.4pt}{}{}
\makeatother
\SetKwInput{KwInput}{Input}              
\SetKwInput{KwOutput}{Output}           
\usepackage{tikz}
\usepackage{environ}

\definecolor{ballblue}{rgb}{0.13, 0.67, 0.8}

\newlength\myindent
\makeatletter
\newcommand{\labeltext}[2]{
	\@bsphack
	\csname phantomsection\endcsname 
	\def\@currentlabel{#1}{\label{#2}}
	\@esphack
}
\makeatother
\newcommand*{\thead}[1]{\multicolumn{1}{|c|}{\bfseries #1}}

\usetikzlibrary{shapes.geometric,arrows,arrows.meta,calc,fit,matrix}
\definecolor{arrowcolor}{rgb}{.27,.45,.77}
\tikzstyle{arrow} = [arrowcolor,opacity=1,thin, {Triangle[angle=60:1.2mm]}-{Triangle[angle=60:1.2mm]}]
\tikzstyle{chart} = [rectangle, minimum width=3cm, minimum height=1cm, text centered,text width =3cm, draw=black, fill=white!30]
\tikzstyle{circlular} = [circle, minimum width=1cm, minimum height=1cm, text centered,text width =1cm, draw=black, fill=white!30]
\tikzstyle{circ} = [circle, minimum width=5mm, minimum height=5mm, text centered,text width =5mm, draw=black, fill=yellow!30]
\tikzstyle{roundrect3} = [rectangle,rounded corners, minimum width=3cm, minimum height=1.6cm, text centered,text width =1.4cm, draw=black, fill=red, text opacity=1]
\tikzstyle{roundrect4} = [rectangle,rounded corners, minimum width=3cm, minimum height=1.6cm, text centered,text width =1.4cm, draw=black, fill=blue, text opacity=1]
\tikzstyle{roundrect6} = [rectangle,rounded corners, minimum width=3cm, minimum height=1.6cm, text centered,text width =1.4cm, draw=black, fill=green, text opacity=1]
\tikzstyle{invisible} = [minimum width=3cm, minimum height=2cm, text centered,text width =1.4cm, draw=black, opacity = 1]
\tikzstyle{tinyarrow} = [thin,->,>=stealth]
\tikzstyle{strippedline} = [thick,dotted,>=stealth]
\BeforeBeginEnvironment{appendices}{\clearpage}

\makeatletter
\newcommand{\removelatexerror}{\let\@latex@error\@gobble}
\makeatother

\begin{document}\vspace{-0.2in}
\title{Efficient Temporal Pattern Mining in Big Time Series Using Mutual Information}

\author{Van Long Ho}
\affiliation{
	\institution{Aalborg University}
	\city{Aalborg}
	\state{Denmark}
}
\email{vlh@cs.aau.dk}

\author{Nguyen Ho}
\affiliation{
	\institution{Aalborg University}
	\city{Aalborg}
	\state{Denmark}
}
\email{ntth@cs.aau.dk}

\author{Torben Bach Pedersen}
	\affiliation{
		\institution{Aalborg University}
		\city{Aalborg}
		\state{Denmark}
	}
\email{tbp@cs.aau.dk}

\begin{abstract}
	Very large time series are increasingly available from an ever wider range of IoT-enabled sensors deployed in different environments. Significant insights can be gained by mining temporal patterns from these time series.
	Unlike traditional pattern mining, temporal pattern mining (TPM) adds event time intervals into extracted patterns, making them more expressive at the expense of increased time and space complexities. 
	Existing TPM methods either cannot scale to large datasets, or work only on pre-processed temporal events rather than on time series.
	This paper presents our Frequent Temporal Pattern Mining from Time Series (FTPMfTS) approach providing: (1) The end-to-end FTPMfTS process taking time series as input and producing frequent temporal patterns as output. (2) The efficient Hierarchical Temporal Pattern Graph Mining (HTPGM) algorithm that uses efficient data structures for fast support and confidence computation, and employs effective pruning techniques for significantly faster mining. (3) An approximate version of HTPGM that uses mutual information, a measure of data correlation, to prune unpromising time series from the search space. (4) An extensive experimental evaluation showing that HTPGM outperforms the baselines in runtime and memory consumption, and can scale to big datasets. The approximate HTPGM is up to two orders of magnitude faster and less memory consuming than the baselines, while retaining high accuracy. 
\end{abstract}

\maketitle
\pagestyle{\vldbpagestyle}
\section{Introduction}
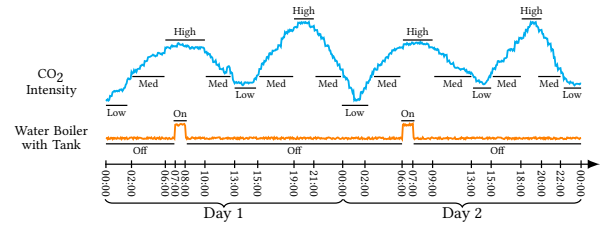
\begin{figure}[!t]
	\resizebox{1\columnwidth}{!}{\begin{tikzpicture}

% sX,eX,Y
% \newcommand{\DrawEvent}[4]{
%     \draw [black,thin] (axis cs: #1, #3+0.15) -- (axis cs: #2, #3+0.15);
%     \node [black, scale=0.48] at (axis cs: #1/2+#2/2, #3+0.35) {\small #4};
% }
% sX,eX,Y
\newcommand{\DrawOn}[3]{
	\draw [black,thin] (axis cs: #1, #3+0.13) -- (axis cs: #2, #3+0.13);
	\node [black, scale=0.5] at (axis cs: #1/2+#2/2, #3+0.32) {\footnotesize On};
}

\newcommand{\DrawHigh}[3]{
	\draw [black,thin] (axis cs: #1, #3+0.13) -- (axis cs: #2, #3+0.13);
	\node [black, scale=0.5] at (axis cs: #1/2+#2/2, #3+0.32) {\footnotesize High};
}

%% sX,eX,Y
\newcommand{\DrawOff}[3]{
	\draw [black,thin] (axis cs: #1, #3-0.12) -- (axis cs: #2, #3-0.12);
	\node [black, scale=0.5] at (axis cs: #1/2+#2/2, #3-0.3) {\footnotesize Off};
}
\newcommand{\DrawLow}[3]{
	\draw [black,thin] (axis cs: #1, #3-0.12) -- (axis cs: #2, #3-0.12);
	\node [black, scale=0.5] at (axis cs: #1/2+#2/2, #3-0.3) {\footnotesize Low};
}
\newcommand{\DrawMedium}[3]{
	\draw [black,thin] (axis cs: #1, #3-0.12) -- (axis cs: #2, #3-0.12);
	\node [black, scale=0.5] at (axis cs: #1/2+#2/2, #3-0.3) {\footnotesize Med};
}
\begin{axis}[axis line style={draw=none},
             tick style={draw=none},
             ymin = -3.0, ymax = 3.5,
             xmin = -130, xmax = 600,
             xticklabels={,,},
             yticklabels={,,},
             yscale=.5
             ]
%time arrow
\draw [black, thin, -latex] (axis cs: 0, -0.85) -- (axis cs: 600,-0.85) node [right] {};             

\addplot [mark=none, color=cyan, line width=0.6pt] table [x=x, y=co2_time, col sep=comma] {data/tikz/data_using.csv};
\addplot [mark=none, color=orange, line width=0.6pt] table [x=x, y=boiler_time, col sep=comma] {data/tikz/data_using.csv};

\node[black, scale=0.9, text width=2cm,align=center, font=\tiny\linespread{0.8}\selectfont] at (axis cs: -65, -0.15) {Water Boiler\\with Tank};
\node[black, scale=0.9, text width=2cm,align=center, font=\tiny\linespread{0.8}\selectfont] at (axis cs: -65, 1.25) {CO$_2$ \\Intensity};
%\node[black, scale=0.9] at (axis cs: -30, 2.67) {\footnotesize Washer};

%tick of time
\pgfplotsinvokeforeach{1, 32, 73, 85, 97, 121, 157, 185, 229, 253, 288, 315, 360, 373, 397, 444, 468, 504, 529, 552, 577}{
    \draw [black, thin] (axis cs: #1, -0.75) -- (axis cs: #1,-0.95);
}

\node [black, scale=0.45,rotate=270] at (axis cs: 1,-1.35) {\textbf \footnotesize 00:00};
\node [black, scale=0.45,rotate=270] at (axis cs: 32,-1.35) {\textbf \footnotesize 02:00};
\node [black, scale=0.45,rotate=270] at (axis cs: 73,-1.35) {\textbf \footnotesize 06:00};
\node [black, scale=0.45,rotate=270] at (axis cs: 85,-1.35) {\textbf \footnotesize 07:00};
\node [black, scale=0.45,rotate=270] at (axis cs: 97,-1.35) {\textbf \footnotesize 08:00};
\node [black, scale=0.45,rotate=270] at (axis cs: 121,-1.35) {\textbf \footnotesize 10:00};
\node [black, scale=0.45,rotate=270] at (axis cs: 157,-1.35) {\textbf \footnotesize 13:00};
\node [black, scale=0.45,rotate=270] at (axis cs: 185,-1.35) {\textbf \footnotesize 15:00};
\node [black, scale=0.45,rotate=270] at (axis cs: 229,-1.35) {\textbf \footnotesize 19:00};
\node [black, scale=0.45,rotate=270] at (axis cs: 253,-1.35) {\textbf \footnotesize 21:00};
\node [black, scale=0.45,rotate=270] at (axis cs: 288,-1.35) {\textbf \footnotesize 00:00};
\node [black, scale=0.45,rotate=270] at (axis cs: 315,-1.35) {\textbf \footnotesize 02:00};
\node [black, scale=0.45,rotate=270] at (axis cs: 360,-1.35) {\textbf \footnotesize 06:00};
\node [black, scale=0.45,rotate=270] at (axis cs: 373,-1.35) {\textbf \footnotesize 07:00};
\node [black, scale=0.45,rotate=270] at (axis cs: 397,-1.35) {\textbf \footnotesize 09:00};
\node [black, scale=0.45,rotate=270] at (axis cs: 444,-1.35) {\textbf \footnotesize 13:00};
\node [black, scale=0.45,rotate=270] at (axis cs: 468,-1.35) {\textbf \footnotesize 15:00};
\node [black, scale=0.45,rotate=270] at (axis cs: 504,-1.35) {\textbf \footnotesize 18:00};
\node [black, scale=0.45,rotate=270] at (axis cs: 529,-1.35) {\textbf \footnotesize 20:00};
\node [black, scale=0.45,rotate=270] at (axis cs: 552,-1.35) {\textbf \footnotesize 22:00};
\node [black, scale=0.45,rotate=270] at (axis cs: 577,-1.35) {\textbf \footnotesize 00:00};
%
%
%\DrawLow{1}{36}{1}
%\DrawLow{37}{72}{1.25}
%\DrawHigh{73}{121}{1.75}
%\DrawLow{122}{144}{1.25}
%\DrawLow{150}{178}{1}
%\DrawLow{181}{205}{1.25}
%\DrawHighLeft{205}{225}{1.75}
%\DrawHigh{229}{242}{2}
%\DrawHighRight{245}{265}{1.75}
%\DrawLow{266}{288}{1.25}
%\DrawLow{293}{332}{1}
%\DrawLow{338}{360}{1.25}
%\DrawHigh{361}{397}{1.75}
%\DrawLow{399}{493}{1.25}
%\DrawHigh{494}{526}{1.75}
%\DrawHigh{529}{541}{2}
%\DrawHighRight{543}{565}{1.75}
%\DrawLow{565}{577}{1}

\DrawLow{1}{27}{0.8}
\DrawMedium{33}{72}{1.55}
\DrawHigh{73}{121}{2.25}
\DrawMedium{122}{152}{1.55}
\DrawLow{157}{183}{1.25}
\DrawMedium{187}{228}{1.55}
\DrawHigh{229}{253}{2.8}
\DrawMedium{255}{286}{1.55}
\DrawLow{287}{319}{0.8}
\DrawMedium{326}{358}{1.55}
\DrawHigh{361}{397}{2.25}
\DrawMedium{399}{444}{1.55}
\DrawLow{446}{468}{1.25}
\DrawMedium{470}{504}{1.55}
\DrawHigh{505}{529}{2.8}
\DrawMedium{525}{552}{1.55}
\DrawLow{556}{577}{1.25}
\DrawOff{1}{84}{-0.2}
\DrawOn{83}{99}{0.15}
\DrawOff{99}{360}{-0.2}
\DrawOn{359}{375}{0.15}
\DrawOff{374}{577}{-0.2}

\draw[decorate,decoration={brace,mirror, raise=5pt}] (axis cs: 0,-1.36) --  node[scale=.9,below=5pt] {\scriptsize Day 1} (axis cs: 288,-1.36);
\draw[decorate,decoration={brace,mirror, raise=5pt}] (axis cs: 289,-1.36) --  node[scale=.9,below=5pt] {\scriptsize Day 2} (axis cs: 577,-1.36);

\end{axis}
\end{tikzpicture}}
	\vspace{-0.4in}
	\caption{CO2 intensity and water boiler electricity usage}
	\label{fig:TimeSeries}
\end{figure}

\vspace{-0.02in}
IoT-enabled sensors have enabled the collection of many big time series, e.g., from smart-meters, -plugs, and -appliances in households, weather stations, and GPS-enabled mobile devices. Extracting patterns from these time series can offer new domain insights for evidence-based decision making and optimization. 
As an example, consider Fig. \ref{fig:TimeSeries} that shows the electricity usage of a water boiler with a hot water tank collected by a $20$ euro  wifi-enabled smart-plug, and accurate CO2 intensity (g/kWh) forecasts of local electricity, e.g., as supplied by the Danish Transmission System Operator \cite{co2}.   
From Fig. \ref{fig:TimeSeries}, we can identify several useful patterns. First, the water boiler switches \textit{On} once a day, for one hour between 6 and 8AM. This indicates that the resident takes only one hot shower per day which starts between 5.30 and 6.30AM. Second, all water boiler \textit{On} events are contained in CO2 \textit{High} events, i.e., the periods when CO2 intensity is high. Third, between two consecutive \textit{On} events of the boiler, there is a CO2 \textit{Low} event lasting for one or more hours which occurs at most 4 hours before the hot shower (so water heated during that event will still be hot at 6AM).  
Pattern mining can be used to extract the relations between CO2 intensity and water boiler events. However, traditional sequential patterns only capture the sequential occurrence of events, e.g., that one boiler \textit{On} event follows after another, but not that there is at least $23$ hours between them; or that there is a CO2 \textit{Low} event between the two boiler \textit{On} events, but not when or for how long it lasts. 
In contrast, \textit{temporal pattern mining} (TPM) adds temporal information into patterns, providing details on when certain relations between events happen, and for how long.  
For example, TPM expresses the above relations as:  
([7:00 - 8:00, Day X] BoilerOn $\rightarrow$ [6:00 - 7:00, Day X+1] BoilerOn) (meaning BoilerOn is followed by BoilerOn), ([6:00 - 10:00, Day X] HighCO2 $\succcurlyeq$ [7:00 - 8:00, Day X] BoilerOn) (meaning HighCO2 contains BoilerOn), and ([7:00 - 8:00, Day X] BoilerOn $\rightarrow$ [0:00 - 2:00, Day X+1] LowCO2 $\rightarrow$ [6:00 - 7:00, Day X+1] BoilerOn). 
As the resident is very keen on reducing her CO2 footprint, we can rely on the above temporal patterns to automatically (using the smart-plug) delay turning on the boiler until the CO2 intensity is low again, saving CO2 without any loss of comfort for the resident.
	
Another example is in the smart city domain in which temporal patterns extracted from vehicle GPS data \cite{torp2019traveltime} can reveal spatio-temporal correlations between traffic jams. For example, if the pattern ([07:30, 08:00] SlowSpeedTunnel $\rightarrow$ [08:00, 08:30] SlowSpeedMainBoulevard) is found with high frequency and high confidence on weekdays, it can be used to advise drivers to take another route for their morning commute.

Although temporal patterns are useful, mining them is much more expensive than sequential patterns. Not only does the temporal information add extra computation to the mining process, the complex relations between events also add an additional exponential factor O($3^{h^2}$) to the complexity O($m^h$) of the search space ($m$ is the number of events and $h$ is the length of temporal patterns), yielding an overall complexity of O($m^h3^{h^2}$) (see Lemma \ref{lem1} in Section \ref{sec:2freq}).  
Existing TPM methods \cite{tpminer,ieminer,hdfs} do not scale on big datasets, i.e., many time series and many sequences, %as the challenge of the exponential search space still exists, 
and/or do not work directly on time series but rather on pre-processed temporal events. 

\textit{Contributions.} In this paper, we present our comprehensive Frequent Temporal Pattern Mining from Time Series (FTPMfTS) approach which overcomes the above limitations.  
Our key contributions are:
(1) We present the first end-to-end FTPMfTS process that receives time series as input, and produces frequent temporal patterns as output. Within this process, a splitting strategy is proposed to convert time series into event sequences while ensuring the preservation of temporal patterns. 
(2) We propose the efficient Hierarchical Temporal Pattern Graph Mining (HTPGM) algorithm that employs: a) efficient data structures, Hierarchical Pattern Graph and \textit{bitmap}, to enable fast support and confidence computation; and b) pruning techniques based on the Apriori principle and the transitivity property of temporal relations to enable faster mining. 
(3) Based on the concept of mutual information which measures the correlation among time series, we propose a novel approximate version of HTPGM that prunes unpromising time series to significantly reduce the search space and can scale on big datasets, i.e., many time series and many sequences. 
(4) We perform extensive experiments on synthetic and real-world datasets which show that HTPGM outperforms the baselines in both runtime and memory usage. The approximate HTPGM is up to two orders of magnitude faster and less memory consumption than  the baselines while retaining high accuracy compared to the exact HTPGM.

\vspace{-0.15in}
\section{Related work}\label{sec:relatedwork}\vspace{-0.02in}
\textit{Temporal pattern mining:} Compared to sequential pattern mining, TPM is rather a new research area. One of the first papers in this area is \cite{kam2000discovering} from Kam et al. that uses a hierarchical representation to manage temporal relations, and based on that mines temporal patterns. However, the approach in \cite{kam2000discovering} suffers from \textit{ambiguity} when presenting temporal relations.  
In \cite{wu2007mining}, Wu et al. develop TPrefix to mine temporal patterns from non-ambiguous temporal relations. However, TPrefix has several inherent limitations: it scans the database repeatedly, and the algorithm does not employ any pruning strategies to reduce the search space. 
In \cite{moskovitch2015fast}, Moskovitch et al. design a TPM algorithm using the transitivity property of temporal relations. They use this property to generate candidates by inferring new relations between events. In comparison, our HTPGM uses the transitivity property for effective pruning. In \cite{batal2012mining}, Iyad et al. propose a TPM framework to detect events in time series. However, their focus is to find irregularities in the data. In \cite{wang2020mining}, Wang et al. propose a temporal pattern mining algorithm HUTPMiner to mine high-utility patterns. Different from our HTPGM which uses \textit{support} and \textit{confidence} to measure the frequency of patterns, HUTPMiner uses \textit{utility} to measure the importance or profit of an event/ pattern, thereby addresses an orthogonal problem. In \cite{sharma2018stipa}, Amit et al. propose STIPA which uses a Hoeppner matrix representation to compress temporal patterns for memory savings. However, STIPA does not use any pruning/ optimization strategies and thus, despite the efficient use of memory, it cannot scale to large datasets, unlike our HTPGM. 
Other work \cite{batal2013temporal}, \cite{campbell2020temporal} proposes TPM algorithms to classify health record data.  
However, these methods are very domain-specific, thus cannot generalize to other domains. 

\begin{table*}
\begin{minipage}{1\textwidth}
	\caption{A Symbolic Database $\mathcal{D}_{\text{SYB}}$}
	\vspace{-0.12in}
	\resizebox{\textwidth}{1.2cm}{
		\setlength{\tabcolsep}{0.5mm} 
		\renewcommand{\arraystretch}{1.5}
		\begin{tabular}{|c|ccccccccc|ccccccccc|ccccccccc|ccccccccc|}
			\hline
			\textbf{Time} & \textbf{10:00} & \textbf{10:05} & \textbf{10:10} & \textbf{10:15} & \textbf{10:20} & \textbf{10:25} & \textbf{10:30} & \textbf{10:35} & \textbf{10:40} & \textbf{10:45} & \textbf{10:50} & \textbf{10:55} & \textbf{11:00} & \textbf{11:05} & \textbf{11:10} & \textbf{11:15} & \textbf{11:20} & \textbf{11:25} & \textbf{11:30} & \textbf{11:35} & \textbf{11:40} & \textbf{11:45} & \textbf{11:50} & \textbf{11:55} & \textbf{12:00} & \textbf{12:05} & \textbf{12:10} & \textbf{12:15} & \textbf{12:20} & \textbf{12:25} & \textbf{12:30} & \textbf{12:35} & \textbf{12:40} & \textbf{12:45} & \textbf{12:50} & \textbf{12:55}\\ 
			\hline 
			\textbf{S} & \normalsize On & \normalsize On & \normalsize On & \normalsize On & \normalsize Off & \normalsize Off & \normalsize Off & \normalsize On & \normalsize On & \normalsize Off & \normalsize Off & \normalsize Off & \normalsize Off & \normalsize Off & \normalsize Off & \normalsize On & \normalsize On & \normalsize On & \normalsize Off & \normalsize Off & \normalsize Off & \normalsize Off & \normalsize On & \normalsize On & \normalsize On & \normalsize Off & \normalsize Off & \normalsize On & \normalsize On & \normalsize Off & \normalsize Off & \normalsize On & \normalsize On & \normalsize On & \normalsize Off & \normalsize Off\\ 
			\hline 
			\textbf{T} & \normalsize Off & \normalsize On & \normalsize On & \normalsize On & \normalsize Off & \normalsize Off & \normalsize Off & \normalsize On & \normalsize On & \normalsize Off & \normalsize Off & \normalsize On & \normalsize On & \normalsize Off & \normalsize Off & \normalsize On & \normalsize On & \normalsize On & \normalsize Off & \normalsize Off & \normalsize Off & \normalsize Off & \normalsize On & \normalsize On & \normalsize On & \normalsize Off & \normalsize Off & \normalsize On & \normalsize On & \normalsize Off & \normalsize Off & \normalsize Off & \normalsize On & \normalsize On & \normalsize On & \normalsize Off\\ 
			\hline 
			\textbf{M} & \normalsize Off & \normalsize Off & \normalsize Off & \normalsize Off & \normalsize On & \normalsize On & \normalsize On & \normalsize Off & \normalsize Off & \normalsize On & \normalsize On & \normalsize On & \normalsize Off & \normalsize On & \normalsize On & \normalsize Off & \normalsize Off & \normalsize Off & \normalsize On & \normalsize On & \normalsize Off & \normalsize On & \normalsize On & \normalsize Off & \normalsize Off & \normalsize On & \normalsize On & \normalsize Off & \normalsize Off & \normalsize On & \normalsize On & \normalsize On & \normalsize Off & \normalsize Off & \normalsize On & \normalsize On\\ 
			\hline 
			\textbf{W} & \normalsize Off & \normalsize Off & \normalsize Off & \normalsize Off & \normalsize On & \normalsize On & \normalsize On & \normalsize Off & \normalsize Off & \normalsize On & \normalsize On & \normalsize Off & \normalsize On & \normalsize On & \normalsize On & \normalsize Off & \normalsize Off & \normalsize Off & \normalsize On & \normalsize On & \normalsize Off & \normalsize On & \normalsize On & \normalsize Off & \normalsize Off & \normalsize On & \normalsize On & \normalsize Off & \normalsize Off & \normalsize On & \normalsize On & \normalsize On & \normalsize Off & \normalsize Off & \normalsize On & \normalsize On\\ 
			\hline 
			\textbf{D} & \normalsize Off & \normalsize Off & \normalsize Off & \normalsize Off & \normalsize Off & \normalsize Off & \normalsize Off & \normalsize Off & \normalsize Off & \normalsize On & \normalsize On & \normalsize Off & \normalsize Off & \normalsize Off & \normalsize Off & \normalsize Off & \normalsize On & \normalsize On & \normalsize Off & \normalsize Off & \normalsize Off & \normalsize Off & \normalsize Off & \normalsize Off & \normalsize Off & \normalsize Off & \normalsize Off & \normalsize On & \normalsize On & \normalsize Off & \normalsize Off & \normalsize Off & \normalsize On & \normalsize On & \normalsize Off & \normalsize Off\\ 
			\hline 
			\textbf{I} & \normalsize Off & \normalsize Off & \normalsize Off & \normalsize Off & \normalsize Off & \normalsize Off & \normalsize Off & \normalsize On & \normalsize On & \normalsize Off & \normalsize Off & \normalsize Off & \normalsize Off & \normalsize Off & \normalsize Off & \normalsize Off & \normalsize Off & \normalsize Off & \normalsize On & \normalsize On & \normalsize Off & \normalsize Off & \normalsize Off & \normalsize Off & \normalsize Off & \normalsize Off & \normalsize Off & \normalsize On & \normalsize On & \normalsize Off & \normalsize Off & \normalsize Off & \normalsize Off & \normalsize Off & \normalsize On & \normalsize On\\ 
			\hline 
			
		\end{tabular}
	}
	\label{tbl:SymbolDatabase}
\end{minipage}
\vspace{-0.1in}
\end{table*}

The state-of-the-art TPM methods that currently achieve the best performance are our baselines: H-DFS \cite{hdfs}, TPMiner \cite{tpminer}, IEMiner \cite{ieminer}, and Z-Miner \cite{lee2020z}. H-DFS is a hybrid algorithm that uses breadth-first and depth-first search strategies to mine frequent arrangements of temporal intervals. H-DFS uses a data structure called ID-List to transform event sequences into vertical representations, and temporal patterns are generated by merging the ID-Lists of different events. This means that H-DFS does not scale well when the number of time series increases. In \cite{ieminer}, Patel et al. design a hierarchical lossless representation to model event relations, and propose IEMiner that uses Apriori-based optimizations to efficiently mine patterns from this new representation. In \cite{tpminer}, Chen et al. propose TPMiner that uses endpoint and endtime representations to simplify the complex relations among events. Similar to \cite{hdfs}, IEMiner and TPMiner do not scale to datasets with many time series. 
Z-Miner \cite{lee2020z}, proposed by Lee et al., is the most recent work addressing TPM. Z-Miner improves the mining efficiency over existing methods by employing two data structures: a hierarchical hash-based structure called Z-Table for time-efficient candidate generation and support count, and Z-Arrangement, a structure to efficiently store event intervals in temporal patterns for efficient memory consumption. Although using efficient data structures, Z-Miner neither employs the transitivity property of temporal relations nor mutual information for pruning. Thus, Z-Miner is less efficient than our exact and approximate HTPGM in both runtimes and memory usage, and does not scale to large datasets with many sequences and many time series (see Section \ref{sec:experiment}).
Our HTPGM algorithm improves on these methods by: 
(1) using efficient data structures and applying pruning techniques based on the Apriori principle and the transitivity property of temporal relations to enable fast mining, (2) the approximate HTPGM can handle datasets with many time series and sequences, and (3), providing an end-to-end FTPMfTS process to mine temporal patterns directly from time series, a feature that is not supported by the baselines.

\textit{Using correlations in TPM:} Different correlation measures such as expected support \cite{ahmed2016mining}, all-confidence \cite{lee2003comine}, and mutual information (MI) \cite{ke2008correlated, ho2019icde, blanchard2005using, ho2019amic, cunjin2015mutual, yao2003information, ho2020timedelay, ho2016corr, ho2015datavalue, energy2014, energy2016, ho2017improving} have been used to optimize the pattern mining process. However, these only support sequential patterns. To the best of our knowledge, our proposed approximate HTPGM is the first that uses MI to optimize TPM.

\section{Preliminaries}\label{sec:preliminary}\vspace{-0.02in}
In this section, we introduce the notations and the main concepts that will be used throughout the paper.

\vspace{-0.1in}
\subsection{Temporal Event of Time Series}\vspace{-0.02in}
\textbf{Definition 3.1} (Time series) A \textit{time series} $X=  x_1, x_2, ..., x_n$ is a sequence of data values that measure the same phenomenon during an observation time period, and are chronologically ordered.

\hspace{-0.15in}\textbf{Definition 3.2} (Symbolic time series) A \textit{symbolic time series} $X_S$ of a time series $X$ encodes the raw values of $X$ into a sequence of symbols. The finite set of permitted symbols used to encode $X$ is called the \textit{symbol alphabet} of $X$, denoted as $\Sigma_X$.

The symbolic time series $X_S$ is obtained using a mapping function $f$$:$ $X$$\rightarrow$$\Sigma_{X}$ that maps each value $x_i \in X$ to a symbol $\omega \in \Sigma_{X}$. For example, let $X$ = 1.61, 1.21, 0.41, 0.0 be a time series representing the energy usage of an electrical device. Using the symbol alphabet $\Sigma_X$ = \{On, Off\}, where On represents that the device is on and operating (e.g., $x_i \ge 0.5$), and Off that the device is off ($x_i < 0.5$), the symbolic representation of $X$ is: $X_S$ = On, On, Off, Off. The mapping function $f$ can be defined using existing time series representation techniques such as SAX \cite{lin2003symbolic} or MVQ \cite{megalooikonomou2005multiresolution}.

\hspace{-0.15in}\textbf{Definition 3.3} (Symbolic database)  
Given a set of time series $\mathcal{X}=\{X_1,...,X_n\}$, the set of symbolic representations of the time series in $\mathcal{X}$ forms a \textit{symbolic database} $\mathcal{D_{\text{SYB}}}$.

An example of the symbolic database $\mathcal{D}_{\text{SYB}}$ is shown in Table \ref{tbl:SymbolDatabase}. There are $6$ time series representing the energy usage of $6$ electrical appliances: \{Stove, Toaster, Microwave, Clothes Washer, Dryer, Iron\}. For brevity, we name the appliances respectively as \{S, T, M, W, D, I\}. 
All appliances have the same alphabet $\Sigma$ = \{On, Off\}.

\hspace{-0.15in}\textbf{Definition 3.4} (Temporal event in a symbolic time series) A \textit{temporal event} $E$ in a symbolic time series $X_S$ is a tuple $E = (\omega, T)$ where $\omega \in \Sigma_X$ is a symbol, and $T=\{[t_{s_i}, t_{e_i}]\}$ is the set of time intervals during which $X_S$ is associated with the symbol $\omega$.   

Given a time series $X$, a temporal event is created by first converting $X$ into symbolic time series $X_S$, and then combining identical consecutive symbols in $X_S$ into one single time interval. For example, consider the symbolic representation of $S$ in Table \ref{tbl:SymbolDatabase}. By combining its consecutive On symbols, we form the temporal event \textit{``Stove is On''} as: (SOn, \{[10:00, 10:15], [10:35, 10:40], [11:15, 11:25], [11:50, 12:00], [12:15, 12:20], [12:35, 12:45]\}). 
 
\hspace{-0.15in}\textbf{Definition 3.5} (Instance of a temporal event) Let $E = (\omega, T)$ be a temporal event, and $[t_{s_i},t_{e_i}] \in T$ be a time interval. The tuple $e = (\omega, [t_{s_i},t_{e_i}])$ is called an \textit{instance} of the event $E$, representing a single occurrence of $E$ during $[t_{s_i},t_{e_i}]$. We use the notation $E_{\triangleright e}$ to denote that event $E$ has an instance $e$. 

\begin{table*}
	\begin{minipage}{.39\textwidth}
		\caption{Temporal Relations between Events}
		\vspace{-0.12in}
		\scriptsize
		\resizebox{\textwidth}{3.2cm}{
			\begin{tabular}{|m{.25\columnwidth}| m{.94\columnwidth}|}
				\hline
				Follows: \hspace{0.2cm}$E_{i_{\triangleright e_i}} \rightarrow E_{j_{\triangleright e_j}}$ & \begin{tikzpicture}
					
					\draw (0,0) -- node[below]{\textbf{e$_{i}$}} ++(0.7,0);
					\filldraw (0,0) circle (2pt)node[above]{\small t$_{s_i}$} ;
					\filldraw (0.7,0) circle (2pt)node[above ]{\small t$_{e_i}{\pm \epsilon}$};
					
					\filldraw (0.9,0) circle (2pt)node[below right]{\small t$_{s_j}$};
					\filldraw (2.3,0) circle (2pt)node[below]{\small t$_{e_j}$};
					\draw (0.9,0) -- node[above]{\textbf{e$_{j}$}} ++(1.4,0);
					
					%%%%%%%%%%%%%%%%%%%%%%%%%%%%%%%%%%%%%%%%%%%%%%%%%%%%%%%%%
					\draw (3.5,0) -- node[below]{\textbf{e$_{i}$}} ++(0.7,0);
					\filldraw (3.5,0) circle (2pt)node[above]{\small t$_{s_i}$} ;
					\filldraw (4.2,0) circle (2pt)node[above]{\small t$_{e_i}{\pm \epsilon}$};
					
					\filldraw (4.7,0) circle (2pt)node[below right]{\small t$_{s_j}$};
					\filldraw (6.1,0) circle (2pt)node[below]{\small t$_{e_j}$};
					\draw (4.7,0) -- node[above]{\textbf{e$_{j}$}} ++(1.4,0);
					
					\node [align=center] at (3.5,-0.7) {\small t$_{e_i}{\pm \epsilon}$ $\le$ t$_{s_j}$};
					
				\end{tikzpicture}  \\ \hline
				
				Contains: \hspace{0.2cm}$E_{i_{\triangleright e_i}} \succcurlyeq E_{j_{\triangleright e_j}}$ & 	\begin{tikzpicture}
					\draw (0,0) -- node[above]{\textbf{e$_i$}} ++(2.0,0);
					\filldraw (0,0) circle (2pt)node[above]{\small t$_{s_i}$} ;
					\filldraw (2,0) circle (2pt)node[above]{\small t$_{e_i}\pm \epsilon$};
					
					\draw (0,-0.5) -- node[above]{\textbf{e$_{j}$}} ++(2.0,0);
					\filldraw (0,-0.5) circle (2pt)node[below]{\small t$_{s_j}$} ;
					\filldraw (2,-0.5) circle (2pt)node[below]{\small t$_{e_j}$};
					
					%%%%%%%%%%%%%%%%%%%%%%%%%%%%%%%
					
					\draw (3.5,0) -- node[above]{\textbf{e$_i$}} ++(2.1,0);
					\filldraw (3.5,0) circle (2pt)node[above]{\small t$_{s_i}$} ;
					\filldraw (5.6,0) circle (2pt)node[above]{\small t$_{e_i}\pm \epsilon$};	
					
					\draw (3.9,-0.5) -- node[above]{\textbf{e$_j$}} ++(1.4,0);
					\filldraw (3.9,-0.5) circle (2pt)node[below]{\small t$_{s_j}$} ;
					\filldraw (5.3,-0.5) circle (2pt)node[below]{\small t$_{e_j}$};	
					
					%%%%%%%%%%%%%%%
					\draw (0,-1.5) -- node[above]{\textbf{e$_i$}} ++(2.0,0);
					\filldraw (0,-1.5) circle (2pt)node[above]{\small t$_{s_i}$} ;
					\filldraw (2.0,-1.5) circle (2pt)node[above]{\small t$_{e_i}\pm \epsilon$};
					
					\draw (-0,-2) -- node[above]{\textbf{e$_{j}$}} ++(1.5,0);
					\filldraw (0,-2) circle (2pt)node[below]{\small t$_{s_j}$} ;
					\filldraw (1.5,-2) circle (2pt)node[below]{\small t$_{e_j}$};
					
					\draw (3.6,-1.5) -- node[above]{\textbf{e$_i$}} ++(2.0,0);
					\filldraw (3.6,-1.5) circle (2pt)node[above]{t$_{s_i}$} ;
					\filldraw (5.6,-1.5) circle (2pt)node[above]{t$_{e_i}\pm \epsilon$};
					
					\draw (4.1,-2) -- node[above]{\textbf{e$_{j}$}} ++(1.5,0);
					\filldraw (4.1,-2) circle (2pt)node[below]{t$_{s_j}$} ;
					\filldraw (5.6,-2) circle (2pt)node[below]{t$_{e_j}$};
					
					\node [align=center] at (3.0,-2.7) {\small (t$_{s_{i}} \le$ t$_{s_j}$)  $\wedge$ (t$_{e_i}{\pm \epsilon}$ $\ge$ t$_{e_j}$)};
					
				\end{tikzpicture} \\ \hline
				
				Overlaps:\hspace{0.2cm} $E_{i_{\triangleright e_i}} \between E_{j_{\triangleright e_j}}$ & \begin{tikzpicture}
					
					\draw (0,0) -- node[above]{\textbf{e$_i$}} ++(2,0);
					\filldraw (0,0) circle (2pt)node[above]{\small t$_{s_i}$} ;
					\filldraw (2,0) circle (2pt)node[above]{\small t$_{e_i}\pm \epsilon$};
					
					\draw (1,-0.75) -- node[above right]{\textbf{e$_{j}$}} ++(2.0,0);
					\filldraw (1,-0.75) circle (2pt)node[below]{\small t$_{s_j}$} ;
					\filldraw (3,-0.75) circle (2pt)node[below]{\small t$_{e_j}$};
					
					\draw[dashed] (1,0) -- (1,-0.75);
					\draw[dashed] (2,0) -- (2,-0.75);
					
					\draw[dashed,>=latex,thin,<->] (1,-0.325) -- node[above]{d$_{o}$} ++(1,0);
					
					\node [align=center] at (1.75,-1.45) {\small (t$_{s_i}<$ t$_{s_j}$) $\wedge$ (t$_{e_i}{\pm \epsilon}$ $<$ t$_{e_j}$) $\wedge$ 
						\small (t$_{e_i}$ $-$ t$_{s_j}$ $\ge$ d${_o}{\pm \epsilon}$)};
					
				\end{tikzpicture} \\
				\hline
		\end{tabular} }
		\label{tbl:relations}
	\end{minipage}
	\begin{minipage}{.59\textwidth}
	\caption{A Temporal Sequence Database $\mathcal{D}_{\text{SEQ}}$}
	\vspace{-0.12in}
	\label{tbl:SequenceDatabase}
	\resizebox{\textwidth}{3.2cm}{
		\begin{tabular}{ |c|p{10.5cm}| }
			\hline  \thead{ID} & \thead{Temporal sequences} \\
			\hline  
			1   &  (SOn,[10:00,10:15]), (TOff,[10:00,10:05]), (MOff,[10:00,10:20]), (WOff,[10:00,10:20]), (DOff,[10:00,10:40]), (IOff,[10:00,10:35]), (TOn,[10:05,10:15]), (SOff,[10:15,10:35]), (TOff,[10:15,10:35]), (MOn,[10:20,10:30]), (WOn,[10:20,10:30]), (WOff,[10:30,10:40]),  (MOff,[10:30,10:40]),  (SOn,[10:35,10:40]), (TOn,[10;35,10:40]), (IOn,[10:35,10:40]) 
			\\
			\hline 
			2   &  (SOff,[10:45,11:15]), (TOff,[10:45,10:55]), (MOn,[10;45,10:55]), (WOn,[10:45,10:50]), (DOn,[10:45,10:50]), (IOff,[10:45,11:25]), (WOff,[10:50,11:00]), (DOff,[10:50,11:20]), (MOff,[10:55,11:05]), (TOn,[10:55,11:00]), (TOff,[11:00,11:15]), (WOn,[11:00,11:10]), (MOn,[11:05,11:10]), (WOff,[11:10,11:25]), (MOff,[11:10,11:25]),   (SOn,[11:15,11:25]), (TOn,[11:15,11:25]), (DOn,[11:20,11:25])
			\\ 
			\hline 
			3   &  (SOff,[11:30,11:50]), (TOff,[11:30,11:50]), (MOn,[11:30,11:35]), (WOn,[11:30,11:35]), (DOff,[11:30,12:10]), (IOn,[11:30,11:35]), (IOff,[11:35,12:10]), (MOff,[11:35,11:45]), (WOff,[11:35,11:45]), (WOn,[11:45,11:50]), (MOn,[11:45,11:50]), (SOn,[11:50,12:00]), (MOff,[11:50,12:05]), (TOn,[11:50,12:00]), (WOff,[11:50,12:05]), (SOff,[12:00,12:10]), (TOff,[12:00,12:10]),  (MOn,[12:05,12:10]), (WOn,[12:05,12:10]) 
			\\
			\hline 
			4   &  (SOn,[12:15,12:20]), (TOn,[12:15,12:20]), (MOff,[12:15,12:25]), (WOff,[12:15,12:25]), (DOn,[12:15,12:20]), (IOn,[12:15,12:20]), (IOff,[12:20,12:50]), (DOff,[12:20,12:40]), (TOff,[12:20,12:40]), (SOff,[12:20,12:35]), (WOn,[12:25,12:35]), (MOn,[12:25,12:35]), (MOff,[12:35,12:50]), (SOn,[12:35,12:45]), (WOff,[12:35,12:50]), (TOn,[12:40,12:50]),  (DOn,[12:40,12:45]), (DOff,[12:45,12:55]), (SOff,[12:45,12:55]),  (TOff,[12:50,12:55]), (MOn,[12:50,12:55]), (WOn,[12:50,12:55]), (IOn,[12:50,12:55])
			\\
			\hline
		\end{tabular} 
	}
	\end{minipage}	
	\vspace{-0.1in}
\end{table*}

\vspace{-0.15in}
\subsection{Relations between Temporal Events}\vspace{-0.02in} 
We adopt the popular Allen's relations model  \cite{allen} and define three
basic temporal relations between events. Furthermore, to avoid the exact time mapping problem in Allen's relations, we adopt the \textit{buffer} idea from \cite{hdfs}, adding a tolerance \textit{buffer} $\epsilon$ to the relation's endpoints. However, we change the way $\epsilon$ is used in \cite{hdfs} to ensure the relations are \textit{mutually exclusive} (proof is in the full paper \cite{ho2020efficient}). 
 
Consider two temporal events $E_i$ and $E_j$, and their corresponding instances, $e_i=(\omega_i,[t_{s_i}, t_{e_i}])$ and $e_j=(\omega_j,[t_{s_j}, t_{e_j}])$. Let $\epsilon$ be a non-negative number ($\epsilon \ge 0$) representing the buffer size. The following relations can be defined between $E_i$ and $E_j$ through $e_i$ and $e_j$.

\hspace{-0.15in}\textbf{Definition 3.6} (Follows)  
$E_i$ and $E_j$ form a \textit{Follows} relation through $e_i$ and $e_j$, denoted as Follows($E_{i_{\triangleright e_i}}$,$E_{j_{\triangleright e_j}}$) or $E_{i_{\triangleright e_i}}$$\rightarrow$$E_{j_{\triangleright e_j}}$, iff \scalebox{0.9}{$t_{e_i}$$\pm$$\epsilon$$\le$$t_{s_j}$}.

\hspace{-0.15in}\textbf{Definition 3.7} (Contains) $E_i$ and $E_j$ form a \textit{Contains} relation through $e_i$ and $e_j$, denoted as Contains($E_{i_{\triangleright e_i}}$, $E_{j_{\triangleright e_j}}$) or $E_{i_{\triangleright e_i}}$$\succcurlyeq$$E_{j_{\triangleright e_j}}$, iff $(t_{s_i}$ $\le$ $t_{s_j})$ $\wedge$ $(t_{e_i} \pm \epsilon \ge t_{e_j})$.
  
\hspace{-0.15in}\textbf{Definition 3.8} (Overlaps) 
$E_i$ and $E_j$ form an \textit{Overlaps} relation through $e_i$ and $e_j$, denoted as Overlaps($E_{i_{\triangleright e_i}}$, $E_{j_{\triangleright e_j}}$) or $E_{i_{\triangleright e_i}}$ $\between$ $E_{j_{\triangleright e_j}}$, iff ${(t_{s_i} < t_{s_j})}$ $\wedge$ $(t_{e_i} \pm \epsilon < t_{e_j})$ $\wedge$ $(t_{e_i}-t_{s_j} \ge d_o \pm \epsilon)$, where $d_o$ is the minimal overlapping duration between two event instances, and $0 \le \epsilon \ll d_o$. 

The \textit{Follows} relation represents sequential occurrences of one event after another. For example, $E_{i_{\triangleright e_i}}$ is followed by $E_{j_{\triangleright e_j}}$ if the end time $t_{e_i}$ of $e_i$ occurs before the start time $t_{s_j}$ of $e_j$. Here, the buffer $\epsilon$ is used as a tolerance, i.e., the \textit{Follows} relation between $E_{i_{\triangleright e_i}}$ and $E_{j_{\triangleright e_j}}$ holds if $(t_{e_i} + \epsilon)$ or $(t_{e_i} - \epsilon)$ occurs before $t_{s_j}$. On the other hand, in a \textit{Contains} relation, one event occurs entirely within the timespan of another event. Finally, in an \textit{Overlaps} relation, the timespans of the two occurrences overlap each other. Table \ref{tbl:relations} illustrates the three temporal relations and their conditions.

\vspace{-0.18in}
\subsection{Temporal Pattern}\vspace{-0.02in}
\textbf{Definition 3.9} (Temporal sequence)
A list of $n$ event instances $S$$=$$<$$e_1,...,e_i,..., e_n$$>$ forms a \textit{temporal sequence} if the instances are chronologically ordered by their start times.  
Moreover, $S$ has size $n$, denoted as $|S| = n$.

\hspace{-0.15in}\textbf{Definition 3.10} (Temporal sequence database)
A set of temporal sequences forms a \textit{temporal sequence database} $\mathcal{D}_{\text{SEQ}}$ where each row $i$ contains a temporal sequence $S_i$. 

Table \ref{tbl:SequenceDatabase} shows the temporal sequence database $\mathcal{D}_{\text{SEQ}}$, created from the symbolic database $\mathcal{D}_{\text{SYB}}$ in Table \ref{tbl:SymbolDatabase}. 

\hspace{-0.15in}\textbf{Definition 3.11} (Temporal pattern) Let $\Re$$=$\{Follows, Contains, Overlaps\} be the set of temporal relations. A \textit{temporal pattern} $P$$=$$<$$(r_{12}, E_{1},$ $E_{2})$,...,$(r_{(n-1)(n)},E_{n-1},E_{n})$$>$ is a list of triples $(r_{\textit{ij}}$,$E_{i}$,$E_{j})$, each representing a relation $r_{\textit{ij}} \in \Re$ between two events $E_i$ and $E_j$.

Note that the relation $r_{\textit{ij}}$ in each triple is formed using the specific instances of $E_i$ and $E_j$. A temporal pattern that has $n$ events is called an $n$-event pattern. We use $E_i \in P$ to denote that the event $E_i$ occurs in $P$, and $P_1 \subseteq P$ to say that a pattern $P_1$ is a sub-pattern of $P$.

\hspace{-0.15in}\textbf{Definition 3.12} (Temporal sequence supports a pattern) Let $S$$=$$<$$e_1$,\\...,$e_i$,...,$e_n$$>$ be a temporal sequence. We say that $S$ \textit{supports} a temporal pattern $P$, denoted as $P \in S$, iff $|S| \ge 2$ $\wedge$ $\forall (r_{\textit{ij}},E_i,E_j) \in P, $ $\exists (e_l, e_m) \in S$ such that $r_{\textit{ij}}$ holds between $E_{i_{\triangleright e_l}}$ and $E_{j_{\triangleright e_m}}$.

If $P$ is supported by $S$, $P$ can be written as $P$$=$$<$$(r_{12}$, $E_{1_{\triangleright e_1}}$, $E_{2_{\triangleright e_2}})$, ..., $(r_{(n-1)(n)}$,$E_{{n-1}_{\triangleright e_{n-1}}}$, $E_{{n}_{\triangleright e_{n}}})$$>$, where the relation between two events in each triple is expressed using the event instances.
 
In Fig. \ref{fig:TimeSeries}, consider the sequence $S=<$$e_1$=(HighCO2, [6:00, 10:00]), $e_2$$=$(BoilerOn, [7:00, 8:00]), $e_3$$=$(LowCO2, [13:00, 15:00])$>$ representing the order of CO2 intensity and boiler events. Here, $S$ supports a 3-event pattern $P$$=$$<$(Contains, HighCO2$_{\triangleright e_1}$, BoilerOn$_{\triangleright e_2}$), (Follows, HighCO2$_{\triangleright e_1}$, LowCO2$_{\triangleright e_3}$), (Follows, BoilerOn$_{\triangleright e_2}$, LowCO2$_{\triangleright e_3}$)$>$. 

\textit{Maximal duration constraint}: Let $P \in S$ be a temporal pattern supported by the sequence $S$. The duration between the start time of the instance $e_1$, and the end time of the instance $e_n$ in $S$ must not exceed the predefined maximal time duration $t_{\max}$: $t_{e_n} - t_{s_1} \leq t_{\max}$.

The maximal duration constraint guarantees that the relation between any two events is temporally valid. This enables the pruning of invalid patterns. For example, under this constraint, a \textit{Follows} relation between a \textit{``Washer On''} event and a \textit{``Dryer On''} event in Table \ref{tbl:SequenceDatabase} happening one year apart should be considered invalid. 

\vspace{-0.12in}
\subsection{Frequent Temporal Pattern}\vspace{-0.02in}
Given a  temporal sequence database $\mathcal{D}_{\text{SEQ}}$, we want to find patterns that occur frequently in $\mathcal{D}_{\text{SEQ}}$. 
We use \textit{support} and \textit{confidence} \cite{omiecinski2003alternative} to measure the frequency and the likelihood of a pattern. 

\hspace{-0.15in}\textbf{Definition 3.13} (Support of a temporal event) 
The \textit{support} of a temporal event $E$ in $\mathcal{D}_{\text{SEQ}}$ is the number of sequences $S \in \mathcal{D}_{\text{SEQ}}$ which contain at least one instance $e$ of $E$. \vspace{-0.05in}
\begin{equation}
\small
\textit{supp}(E) = \lvert \{S \in \mathcal{D}_{\text{SEQ}}  \textit{ s.t. } \exists e \in S: E_{\triangleright e}  \} \rvert 
\label{eq:support1}
\vspace{-0.05in}
\end{equation}

\hspace{-0.15in}The \textit{relative support} of $E$ is the fraction between $\textit{supp}(E)$ and the size of $\mathcal{D}_{\text{SEQ}}$: \vspace{-0.07in}
\begin{equation}
\small
\textit{rel-supp}(E)= \textit{supp}(E) / |\mathcal{D}_{\text{SEQ}}| 
\label{eq:support2}
%\vspace{-0.05in}
\end{equation}

\hspace{-0.15in}Similarly, the support of a group of events $(E_1,..., E_n)$, denoted as $\textit{supp}(E_1,..., E_n)$, is the number of sequences $S \in \mathcal{D}_{\text{SEQ}}$ which contain at least one instance $(e_1,...,  e_n)$ of the event group. 

\hspace{-0.15in}\textbf{Definition 3.14} (Support of a temporal pattern)
The \textit{support} of a pattern $P$ is the number of sequences $S \in \mathcal{D}_{\text{SEQ}}$ that support $P$. \vspace{-0.02in}
\begin{equation}
\vspace{-0.05in}
\small
\textit{supp}(P) = \lvert \{S \in \mathcal{D}_{\text{SEQ}} \textit{ s.t. } P \in S\} \rvert 
\label{eq:support3}
\vspace{-0.02in}
\end{equation}

\hspace{-0.15in}The \textit{relative support} of $P$ in $\mathcal{D}_{\text{SEQ}}$ is the fraction \vspace{-0.05in} 
\begin{equation}
\vspace{-0.05in}
\small
\textit{rel-supp}(P)= \textit{supp(P)} / |\mathcal{D}_{\text{SEQ}}| 
\label{eq:support4}
\end{equation}   

\hspace{-0.15in}\textbf{Definition 3.15} (Confidence of an event pair)  
The \textit{confidence} of an event pair $(E_i, E_j)$ in $\mathcal{D}_{\text{SEQ}}$ is the fraction between $\textit{supp}(E_i, E_j)$ and the support of its most frequent event: \vspace{-0.05in} 
\begin{equation}
\vspace{-0.05in}
\small
\textit{conf}(E_i,E_j) = \frac{\textit{supp}(E_i, E_j)} {\max\{\textit{supp}(E_i), \textit{supp}(E_j) \}} 
\label{eq:eventpairconf}
\end{equation}

\hspace{-0.15in}\textbf{Definition 3.16} (Confidence of a temporal pattern)
The \textit{confidence} of a temporal pattern $P$ in $\mathcal{D}_{\text{SEQ}}$ is the fraction between $\textit{supp}(P)$ and the support of its most frequent event: \vspace{-0.05in} 
\begin{equation}
\vspace{-0.05in}
\small
\textit{conf}(P) = \frac{\textit{supp}(P)}{\max_{1 \leq k \leq |P|}\{\textit{supp}(E_k) \}} 
\label{eq:confidence}
\end{equation}
where $E_k \in P$ is a temporal event. 
Since the denominator in Eq. \eqref{eq:confidence} is the maximum support of the events in $P$, the confidence computed in Eq. \eqref{eq:confidence} is the \textit{minimum confidence} of a pattern $P$ in $\mathcal{D}_{\text{SEQ}}$,  
which is also called the \textit{all-confidence} as in \cite{omiecinski2003alternative}. 

Note that unlike association rules, temporal patterns do not have antecedents and consequents. Instead, they represent pair-wise temporal relations between events based on their temporal occurrences. Thus, while the \textit{support} and \textit{relative support} of event(s)/ pattern(s) defined in Eqs. \eqref{eq:support1} $-$ \eqref{eq:support4} follow the same intuition as the traditional support concept, indicating how frequently an event/ pattern occurs in a given database, the \textit{confidence} computed in Eqs. \eqref{eq:eventpairconf} $-$ \eqref{eq:confidence} instead represents the minimum likelihood of an event pair/ pattern, knowing the likelihood of its most frequent event.

\textbf{Frequent Temporal Pattern Mining from Time Series} \textbf{(FTP\\MfTS).} Given a set of univariate time series $\mathcal{X}=\{X_1,...,X_n\}$, let $\mathcal{D}_{\text{SEQ}}$ be the temporal sequence database obtained from ${\mathcal{X}}$, and $\sigma$ and $\delta$ be the support and  confidence thresholds, respectively. The FTPMfTS problem aims to find all temporal patterns $P$ that have high enough support and confidence in $\mathcal{D}_{\text{SEQ}}$: $\textit{supp}(P) \geq \sigma$ $\wedge$ $\textit{conf}(P) \geq \delta$. 
\vspace{-0.1in}
\section{Frequent Temporal Pattern Mining} \label{sec:FTPMfTSMining}
Fig. \ref{fig:framework} gives an overview of the FTPMfTS process which consists of $2$ phases. The first phase, \textit{Data Transformation}, converts a set of time series $\mathcal{X}$ into a symbolic database $\mathcal{D}_{\text{SYB}}$, and then converts $\mathcal{D}_{\text{SYB}}$ into a temporal sequence database $\mathcal{D}_{\text{SEQ}}$. The second phase, \textit{Frequent Temporal Pattern Mining}, mines frequent patterns which includes $3$ steps: (1) \textit{Frequent Single Event Mining}, (2) \textit{Frequent 2-Event Pattern Mining}, and (3) \textit{Frequent k-Event Pattern Mining} (\text{k}$>$$2$). The final output is a set of all frequent patterns in $\mathcal{D}_{\text{SEQ}}$.

\vspace{-0.1in}
\subsection{Data Transformation}\vspace{-0.02in}
\subsubsection{Symbolic Time Series Representation}
Given a set of time series $\mathcal{X}$, the symbolic representation of each time series $X \in \mathcal{X}$ is obtained by using a mapping function as in Def. 3.2. 

\begin{figure*}
	\setlength{\tabcolsep}{0pt}
	\begin{tabularx}{\linewidth}{ll}
		\begin{minipage}{.3\linewidth}
			\begin{minipage}{\linewidth}
				\resizebox{\linewidth}{.8\linewidth}{%
					\begin{tikzpicture}[node distance=-1.5cm, fill opacity=0.2]
\tikzstyle{arrow} = [black,opacity=1,thin,-{Triangle[angle=60:1.2mm]}]
\tikzstyle{line} = [black,opacity=1,thin,dotted]
\tikzstyle{box} = [rectangle,rounded corners, minimum width=5.6cm, minimum height=0.5cm, text centered,text width = 5.6cm, draw=black, text opacity=1]
\tikzstyle{bounding1} = [rectangle, dotted, minimum width=7cm, minimum height=2cm, draw=black, fill=blue, text opacity=1]
\tikzstyle{bounding2} = [rectangle,dotted, minimum width=7cm, minimum height=2cm, draw=black, fill=green, text opacity=1]
\tikzstyle{bounding3} = [rectangle,dotted, minimum width=8cm, minimum height=4.8cm, draw=black, fill=teal, text opacity=1]
\tikzstyle{textlabel} = [text centered,text width = 3cm, text opacity=1, rotate=270, yshift = -12]
\tikzstyle{textlabel2} = [text centered,text width = 3cm, text opacity=1, rotate=90, yshift = -10]

\node (A) [box, fill=red] {\small Set of Time Series $\mathcal{X}$};
%		\node [boundingA, left of=boundA] {\small {Data Transformation}} ;
%

\node (B) [box, below of=A, yshift=-2.4cm] {\small Symbolic Time Series Representation};
\node (C) [box, below of=B, yshift=-2.4cm] {\small Temporal Sequence Database Conversion};
%		\node (boundB) [bounding2, above of=A, yshift=-2.8cm] {};
\node (D) [box, below of=C, yshift=-2.5cm] {\small Frequent Single Event Mining};
\node (E) [box, below of=D, yshift=-2.4cm] {\small Frequent 2-Event Pattern Mining};
\node (F) [box, below of=E, yshift=-2.4cm] {\small Frequent k-Event Pattern Mining (k $>2$)};

\node (boundingC) [bounding3, fit=(B) (C) (D) (E) (F),  yshift=.05cm] {};
\draw [line] ($(boundingC.north) + (3.2cm,0)$) -- node [black, textlabel2] {\small FTPMfTS Process}  ($(boundingC.south) + (3.2cm,0)$);

\node (boundingA)[bounding1, fit=(B) (C), xshift=-.4cm] {};
\draw [line] ($(boundingA.north) - (2.6cm,0)$) -- node [black, textlabel] {\small \phantom{abc}Data\phantom{abc} Transformation }  ($(boundingA.south) - (2.6cm,0)$);

\node (boundingB)[bounding2, fit=(D) (E) (F), xshift=-.4cm] {};
\draw [line] ($(boundingB.north) - (2.6cm,0)$) -- node [black, textlabel] {\small Temporal Patterns Mining (HTPGM)}  ($(boundingB.south) - (2.6cm,0)$);

\node (G) [box, fill=red, below of=F, yshift=-2.4cm] {\small Frequent Temporal Patterns};
\draw[arrow] (A) -- (B) ;
\draw[arrow] (B) -- node[anchor=west] {$\mathcal{D}_{\text{SYB}}$} (C) ;
\draw[arrow] (C) -- node[anchor=west] {$\mathcal{D}_{\text{SEQ}}$} (D) ;
\draw[arrow] (D) -- (E) ;
\draw[arrow] (E) -- (F) ;
\draw[arrow] (F) -- (G) ;
\end{tikzpicture}
				}
				\caption{The FTPMfTS process}
				\label{fig:framework}
			\end{minipage}
			\\
			\\
			\begin{minipage}{\linewidth}
				\begin{subfigure}[b]{\linewidth}
					\resizebox{0.9\linewidth}{.9cm}{%
						\begin{tikzpicture}[fill opacity=.2]
\definecolor{arrowcolor}{rgb}{.27,.45,.77}
\tikzstyle{arrow} = [arrowcolor,opacity=1,thin, {Triangle[angle=60:1.2mm]}-{Triangle[angle=60:1.2mm]}]
\tikzstyle{roundrect} = [rectangle,rounded corners, minimum width=6cm, minimum height=0.8cm, text centered,text width =1.4cm, draw=white, fill=white, text opacity=1]
\tikzstyle{roundrect3} = [rectangle,rounded corners, minimum width=3cm, minimum height=0.8cm, text centered,text width =1.4cm, draw=black, fill=red, text opacity=1, fill opacity=.15]
\tikzstyle{roundrect4} = [rectangle,rounded corners, minimum width=3cm, minimum height=0.8cm, text centered,text width =1.4cm, draw=black, fill=blue, text opacity=1, fill opacity=.3]
\tikzstyle{line} = [black,opacity=1,thin, {Circle[scale=0.4]}-{Circle[scale=0.4]}]

\node (bbox) [roundrect] {};
\node (anode) [roundrect3, left of=bbox, xshift=-.5cm] {};
\node (cnode) [roundrect4,  left of=anode, xshift = 4cm] {};
\node [text opacity=1] at ($(anode.west)!0.5!(cnode.west)$) {$S_1$} ;
\node [text opacity=1] at ($(anode.east)!0.5!(cnode.east)$) {$S_2$} ;
%\draw [arrow] ($(cnode.west)+(0,1.2)$) -- node [black, below,yshift=2] {\small t\textsubscript{max}}  ($(anode.east)+(0,1.2)$);
\draw [arrow] ($(anode.west)+(0,.6)$) -- node [black, above, yshift=-2] {\small $t$}   ($(anode.east)+(0,.6)$);
\draw [arrow] ($(cnode.west)+(0,.6)$) -- node [black, above, yshift=-2] {\small $t$}   ($(cnode.east)+(0,.6)$);

\draw [line] ($(anode.east)+(-.01,.2)$) -- node [black, above,yshift=-2.5] {\tiny SOn}  ($(anode.east)+(-1,.2)$);
\draw [line] ($(anode.east)+(-.01,.0)$) -- node [black, above,yshift=-2.5] {\tiny TOn}  ($(anode.east)+(-1,.0)$);
\draw [line] ($(cnode.west)+(.01,-.1)$) -- node [black, above,yshift=-2.5] {\tiny MOn}  ($(cnode.west)+(.8,-.1)$);
\draw [line] ($(cnode.west)+(.01,-.3)$) -- node [black, above,yshift=-2.5] {\tiny WOn}  ($(cnode.west)+(.8,-.3)$);
\end{tikzpicture}
					}
				\vspace{0.05in}
					\caption{With no overlapping}
					\label{fig:basicSplit}
				\end{subfigure}
				\begin{subfigure}[b]{\linewidth}
					\resizebox{0.9\linewidth}{1.1cm}{%
						\begin{tikzpicture}[fill opacity=.2]
\definecolor{arrowcolor}{rgb}{.27,.45,.77}
\tikzstyle{arrow} = [arrowcolor,opacity=1,thin, {Triangle[angle=60:1.2mm]}-{Triangle[angle=60:1.2mm]}]
\tikzstyle{roundrect} = [rectangle,rounded corners, minimum width=6cm, minimum height=.8cm, text centered,text width =1.4cm, draw=white, fill=white, text opacity=1]
\tikzstyle{roundrect3} = [rectangle, rounded corners, minimum width=3cm, minimum height=.8cm, text centered,text width =1.4cm, draw=black, fill=red, text opacity=1, dashed, fill opacity=.15]
\tikzstyle{roundrect4} = [rectangle,rounded corners, minimum width=3cm, minimum height=.8cm, text centered,text width =1.4cm, draw=black, fill=blue, text opacity=1, fill opacity=.3]
\tikzstyle{line} = [black,opacity=1,thin, {Circle[scale=0.4]}-{Circle[scale=0.4]}]
	
\node (bbox) [roundrect] {};
\node (anode) [roundrect3, left of=bbox, xshift=-.5cm] {};
%\draw [rounded corners] (0,0) -- (4,0) -- (4,4) -- (0,4);
\draw [rounded corners] ($(anode.south east) + (-0.5,.007)$) --  ($(anode.south west) + (.007,.007)$) -- ($(anode.north west) + (.007,-.007)$)--  ($(anode.north east) - (-0.5,.007)$);

\node (cnode) [roundrect4,  left of=anode, xshift = 3cm] {};
\node [text opacity=1] at ($(anode.west)!0.6!(cnode.west)$) {$S_1$} ;
\node [text opacity=1] at ($(anode.east)!0.7!(cnode.east)$) {$S_2$} ;
\draw [arrow] ($(cnode.west)+(0,.48)$) -- node [black, above,yshift=-2] {\small $t_{\text{ov}}$}  ($(anode.east)+(0,.48)$);
\draw [arrow] ($(anode.west)+(0,.8)$) -- node [black, above, yshift=-2] {\small $t$}   ($(anode.east)+(0,.8)$);
\draw [arrow] ($(cnode.west)+(0,1)$) -- node [black, above, yshift=-2] {\small $t$}   ($(cnode.east)+(0,1)$);

\draw [line] ($(cnode.west)+(.01,.2)$) -- node [black, above,yshift=-2.5] {\tiny SOn}  ($(cnode.west)+(1,.2)$);
\draw [line] ($(cnode.west)+(.01,0)$) -- node [black, above,yshift=-2.5] {\tiny TOn}  ($(cnode.west)+(1,0)$);
\draw [line] ($(anode.east)+(.01,-.1)$) -- node [black, above,yshift=-2.5] {\tiny MOn}  ($(anode.east)+(.8,-.1)$);
\draw [line] ($(anode.east)+(.01,-.3)$) -- node [black, above,yshift=-2.5] {\tiny WOn}  ($(anode.east)+(.8,-.3)$);
\end{tikzpicture}
					}
				\vspace{0.05in}
					\caption{With overlapping}
					\label{fig:overlaptransaction2}
					\vspace{-0.05in}
				\end{subfigure}
				\caption{Splitting strategy}
				\label{fig:splitting}
			\end{minipage}
		\end{minipage}&
		\hspace{0.3in}
		\begin{minipage}{.65\linewidth}
				\includegraphics[width=\textwidth,height=8.0cm]{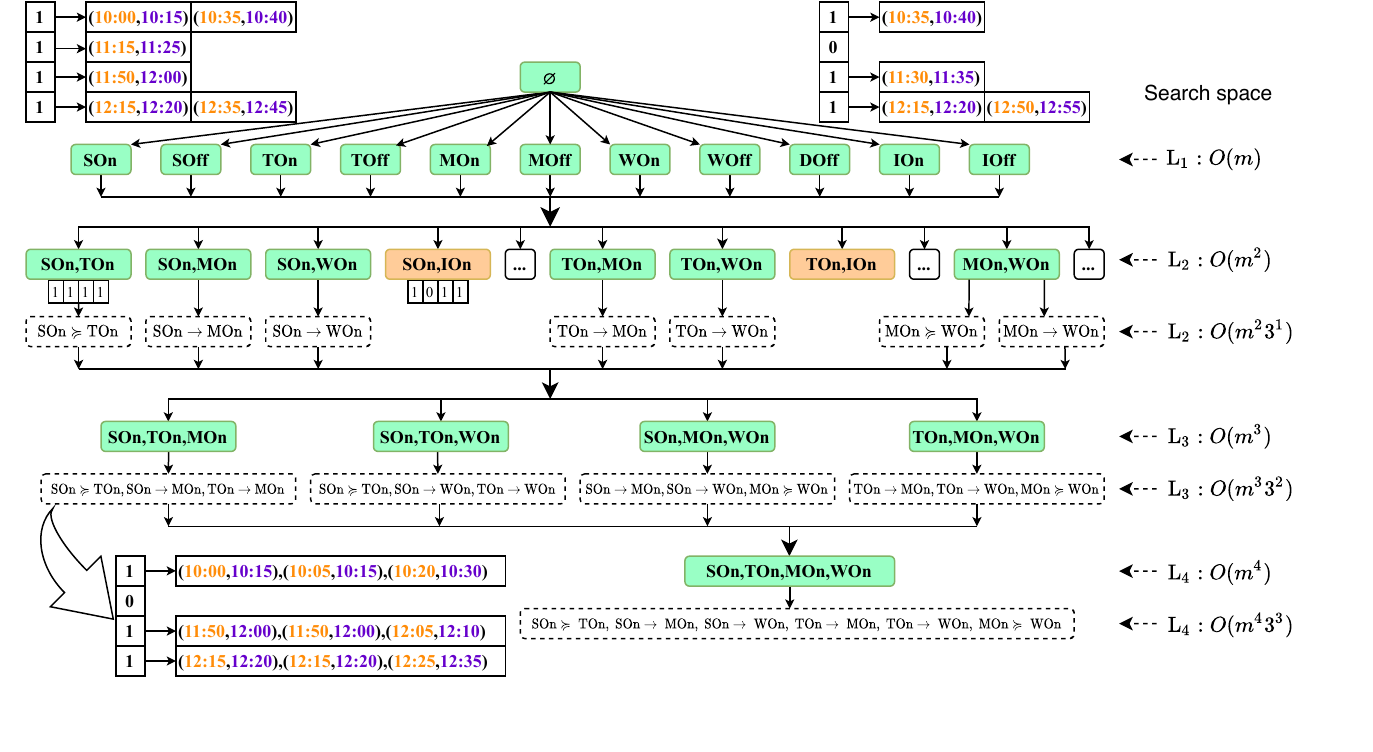}
			\vspace{-0.15in}
				\caption{A Hierarchical Pattern Graph for Table  \ref{tbl:SequenceDatabase}}
				\label{fig:patternTree}
		\end{minipage}
	\end{tabularx}
\vspace{-0.2in}
\end{figure*}

\vspace{-0.05in}\subsubsection{Temporal Sequence Database Conversion}
To convert $\mathcal{D}_{\text{SYB}}$ to $\mathcal{D}_{\text{SEQ}}$, a straightforward approach is to split the symbolic series in $\mathcal{D}_{\text{SYB}}$ into equal-length sequences, each belongs to a row in $\mathcal{D}_{\text{SEQ}}$. For example, if each symbolic series in Table \ref{tbl:SymbolDatabase} is split into $4$ sequences, then each sequence will last for $40$ minutes. The first sequence $S_1$ of $\mathcal{D}_{\text{SEQ}}$ therefore contains temporal events of S, T, M, W, D, and I from 10:00 to 10:40. The second sequence $S_2$ contains events from 10:45 to 11:25, and similarly for $S_3$ and $S_4$.

However, the splitting can lead to a potential loss of temporal patterns. The loss happens when a \textit{splitting point} accidentally divides a temporal pattern into different sub-patterns, and places these into separate sequences. We explain this situation in Fig. \ref{fig:basicSplit}.  
Consider $2$ sequences $S_1$ and $S_2$, each of length $t$. Here, the splitting point divides a pattern of $4$ events, \{SOn, TOn, MOn, WOn\}, into two sub-patterns, in which SOn and TOn are placed in $S_1$, and MOn and WOn in $S_2$. This results in the loss of this $4$-event pattern which can be identified only when all $4$ events are in the same sequence.

To prevent such a loss, we propose a \textit{splitting strategy} using overlapping sequences. Specifically, two consecutive sequences are overlapped by a duration $t_{\text{ov}}$: $0 \le t_{\text{ov}} \le t_{\max}$, where $t_{\max}$ is the \textit{maximal duration} of a temporal pattern. The value of $t_{\text{ov}}$ decides how large the overlap between $S_i$ and $S_{i+1}$ is: $t_{\text{ov}}=0$ results in no overlap, i.e., no redundancy, but with a potential loss of patterns, while $t_{\text{ov}}=t_{\max}$ creates large overlaps between sequences, i.e., high redundancy, but all patterns are preserved.  
As illustrated in Fig. \ref{fig:overlaptransaction2}, the overlapping between $S_1$ and $S_2$ keeps the $4$ events together in the same sequence $S_2$, and thus helps preserve the pattern. 

\vspace{-0.15in}
\subsection{Frequent Temporal Patterns Mining}\vspace{-0.02in}
We now present our method, called Hierarchical Temporal Pattern Graph Mining (HTPGM), to mine frequent temporal patterns from $\mathcal{D}_{\text{SEQ}}$. The main novelties of HTPGM are: a) the use of efficient data structures, i.e., the proposed Hierarchical Pattern Graph and \textit{bitmap indexing}, to enable fast computations of support and confidence, and b) the proposal of two groups of pruning techniques based on the Apriori principle and the temporal transitivity property of temporal events. 
In Section \ref{sec:MI}, we introduce an approximate version of HTPGM based on mutual information to further optimize the mining process. We first discuss the data structures used in HTPGM.

\textbf{Hierarchical Pattern Graph (HPG):} We use a hierarchical graph structure, called the \textit{Hierarchical Pattern Graph}, to keep track of the frequent events and patterns found in each mining step. The HPG allows HTPGM to mine iteratively (e.g., 2-event patterns are mined based on frequent single events, 3-event patterns are mined based on 2-event patterns, and so on) and perform effective pruning. Fig. \ref{fig:patternTree} shows the HPG built from $\mathcal{D}_{\text{SEQ}}$ in Table \ref{tbl:SequenceDatabase}: the root is the empty set $\emptyset$, and each level L$_k$ maintains frequent $k$-event patterns. As HTPGM proceeds, HPG is constructed gradually. We explain this process for each mining step.

\textbf{Efficient bitmap indexing:} We use \textit{bitmaps} to index the occurrences of events and patterns in $\mathcal{D}_{\text{SEQ}}$, enabling fast computations of support and confidence. Specifically, each event $E$ or pattern $P$ found in $\mathcal{D}_{\text{SEQ}}$ is associated with a \textit{bitmap} indicating where $E$ or $P$ occurs. Each \textit{bitmap} $b$ has length $\mid$$\mathcal{D}_{\text{SEQ}}$$\mid$ (i.e., the number of sequences), and has value $b[i]=1$ if $E$ or $P$ is present in sequence $i$ of $\mathcal{D}_{\text{SEQ}}$, or $b[i]=0$ otherwise. An example \textit{bitmap} can be seen at L$_1$ in Fig. \ref{fig:patternTree}. The event IOn has the \textit{bitmap} $b_{\text{IOn}}=$ [1,0,1,1], indicating that IOn occurs in all but the second sequence of $\mathcal{D}_{\text{SEQ}}$. 

Constructing the \textit{bitmap} is also done step by step. For single events in $\mathcal{D}_{\text{SEQ}}$, \textit{bitmaps} are built by scanning $\mathcal{D}_{\text{SEQ}}$ only once. Algorithm \ref{algorithmHTPGM} provides the pseudo-code of HTPGM. The details are explained in each mining step.  

\SetNlSty{}{}{:} 	\vspace{-0.1in}	
\begin{algorithm}
	\algsetup{linenosize=\tiny}
	\SetInd{0.5em}{0.5em}
	\small
	\DontPrintSemicolon
	\caption{Hierarchical Temporal Pattern Graph Mining} 
	\label{algorithmHTPGM}
	
	\KwInput{Temporal sequence database $\mathcal{D_{\text{SEQ}}}$, a support threshold $\sigma$, a confidence threshold $\delta$}
	\KwOutput{The set of frequent temporal patterns $P$}
	
		\nonl // Mining frequent single events \;
		\ForEach{\textit{event} $E_i \in \mathcal{D_{\text{SEQ}}}$}{
			$\textit{supp}(E_i) \leftarrow \text{countBitmap}(b_{E_i})$;\;  
			\If{$\textit{supp}(E_i) \geq \sigma$}{
				Insert $E_i$ to \textit{1Freq};
			}
		}	
		
		\nonl // Mining frequent 2-event patterns \;
		EventPairs $\leftarrow$ Cartesian(\textit{1Freq},\textit{1Freq});\;  
		FrequentPairs $\leftarrow$ $\emptyset$;\;
		\ForEach{$(E_i,E_j)$ in EventPairs}{
			$b_{ij}$ $\leftarrow$ AND($b_{E_i},$$b_{E_j}$);\;
			$\textit{supp}(E_i,E_j) \leftarrow \text{countBitmap}(b_{ij})$;\;  
			\If{$\textit{supp}(E_i,E_j) \geq \sigma$}{
				FrequentPairs $\leftarrow$ Apply\_Lemma3($E_i,E_j$);\;
			}
		}
		\ForEach{$(E_i,E_j)$ in FrequentPairs}{
			Retrieve event instances;\;  
			Check frequent relations;\;  
		}	
		
	\nonl // Mining frequent k-event patterns \;
		Filtered1Freq $\leftarrow$ Transitivity\_Filtering(1Freq); \hspace{0.1in}//Lemmas \ref{lem:transitivity}, \ref{lem:filter}\;
		kEventCombinations $\leftarrow$ Cartesian(\textit{Filtered1Freq},\textit{(k-1)Freq});\;  
		FrequentkEvents $\leftarrow$ Apriori\_Filtering(kEventCombinations);\;
		\ForEach{\textit{kEvents} in FrequentkEvents}{
			Retrieve relations;\;  
			Iteratively check frequent relations;  \hspace{0.35in} //Lemmas \ref{lem:transitivity}, \ref{lem5}, \ref{lem6}\;
		}	
\end{algorithm}

\vspace{-0.05in}
\subsection{Mining Frequent  Single Events}\vspace{-0.02in}
The first step in HTPGM is to find frequent single events (Alg. \ref{algorithmHTPGM}, lines 1-4) which is easily done using the \textit{bitmap}. For each event $E_i$ in $\mathcal{D}_{\text{SEQ}}$, the support \textit{supp}($E_i$) is computed by counting the number of set bits in \textit{bitmap} $b_{E_i}$, and comparing against $\sigma$. Note that for single events, \textit{confidence} is not considered since it is always $1$. 

After this step, the set \textit{1Freq} containing frequent single events is created to build L$_1$ of HPG. We illustrate this process using Table \ref{tbl:SequenceDatabase}, with $\sigma=0.7$ and $\delta=0.7$. Here, \textit{1Freq} contains $11$ frequent events, each belongs to one node in L$_1$. The event DOn is not frequent (only appears in sequences $2$ and $4$), and is thus omitted. Each L$_1$ node has a unique event name, a \textit{bitmap}, and a list of instances corresponding to that event (see SOn at L$_1$).  

\textbf{Complexity:} The complexity of finding frequent single events is $O(m \cdot$$\mid$$\mathcal{D}_{\text{SEQ}}$$\mid)$, where $m$ is the number of distinct events. 

\textbf{Proof.} \textit{Detailed proofs of all complexities, lemmas and theorems in this article can be found in the Appendix of the full paper \cite{ho2020efficient}.}

\vspace{-0.1in}
\subsection{Mining Frequent  2-event Patterns}\label{sec:2freq} \vspace{-0.02in}
\subsubsection{Search space of HTPGM}
The next step in HTPGM is to mine frequent 2-event patterns. A straightforward approach would be to enumerate all possible event pairs, and check whether each pair can form frequent patterns. However, this \textit{naive} approach is very expensive. Not only does it need to repeatedly scan $\mathcal{D}_{\text{SEQ}}$ to check each combination of events, the complex relations between events also add an extra exponential factor $3^{h^2}$ to the $m^h$ number of possible candidates, creating a very large search space that makes the approach infeasible.

\begin{lem}\label{lem1}\vspace{-0.05in}
Let $m$ be the number of distinct events in $\mathcal{D}_{\text{SEQ}}$, and $h$ be the longest length of a temporal pattern. The total number of temporal patterns in HPG from L$_1$ to L$_h$ is $O(m^h3^{h^2})$. 
\vspace{-0.05in}
\end{lem}
Lemma \ref{lem1} shows the driving factors of HTPGM's exponential search space (proof in \cite{ho2020efficient}): the number of events ($m$), the max pattern length ($h$), and the number of temporal relations ($3$). A dataset of just a few hundred events can create a search space with billions of candidate patterns. The optimizations and approximation proposed in the following sections help mitigate this problem.

\vspace{-0.02in}\subsubsection{Two-steps filtering approach}
Given the huge set of pattern candidates, it is expensive to check their support and confidence. We propose a \textit{filtering approach} to reduce the unnecessary candidate checking. Specifically, at any level $l$ $(l \ge 2)$ in HPG, the mining process is divided into two steps: (1) it first finds frequent nodes (i.e., remove infrequent combinations of events), (2) it then generates temporal patterns only from frequent nodes. The correctness of this filtering approach is based on the Apriori-inspired lemmas below.
\begin{lem}\label{lem2}\vspace{-0.05in}
	Let $P$ be a 2-event pattern formed by an event pair $(E_i, E_j)$. Then, $\textit{supp}(P) \le \textit{supp}(E_i,E_j)$. 
	\vspace{-0.05in}
\end{lem} 

From Lemma \ref{lem2}, the support of a pattern is at most the support of its events. Thus, infrequent event pairs cannot form frequent patterns and thereby, can be safely pruned.

\begin{lem}\label{lem3}	\vspace{-0.05in}
	Let $(E_i, E_j)$ be a pair of events occurring in a 2-event pattern $P$.  Then \textit{conf}($P$) $\le$ \textit{conf}($E_i,E_j$).
\vspace{-0.05in}
\end{lem}

From Lemma \ref{lem3}, the confidence of a pattern $P$ is always at most the confidence of its events. Thus, a low-confidence event pair cannot form any high-confidence patterns and therefore, can be safely pruned.
We note that the Apriori principle has already been used in other work, e.g., \cite{tpminer,hdfs}, for mining optimization. However, they only apply this principle to the support (Lemma \ref{lem2}), while we further extend it to the confidence (Lemma \ref{lem3}). 
Applying Lemmas \ref{lem2} and \ref{lem3} to the first filtering step will remove infrequent or low-confidence event pairs,  
reducing the candidate patterns of HTPGM. We detail this filtering below. 

\textbf{Step 2.1. Mining frequent event pairs:}
This step finds frequent event pairs in $\mathcal{D}_{\text{SEQ}}$, using the set \textit{1Freq} found in L$_1$ of HPG (Alg. \ref{algorithmHTPGM}, lines 5-11). 
First, HTPGM generates all possible event pairs by calculating the Cartesian product \textit{1Freq} $\times$ \textit{1Freq}. Next, for each pair $(E_i, E_j)$, the joint \textit{bitmap} $b_{\textit{ij}}$ (representing the set of sequences where both events occur) is computed by \textit{ANDing} the two individual bitmaps: $b_{\textit{ij}}=\textit{AND}(b_{E_i},b_{E_j})$. Finally, HTPGM computes the support $\textit{supp}(E_i, E_j)$ by counting the set bits in $b_{\textit{ij}}$, and comparing against $\sigma$. 
If $\textit{supp}(E_i, E_j) \ge \sigma$, $(E_i, E_j)$ has high enough support. Next, $(E_i, E_j)$ is further filtered using Lemma \ref{lem3}: $(E_i, E_j)$ is selected only if its confidence is at least $\delta$. After this step, only frequent and high-confidence event pairs remain and form the nodes in L$_2$. 

\textbf{Step 2.2. Mining frequent 2-event patterns:} This step finds frequent 2-event patterns from the nodes in L$_2$ (Alg. \ref{algorithmHTPGM}, lines 12-14). For each node $(E_i,E_j) \in$ L$_2$, we use the \textit{bitmap} $b_{\textit{ij}}$ to retrieve the set of sequences $\mathcal{S}$ where both events are present. 
Next, for each sequence $S \in \mathcal{S}$, the pairs of event instances $(e_i,e_j)$ are extracted, and the relations between them are verified. The support and confidence of each relation $r(E_{i_{\triangleright e_i}},E_{j_{\triangleright e_j}})$ are computed and compared against the thresholds, after which only frequent relations are selected and stored in the corresponding node in L$_2$. Examples of the relations in L$_2$ can be seen in Fig. \ref{fig:patternTree}, e.g., node (SOn, TOn). 

Step 2.2 results in two different sets of nodes in L$_2$. The first set contains nodes that have frequent events but do not have any frequent patterns. These nodes (colored in brown in Fig. \ref{fig:patternTree}) are removed from L$_2$. The second set contains nodes that have both frequent events and frequent patterns (colored in green), which remain in L$_2$ and are used in the subsequent mining steps.

\textbf{Complexity:} Let $m$ be the number of frequent single events in L$_1$, and $i$ be the average number of event instances of each frequent event. The complexity of frequent 2-event pattern mining is $O(m^2 i^2\mid$$\mathcal{D}_{\text{SEQ}}$$\mid$$^2)$. 

\vspace{-0.05in}
\subsection{Mining Frequent k-event Patterns}\vspace{-0.02in}
Mining frequent k-event patterns ($k \ge 3$) follows a similar process as 2-event patterns, with additional prunings based on the transitivity property of temporal relations.     

\textbf{Step 3.1. Mining frequent k-event combinations:} This step finds frequent k-event combinations in L$_k$ (Alg. \ref{algorithmHTPGM}, lines 15-17).

Let \textit{(k-1)Freq} be the set of frequent (k-1)-event combinations found in L$_{k-1}$, and \textit{1Freq} be the set of frequent single events in L$_1$. To generate all k-event combinations, the typical process is to compute the Cartesian product: \textit{(k-1)Freq} $\times$ \textit{1Freq}. However, we observe that using \textit{1Freq} to generate k-event combinations at L$_k$ can create redundancy, since \textit{1Freq} might contain events that when combined with nodes in L$_{k-1}$, result in combinations that clearly cannot form any frequent patterns. To illustrate this observation, consider node IOn at L$_1$ in Fig. \ref{fig:patternTree}. Here, IOn is a frequent event, and thus, can be combined with frequent nodes in L$_2$ such as (SOn, TOn) to create a 3-event combination (SOn, TOn, IOn). However, (SOn, TOn, IOn) cannot form any frequent 3-event patterns, since IOn is not present in any frequent 2-event patterns in L$_2$. To reduce the redundancy, the combination (SOn, TOn, IOn) should not be created in the first place. We rely on the \textit{transitivity property} of temporal relations to identify such event combinations.  

\begin{lem} \label{lem:transitivity}\vspace{-0.05in}
Let $S=<e_1$,..., $e_{n-1}>$ be a temporal sequence that supports an (n-1)-event pattern $P=<(r_{12}$, $E_{1_{\triangleright e_1}}$, $E_{2_{\triangleright e_2}})$,..., $(r_{(n-2)(n-1)}$, $E_{{n-2}_{\triangleright e_{n-2}}}$, $E_{{n-1}_{\triangleright e_{n-1}}})>$. Let $e_n$ be a new event instance added to $S$ to create the temporal sequence $S^{'}$$=$$<e_1, ..., e_{n}>$. 

The set of temporal relations $\Re$ is transitive on $S^{'}$: $\forall e_i \in S^{'}$, $i < n$, $\exists r \in \Re$ s.t. $r(E_{i_{\triangleright e_i}}$,$E_{n_{\triangleright e_n}})$ holds.
		
\vspace{-0.05in}
\end{lem}
Lemma \ref{lem:transitivity} says that given a temporal sequence $S$, a new event instance added to $S$ will always form at least one temporal relation with existing instances in $S$. This is due to the temporal transitivity property, which can be used to prove the following lemma.

\begin{lem}\label{lem:filter}\vspace{-0.05in}
	Let $N_{k-1}=(E_1,...,E_{k-1})$ be a frequent (\textit{k-1})-event combination, and $E_k$ be a frequent single event. The combination $N_k= N_{k-1} \cup E_k$ can form frequent k-event temporal patterns if $\forall E_i \in N_{k-1}$, $\exists r \in \Re$ s.t. $r(E_i,E_k)$ is a frequent temporal relation. 
	\vspace{-0.05in}
\end{lem}
From Lemma \ref{lem:filter}, 
only single events in L$_1$ that occur in L$_{k-1}$ should be used to create k-event combinations.
Using this result, a filtering on \textit{1Freq} is performed before calculating the Cartesian product. Specifically, from the nodes in L$_{k-1}$, we extract the distinct single events $D_{k-1}$, and \textit{intersect} them with \textit{1Freq} to remove redundant single events: \textit{Filtered1Freq} = $D_{k-1}$ $\cap$ \textit{1Freq}. 
Next, the Cartesian product \textit{(k-1)Freq} $\times$ \textit{Filtered1Freq} is calculated to generate k-event combinations. Finally, we apply Lemmas \ref{lem2} and \ref{lem3} to select frequent and high-confidence k-event combinations \textit{kFreq} to form L$_{k}$.

\textbf{Step 3.2 Mining frequent k-event patterns:} This step finds frequent k-event patterns from the nodes in L$_k$ (Alg. \ref{algorithmHTPGM}, lines 18-20). Unlike 2-event patterns, determining the relations in a k-event combination ($k \ge 3$) is much more expensive, as it requires to verify the frequency of $\frac{1}{2}k(k-1)$ triples. To reduce the cost of relation checking, we propose an iterative verification method that relies on the \textit{transitivity property} and the Apriori principle.  

\begin{lem}\label{lem5}\vspace{-0.05in}
	Let $P$ and $P^{'}$ be two temporal patterns. If $P^{'} \subseteq P$, then \textit{conf}($P^{'}$) $ \geq$ \textit{conf}($P$).
	\vspace{-0.05in}
\end{lem} 

\begin{lem} \label{lem6}\vspace{-0.05in}
Let $P$ and $P^{'}$ be two temporal patterns. If $P^{'} \subseteq P$ and $\frac{\textit{supp}(P^{'})}{\max_{1 \leq k \leq \mid P \mid}\{\textit{supp}(E_k)\}}_{E_k \in P} \leq \delta$, then \textit{conf($P$)} $\leq \delta$.
\vspace{-0.05in}
\end{lem}

Lemma \ref{lem5} says that, the confidence of a pattern $P$ is always at most the confidence of its sub-patterns. Consequently, from Lemma \ref{lem6}, a temporal pattern $P$ cannot be high-confidence if any of its sub-patterns are low-confidence.

Let $N_{k-1}=(E_1,...,E_{k-1})$ be a node in L$_{k-1}$, $N_1=(E_k)$ be a node in L$_1$, and $N_k=N_{k-1} \cup N_1 = (E_1,...,E_k)$ be a node in L$_k$. To find k-event patterns for $N_k$, we first retrieve the set $P_{k-1}$ containing frequent (k-1)-event patterns in node $N_{k-1}$. 
Each $p_{k-1} \in P_{k-1}$ is a list of $\frac{1}{2}(k-1)(k-2)$ triples: $\{(r_{12}$, $E_{1_{\triangleright e_1}}$, $E_{2_{\triangleright e_2}})$,...,$(r_{(k-2)(k-1)}$, $E_{{k-2}_{\triangleright e_{k-2}}}$, $E_{{k-1}_{\triangleright e_{k-1}}})\}$. We iteratively verify the possibility of $p_{k-1}$ forming a frequent k-event pattern with $E_k$ as follows. 

We first check whether the triple $(r_{(k-1)k}$, $E_{{k-1}_{\triangleright e_{k-1}}}$, $E_{k_{\triangleright e_{k}}})$ is frequent and high-confidence by accessing the node $(E_{k-1},E_k)$ in L$_2$. If the triple is not frequent (using Lemmas \ref{lem:transitivity} and \ref{lem:filter}) or high-confidence (using Lemmas \ref{lem:transitivity}, \ref{lem5}, and \ref{lem6}), the verifying process stops immediately for $p_{k-1}$. Otherwise, it continues on the triple $(r_{(k-2)k}$, $E_{{k-2}_{\triangleright e_{k-2}}}$, $E_{k_{\triangleright e_{k}}})$, until it reaches $(r_{1k}$, $E_{1_{\triangleright e_1}}$, $E_{k_{\triangleright e_{k}}})$. 

We note that the transitivity property of temporal relations has been exploited in \cite{moskovitch2015fast} to generate new relations. Instead, we use this property to prune unpromising candidates (Lemmas \ref{lem:transitivity}, \ref{lem:filter}, \ref{lem5}, \ref{lem6}).

\textbf{Complexity:} 
Let $r$ be the average number of frequent (k-1)-event patterns in L$_{k-1}$. The complexity of frequent k-event pattern mining is $O($$\mid$$\textit{1Freq}$$\mid$ $\cdot$ $\mid$$L_{k-1}$$\mid$ $\cdot$ $r$ $\cdot$ $k^2$$\cdot$$\mid$$\mathcal{D}_{\text{SEQ}}$$\mid$$)$.

\textbf{HTPGM overall complexity:} Throughout this section, we have seen that HTPGM complexity depends on the size of the search space $(O(m^h3^{h^2}))$ and the complexity of the mining process itself, i.e., $O(m \cdot$$\mid$$\mathcal{D}_{\text{SEQ}}$$\mid)$ $+$ $O(m^2 i^2\mid$$\mathcal{D}_{\text{SEQ}}$$\mid$$^2)$ $+$ $O($$\mid$$\textit{1Freq}$$\mid$ $\cdot$ $\mid$$L_{k-1}$$\mid$ $\cdot$ $r$ $\cdot$ $k^2$$\cdot$$\mid$$\mathcal{D}_{\text{SEQ}}$$\mid$$)$.  
While the parameters $m$, $h$, $i$, $r$ and $k$ depend on the number of time series, others such as $\mid$$\textit{1Freq}$$\mid$, $\mid$$L_{k-1}$$\mid$ and $\mid$$\mathcal{D}_{\text{SEQ}}$$\mid$ also depend on the number of temporal sequences. Thus, given a dataset, HTPGM complexity is driven by two main factors: the number of time series and the number of temporal sequences.

\vspace{-0.05in}
\section{Approximate HTPGM}\label{sec:MI}\vspace{-0.02in}
\subsection{Correlated Symbolic Time Series}\vspace{-0.02in}
Let $X_S$ and $Y_S$ be the symbolic series representing the time series $X$ and $Y$, respectively, and $\Sigma_X$, $\Sigma_Y$ be their symbolic alphabets.

\hspace{-0.15in}\textbf{Definition 5.1} (Entropy) The \textit{entropy} of $X_S$, denoted as $H(X_S)$, is defined as\vspace{-0.1in}
\begin{equation}
\small
H(X_S)= - \sum_{x \in \Sigma_X} p(x) \cdot \log p(x) 
\vspace{-0.05in}
\end{equation}	
Intuitively, the entropy measures the amount of information or the inherent uncertainty in the possible outcomes of a random variable. The higher the $H(X_S)$, the more uncertain the outcome of $X_S$.

The conditional entropy $H(X_S\vert Y_S)$ quantifies the amount of information needed to describe the outcome of $X_S$, given the value of $Y_S$, and is defined as\vspace{-0.07in}
\begin{equation}
\small
H(X_S\vert Y_S) =  - \sum_{x \in \Sigma_X} \sum_{y \in \Sigma_Y} p(x,y) \cdot \log \frac{p(x,y)}{p(y)} 
\end{equation}
 
\hspace{-0.15in}\textbf{Definition 5.2} (Mutual information) The \textit{mutual information} of two symbolic series $X_S$ and $Y_S$, denoted as $I(X_S;Y_S)$, is defined as\vspace{-0.02in}
\begin{equation}	
\small
I(X_S;Y_S)=\sum_{x \in \Sigma_X} \sum_{y \in \Sigma_Y} p(x,y) \cdot \log \frac{p(x,y)}{p(x) \cdot p(y)} 
\label{eq:MI}
\end{equation}
The MI represents the reduction of uncertainty of one variable (e.g., $X_S$), given the knowledge of another variable (e.g., $Y_S$). The larger $I(X_S;Y_S)$, the more information is shared between $X_S$ and $Y_S$, and thus, the less uncertainty about one variable given the other. 

We demonstrate how to compute the MI between the symbolic series $S$ and $T$ in Table \ref{tbl:SymbolDatabase}. We have: p(SOn)=$\frac{17}{36}$, p(SOff)=$\frac{19}{36}$, p(TOn)=$\frac{18}{36}$, and p(TOff)=$\frac{18}{36}$. We also have the joint probabilities: p(SOn,TOn)=$\frac{15}{36}$, p(SOff,TOff)=$\frac{16}{36}$, p(SOn,TOff)=$\frac{2}{36}$, and p(SOff,TOn) =$\frac{3}{36}$. Applying Eq. \ref{eq:MI}, we have  $I(S;T)=0.29$.  

Since $0 \le I(X_S;Y_S) \leq \min(H(X_S), H(Y_S))$ \cite{thomas}, the MI value has no upper bound. To scale the MI into the range $[0-1]$, we use normalized mutual information as defined below. 

\hspace{-0.15in}\textbf{Definition 5.3} (Normalized mutual information) The \textit{normalized mutual information} (NMI) of two symbolic time series $X_S$ and $Y_S$, denoted as $\widetilde{I}(X_S;Y_S)$, is defined as\vspace{-0.05in}
\begin{equation}	
\small
\widetilde{I}(X_S;Y_S)= \frac{I(X_S;Y_S)}{H(X_S)} =1-\frac{H(X_S|Y_S)}{H(X_S)}
\label{eq:NMI}
\vspace{-0.05in}
\end{equation}
$\widetilde{I}(X_S;Y_S)$ represents the reduction (in percentage) of the uncertainty of $X_S$ due to knowing $Y_S$. Based on Eq. \eqref{eq:NMI}, a pair of variables $(X_S,Y_S)$ holds a mutual dependency if $\widetilde{I}(X_S;Y_S) > 0$. Eq. \eqref{eq:NMI} also shows that NMI is not symmetric, i.e., $\widetilde{I}(X_S;Y_S)  \neq \widetilde{I}(Y_S;X_S)$.

Using Table \ref{tbl:SymbolDatabase}, we have $I(S;T)=0.29$. However, we do not know what the $0.29$ reduction means in practice. Applying Eq. \eqref{eq:NMI}, we can compute NMI $\widetilde{I}(S;T)=0.43$, which says that the uncertainty of $S$ is reduced by $43$\% given $T$. Moreover, we also have $\widetilde{I}(T;S)=0.42$, showing that $\widetilde{I}(S;T) \neq  \widetilde{I}(T;S)$.

\hspace{-0.15in}\textbf{Definition 5.4} (Correlated symbolic time series) Let $\mu$ ($0 < \mu \le 1$) be the mutual information threshold. We say that the two symbolic series $X_S$ and $Y_S$ are \textit{correlated} iff $\widetilde{I}(X_S; Y_S) \geq \mu \vee \widetilde{I}(Y_S; X_S) \geq \mu$, and \textit{uncorrelated} otherwise.

\begin{figure}[!t]
	\hspace{-0.15in}
	\begin{tabularx}{\linewidth}{ll}
		\begin{minipage}{.58\linewidth}
			\raggedleft
			\includegraphics[width=1\linewidth]{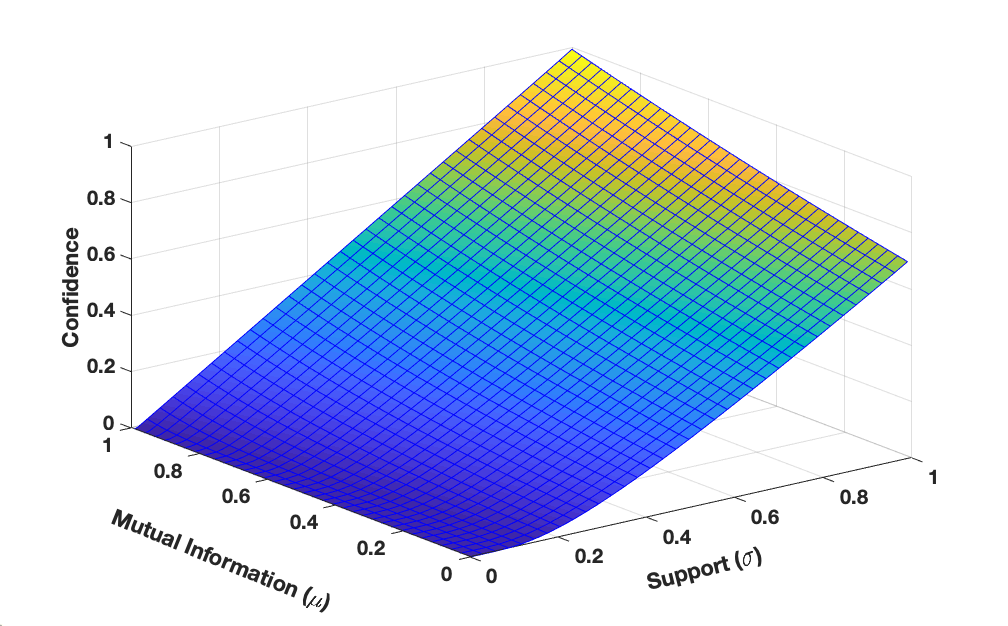}
			\vspace{-0.3in}
			\caption{Shape of the lower bound}
			\label{fig:shapeLB}
		\end{minipage}
		\hspace{0.2in}		
		\begin{minipage}{.35\linewidth}
			\vspace{0.45in}
			\centering
			\includegraphics[width=0.8\linewidth, height=1.5cm]{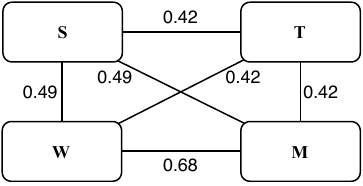}
			\caption{Corr. graph}
			\label{fig:MIGraph}
		\end{minipage}
	\end{tabularx}
	\vspace{-0.1in}
\end{figure}
\vspace{-0.05in}

\vspace{-0.05in}
\subsection{Lower Bound of the Confidence}\vspace{-0.02in}
\subsubsection{Derivation of the lower bound}
Consider $2$ symbolic series $X_S$ and $Y_S$. Let $X_1$ be a temporal event in $X_S$, $Y_1$ be a temporal event in $Y_S$, and $\mathcal{D}_{\text{SYB}}$ and $\mathcal{D}_{\text{SEQ}}$ be the symbolic and the sequence databases created from $X_S$ and $Y_S$, respectively. 
We first study the relationship between the support of $(X_1,Y_1)$ in $\mathcal{D}_{\text{SYB}}$ and $\mathcal{D}_{\text{SEQ}}$. 	\vspace{-0.05in}

\begin{lem}\label{lem:supportconnection1}
	Let $\textit{supp}(X_1,Y_1)_{\mathcal{D}_{\text{SYB}}}$ and $\textit{supp}(X_1,Y_1)_{\mathcal{D}_{\text{SEQ}}}$ be the support of $(X_1,Y_1)$ in $\mathcal{D}_{\text{SYB}}$ and $\mathcal{D}_{\text{SEQ}}$, respectively. We have the following relation:  $\textit{supp}(X_1,Y_1)_{\mathcal{D}_{\text{SYB}}} \leq \textit{supp}(X_1,Y_1)_{\mathcal{D}_{\text{SEQ}}}$.
	\vspace{-0.05in}
\end{lem}

From Lemma \ref{lem:supportconnection1}, if an event pair is frequent in $\mathcal{D}_{\text{SYB}}$, it is also frequent in $\mathcal{D}_{\text{SEQ}}$. We now investigate the connection between $\widetilde{I}(X_S;Y_S)$ in $\mathcal{D}_{\text{SYB}}$, and the confidence of $(X_1,Y_1)$ in $\mathcal{D}_{\text{SEQ}}$. \vspace{-0.05in}

\begin{theorem} (Lower bound of the confidence) Let $\sigma$ and $\mu$ be the minimum support and mutual information thresholds, respectively. Assume that $(X_1,Y_1)$ is frequent in $\mathcal{D}_{\text{SEQ}}$, i.e., $ \textit{supp}(X_1,Y_1)_{\mathcal{D}_{\text{SEQ}}} \ge \sigma$. If the NMI\hspace{0.04in} $\widetilde{I}(X_S$;$Y_S) \ge \mu$, then the confidence of $(X_1,Y_1)$ in $\mathcal{D}_{\text{SEQ}}$ has a lower bound: \vspace{-0.15in} 
\begin{align} \vspace{-0.15in}
\small
\textit{conf}(X_1,Y_1)_{\mathcal{D}_{\text{SEQ}}} \ge \sigma  \cdot  \lambda_1^{\frac{1-\mu}{\sigma} } \cdot \left(\frac{n_x - 1}{1 - \sigma}\right)^{\frac{\lambda_2}{\sigma}}   
\vspace{-0.05in}
\label{eq:lowerbound}
\end{align}
where: $n_x$ is the number of symbols in $\Sigma_X$, $\lambda_1$ is the minimum support of $X_i \in X_S$, and $\lambda_2$ is the support of $(X_i, Y_j) \in (X_S, Y_S)$ such that $p(X_i|Y_j)$ is minimal, $\forall (i \neq 1$ $\&$ $j \neq 1)$.  
\label{theorem:bound}   
\vspace{-0.05in}
\end{theorem}
\vspace{-0.15in}
\begin{proof} (Sketch - Detailed proof in \cite{ho2020efficient}). From Eq. \eqref{eq:NMI}, we have:\vspace{-0.05in}
\begin{align}
\small
\widetilde{I}(X_S;Y_S)= 1 - \frac{H(X_S \vert Y_S)}{H(X_S)} \ge \mu 
\end{align}
\vspace{-0.15in}
\begin{align}
\small
\Rightarrow \frac{H(X_S \vert Y_S)}{H(X_S)} &= \frac{p(X_1, Y_1) \cdot \log p(X_1 \vert Y_1) }   {\sum_{i} p(X_i) \cdot \log p(X_i)} \nonumber \\ &+ \frac{\sum_{i \neq 1 \& j \neq 1} p(X_i, Y_j) \cdot \log \frac{p(X_i , Y_j)} {p(Y_j)}}{\sum_{i} p(X_i) \cdot \log p(X_i)}
 \le  1-\mu
\label{eq:sketch30}
\end{align}
Let $\lambda_1 = p(X_k)$ such that $p(X_k) = \min \lbrace p(X_i)\rbrace \forall i$,  
and $\lambda_2=p(X_m,Y_n)$ such that $p(X_m\vert Y_n) = \min \lbrace p(X_i\vert Y_j) \rbrace, \forall (i \neq 1 \& j \neq 1)$. 
Then, by applying the min-max inequality theorem for the sum of ratio \cite{beckenbach1961introduction} to the numerator of Eq. \eqref{eq:sketch30}, we obtain:\vspace{-0.05in}
\begin{align}
\small
\frac{H(X_S \vert Y_S)}{H(X_S)} &\geq  \frac{p(X_1, Y_1) \cdot \log p(X_1 \vert Y_1) + \lambda_2 \cdot \log \frac{1-p(X_1,Y_1)}{n_x-p(Y_1)}}{\log \lambda_1}
 \nonumber \\
&\geq \frac{\sigma \cdot \log \frac{p(X_1,Y_1)}{p(Y_1)} + \lambda_2 \cdot \log \frac{1-\sigma}{n_x - 1}} {\log \lambda_1} 
\label{eq:sketch33}
\end{align}	

Next, assume that $\textit{supp}(Y_{1})_{\mathcal{D}_{\text{SYB}}} \ge \textit{supp}(X_{1})_{\mathcal{D}_{\text{SYB}}}$. From Eqs. \eqref{eq:sketch30}, \eqref{eq:sketch33}, the confidence lower bound of $(X_1,Y_1)$ in $\mathcal{D}_{\text{SYB}}$ is derived as:\vspace{-0.05in}
\begin{align}
	\small
	\textit{conf}(X_1,Y_1)_{\mathcal{D}_{\text{SYB}}} &= 
	\frac{\textit{supp}(X_1,Y_1)_{\mathcal{D}_{\text{SYB}}}}{\textit{supp}(Y_{1})_{\mathcal{D}_{\text{SYB}}}} \geq  \lambda_1^{\frac{1-\mu}{\sigma} } \cdot \left(\frac{n_x - 1}{1-\sigma}\right)^\frac{\lambda_2}{\sigma}
\end{align}	

Since: 
\vspace{-0.24in}
\begin{align}
	\vspace{-0.2in}
	\small
	\textit{conf}(X_1,Y_1)_{\mathcal{D}_{\text{SEQ}}} \ge \sigma \cdot \textit{conf}(X_1,Y_1)_{\mathcal{D}_{\text{SYB}}}  
	\vspace{-0.05in}
\end{align}

It follows that:\vspace{-0.15in}
\begin{align}
	\small
	\textit{conf}(X_1,Y_1)_{\mathcal{D}_{\text{SEQ}}} \ge \sigma \cdot \lambda_1^{\frac{1-\mu}{\sigma} } \cdot \left(\frac{n_x - 1}{1-\sigma}\right)^\frac{\lambda_2}{\sigma} 
\end{align}
\vspace{-0.15in}
\end{proof}

\textit{Interpretation of the confidence lower bound:} 
Theorem \ref{theorem:bound} says that, given an MI threshold $\mu$, if the two symbolic series $X_S$ and $Y_S$ are correlated, then the confidence of a frequent event pair in ($X_S$,$Y_S$) is at least the lower bound in Eq. \eqref{eq:lowerbound}. 
Combining Theorem \ref{theorem:bound} and Lemma \ref{lem3}, we can conclude that given ($X_S$,$Y_S$), if its event pair has a confidence less than the lower bound, then any pattern $P$ formed by that event pair also has a confidence less than that lower bound. This allows to approximate HTPGM (discussed in Section \ref{sec:approximateHTPGM}).

\vspace{-0.05in}\subsubsection{Shape of the confidence lower bound} 
To understand how the confidence changes w.r.t. the support $\sigma$ and the MI $\mu$, we analyze its shape, shown in Fig. \ref{fig:shapeLB} ($\sigma$ and $\mu$ vary between $0$ and $1$). First, it can be seen that the confidence lower bound has a \textit{direct relationship} with $\sigma$ and $\mu$ (one increases if the other increases and vice versa).  
While the \textit{direct relationship} between the confidence and $\sigma$ can be explained using Eq. \eqref{eq:eventpairconf}, it is interesting to observe the connection between $\mu$ and the confidence. As the MI represents the correlation between two symbolic series, the larger the value of $\mu$, the more correlated the two series. Thus, when the confidence increases together with $\mu$, it implies that patterns with high confidence are more likely to be found in highly correlated series, and vice versa.

 \removelatexerror
\begin{minipage}{\columnwidth}
	\hspace{-0.2in}
	\SetNlSty{}{}{:}
	\begin{algorithm}[H]
		\algsetup{linenosize=\tiny}
		\SetInd{0.5em}{0.5em}
		\small
		\DontPrintSemicolon
		\caption{Approximate HTPGM using MI}
		\label{algorithmMI-TPM}
		\KwInput{A set of time series $\mathcal{X}$, an MI threshold $\mu$, support threshold $\sigma$, confidence threshold $\delta$}
		\KwOutput{The set of frequent temporal patterns $P$}
		
		convert $\mathcal{X}$ to $\mathcal{D}_{\text{SYB}}$ and $\mathcal{D}_{\text{SEQ}}$;\;
		scan $\mathcal{D}_{\text{SYB}}$ to compute the probability of each event and event pair;\;
		\ForEach{\textit{pair of symbolic time series $(X_S,Y_S)$} $\in \mathcal{D}_{\text{SYB}}$}{
			compute $\widetilde{I}(X_S;Y_S)$ and $\widetilde{I}(Y_S;X_S)$;\;
			compute $\mu$;\;
			\If{$\widetilde{I}(X_S$;$Y_S) \ge \mu$ $\vee$ $\widetilde{I}(Y_S;X_S) \ge \mu$}{
				insert $X_S$ and $Y_S$ into $\mathcal{X}_C$;\;
				create an edge between $X_S$ and $Y_S$ in $G_{C}$;\;	
			} 	
		}
		\ForEach{\textit{$X_{S}$} $\in \mathcal{X}_C$}{
			mine frequent single events from $X_{S}$;\;
		}
		\ForEach{event pair $(E_i, E_j)$ in L$_1$}{
			\If{\text{there is an edge between} $X_{S}$ and $Y_{S}$ in $G_{C}$}{
				mine frequent patterns for $(E_i, E_j)$;
			}
		}
		\If{$k \ge 3$}{
			perform HTPGM using L$_1$ and L$_2$;
		}
	\end{algorithm}
\end{minipage}

Fig. \ref{fig:shapeLB} also shows that, when $\sigma$ is low, e.g., $\sigma < 0.1$, we obtain a very low value of the confidence lower bound regardless of $\mu$ value. This implies that the confidence is less sensitive to $\mu$ when the support is low. The opposite is obtained when the support is high, e.g., $\sigma > 0.1$, where we see a visible increase of the confidence lower bound as $\mu$ increases. This indicates that the "insensitive" area of the lower bound (when $\sigma \leq 0.1$) is less accurate than the "sensitive" area ($\sigma > 0.1$) when performing the approximate mining, as we will discuss in Section \ref{sec:experiment}. 

\subsection{Using the Bound to Approximate HTPGM}  \label{sec:approximateHTPGM}\vspace{-0.02in}
\subsubsection{Correlation graph}
Using Theorem \ref{theorem:bound}, we propose to approximate HTPGM by performing the mining only on \textit{the set of correlated symbolic series} $\mathcal{X}_C \subseteq \mathcal{X}$. We first define the correlation graph. 

\hspace{-0.15in}\textbf{Definition 5.5} (Correlation graph) A \textit{correlation} graph is an \textit{undirected} graph $G_{C}=(V,E)$ where $V$ is the set of vertices, and $E$ is the set of edges. Each vertex $v \in V$ represents one symbolic series $X_{S} \in \mathcal{X}_C$. There is an edge $e_{uv}$ between a vertex $u$ containing $X_{S}$, and a vertex $v$ containing $Y_{S}$ iff $\widetilde{I}(X_{S};Y_{S}) \ge \mu$ $\vee$ $\widetilde{I}(Y_{S};X_{S}) \geq \mu$.

Fig. \ref{fig:MIGraph} shows an example of the correlation graph $G_{C}$ built from $\mathcal{D_{\text{SYB}}}$ in Table \ref{tbl:SymbolDatabase}. Here, each node corresponds to one electrical appliance. There is an edge between two nodes if their NMI is at least $\mu$. The number on each edge is the NMI  between two nodes.

\textbf{Constructing the correlation graph:} 
Given a symbolic database $\mathcal{D}_{\text{SYB}}$, the correlation graph $G_{C}$ can easily be constructed by computing the NMI for each symbolic series pair, and comparing their NMI against the threshold $\mu$. A symbolic series pair is included in $G_{C}$ if their NMI is at least $\mu$, and vice versa.

\textbf{Setting the value of $\mu$:}
While NMI can easily be computed using Eq. \eqref{eq:NMI}, it is not trivial how to set the value for $\mu$. Here, we propose a method to determine $\mu$ using the lower bound in Eq. \eqref{eq:lowerbound}. 

Recall that HTPGM relies on two user-defined parameters, the support threshold $\sigma$ and the confidence threshold $\delta$, to look for frequent temporal patterns. Based on the confidence lower bound in Theorem \ref{theorem:bound}, we can derive $\mu$ using $\sigma$ and $\delta$ as the following. \vspace{-0.05in}
\begin{corollary} \label{corollary:lowerbound} 
	The confidence of an event pair $(X_1,Y_1) \in (X_S,Y_S)$ in $\mathcal{D}_{\text{SEQ}}$ is at least $\delta$ if \hspace{0.01in} $\widetilde{I}(X_S;Y_S)$ is at least $\mu$, where: \vspace{-0.07in}
	\begin{align}
	\vspace{-0.1in}
	\small
	\mu &\geq 1-\sigma \cdot  \log_{\lambda_1}{ \left(\frac{\delta}{\sigma} \cdot \left(\frac{1-\sigma}{n_x - 1}\right)^{\frac{\lambda_2}{\sigma}}\right)}
	\label{eq:muSetting}
	\end{align}
	\vspace{-0.15in}
\end{corollary}

Note that $\mu$ in Eq. \eqref{eq:muSetting} only ensures that the event pair $(X_1,Y_1)$ has a minimum confidence of $\delta$. Thus, given $(X_S,Y_S)$, $\mu$ has to be computed for each event pair in $(X_S,Y_S)$. The final chosen $\mu$ value to be compared against $\widetilde{I}(X_S;Y_S)$ is the minimum $\mu$ value among all the event pairs in $(X_S,Y_S)$.

\vspace{-0.05in}\subsubsection{Approximate HTPGM using the correlation graph}
Using the correlation graph $G_{C}$, the approximate HTPGM is described in Algorithm \ref{algorithmMI-TPM}. First, $\mathcal{D_{\text{SYB}}}$ is scanned once to compute the probability of each single event and pair of events (line 2). Next, NMI and $\mu$ are computed for each pair of symbolic series $(X_S,Y_S)$ in $\mathcal{D_{\text{SYB}}}$ (lines 4-5). Then, only pairs whose $\widetilde{I}(X_S;Y_S)$ or $\widetilde{I}(Y_S;X_S)$ is at least $\mu$ are inserted into $\mathcal{X}_C$, and an edge between $X_S$ and $Y_S$ is created (lines 6-8). 
Next, at L$_1$ of HPG, only the correlated symbolic series in $\mathcal{X}_C$ are used to mine frequent single events (lines 9-10). At L$_2$, $G_{C}$ is used to filter 2-event combinations: for each event pair $(E_i, E_j)$, we check whether there is an edge between their corresponding symbolic series in $G_{C}$. If so, we proceed by verifying the support and confidence of $(E_i, E_j)$ as in the exact HTPGM (lines 11-13). Otherwise, $(E_i,E_j)$ is eliminated from the mining of L$_2$. From level L$_k$ ($k \ge 3$) onwards, the exact HTPGM is used (lines 14-15). \vspace{-0.05in} 

\subsubsection{Complexity analysis} To compute NMI and $\mu$, we only have to scan $\mathcal{D}_{\text{SYB}}$ once to calculate the probability for each single event and pair of events. Thus, the cost of NMI and $\mu$ computations is $\mid$$\mathcal{D}_{\text{SYB}}$$\mid$. 
On the other hand, the complexity of the exact HTPGM at L$_1$ and L$_2$ are $O(m^2 i^2\mid$$\mathcal{D}_{\text{SEQ}}$$\mid$$^2) + O(m \cdot$$\mid$$\mathcal{D}_{\text{SEQ}}$$\mid)$ (Section \ref{sec:2freq}). Thus, the approximate HTPGM is significantly faster than HTPGM.

\vspace{-0.05in}
\section{Experimental Evaluation}\label{sec:experiment}\vspace{-0.02in}
We evaluate HTPGM (both exact and approximate), using real-world datasets from three application domains: smart energy, smart city, and sign language. 
Due to space limitations, we only present here the most important results, and discuss other findings in \cite{ho2020efficient}.

\vspace{-0.1in}\subsection{Experimental Setup}\vspace{-0.02in}
\hspace{0.125in}\textbf{Datasets:}
We use $3$ \textit{smart energy} datasets, NIST \cite{nist}, UKDALE \cite{ukdale}, and DataPort \cite{pecan}, all of which measure the energy/power consumption of electrical appliances in residential households. 
For the \textit{smart city}, we use weather and vehicle collision data obtained from NYC Open Data Portal \cite{smartcity}. 
For \textit{sign language}, we use the American Sign Language (ASL) datasets \cite{neidle2018new} containing annotated video sequences of different ASL signs and gestures. 
Table \ref{tbl:datasetCharacteristic} summarizes their characteristics. 

\textbf{Baseline methods:} Our exact method is referred to as E-HTPGM, and the approximate one as A-HTPGM. We use $4$ baselines (described in Section \ref{sec:relatedwork}): Z-Miner \cite{lee2020z}, TPMiner \cite{tpminer}, IEMiner \cite{ieminer}, and H-DFS \cite{hdfs}. Since E-HTPGM and the baselines provide the same exact solutions, we use the baselines only for the quantitative evaluation, and compare only E-HTPGM and A-HTPGM qualitatively. 

\textbf{Infrastructure:}
The experiments are run on virtual machines (VM) with AMD EPYC Processor 32 cores (2GHz) CPU, 256 GB main memory, and 1 TB storage. For scalability evaluation, we use VMs with 512 GB main memory. 

\textbf{Parameters:} Table \ref{tbl:params} lists the parameters and their values used in our experiments.

\begin{table}[!t]
	\vspace{-0.1in}
	\begin{minipage}{\columnwidth}
		\caption{Parameters and values}
		\vspace{-0.15in}
		\small
			\begin{tabular}{ |m{2.0cm}|m{5.5cm}| }
				\hline {\bfseries Params} & {\bfseries Values} \\ 
				\hline 
				Support $\sigma$ & User-defined: $\sigma$ $=$ 0.5\%, 1\%, 10\%, 20\%, ... \\
				\hline 
				Confidence $\delta$ & User-defined: $\delta$ $=$ 0.5\%, 1\%, 10\%, 20\%, ... \\
				\hline
					\multirow{3}{2.0 cm}{Overlapping duration $t_{\text{ov}}$} & User-defined: \\
				& $t_{\text{ov}}$ (hours) $=$ 0, 1, 2, 3 (NIST, UKDALE, DataPort, and Smart City) \\
				& $t_{\text{ov}}$ (frames) $=$ 0, 150, 300, 450 (ASL) \\
				\hline
					\multirow{4}{2.0 cm}{Tolerance \\buffer $\epsilon$} & User-defined:\\
					& $\epsilon$ (mins) $=$ 0, 1, 2, 3 (NIST, UKDALE, DataPort) \\
					& $\epsilon$ (mins) $=$ 0, 5, 10, 15 (Smart City) \\										& $\epsilon$ (frames) $=$ 0, 30, 45, 60 (ASL) \\
				\hline
			\end{tabular}	
		\label{tbl:params}
	\end{minipage}
\end{table}

\vspace{-0.1in}\subsection{Qualitative Evaluation}\vspace{-0.02in}
Our goal is to make sense and learn insights from extracted patterns.
Table \ref{tbl:interestingPatterns} lists some interesting patterns found in the datasets. 

Patterns P1 - P9 are extracted from the energy datasets, showing how the residents interact with electrical devices in their houses.  
Patterns P10 - P15 extracted from the \textit{smart city} datasets, while patterns P16 - P19 are from the ASL dataset.

\vspace{-0.1in}
\subsection{Quantitative Evaluation}\vspace{-0.02in}

\subsubsection{Baselines comparison on real world datasets}\label{sec:baselines}
We compare E-HTPGM and A-HTPGM with the baselines in terms of the runtime and memory usage. 
Tables \ref{tbl:runtimeBaselines} and \ref{tbl:memoryBaselines} show the experimental results on the energy and the smart city datasets. The quantitative results of other datasets are reported in the full paper \cite{ho2020efficient}.

\begin{table*}[!t]
	\centering
	\begin{minipage}{.5\linewidth}
		\caption{Characteristics of the Datasets}
		\vspace{-0.15in}
		\resizebox{\columnwidth}{1cm}{
			\begin{tabular}{ |c|c|c|c|c|c|}
			\hline &&&&&\\[0.01em]  \thead{} & {\bfseries NIST} & {\bfseries UKDALE} & {\bfseries DataPort} & {\bfseries Smart City} & {\bfseries ASL} \\ &&&&&\\[0.01em]
			\hline %&&&&&\\[-1em]  
			\centering \# sequences & 1460 & 1520 & 1460 & 1216 & 1908\\
			\hline %&&&&&\\[-1em] 
			\centering \# variables & 49 & 24 & 21 & 26 & 25\\
			\hline %&&&&&\\[-1em] 
			\centering \# distinct events & 98 & 48 & 42 & 130 & 173\\
			\hline %&&&&&\\[-1em] 
			\centering \# instances/seq. & 55 & 190 & 49 & 162 & 20\\
			\hline
		\end{tabular} %}
		}
		\label{tbl:datasetCharacteristic}
	\end{minipage}%
	\begin{minipage}{.5\linewidth}
		\caption{The Accuracy of A-HTPGM (\%)}
		\vspace{-0.15in}
		\resizebox{\linewidth}{1cm}{
			\small
			\begin{tabular}{|c|c|c|c|c|c|c|c|c|}
				\hline 
				\multirow{3}{*}{Supp. (\%)} & \multicolumn{8}{c|}{\bfseries Conf. (\%)} \\  
				\cline{2-9}  
				& \multicolumn{4}{c|}{\bfseries NIST} & \multicolumn{4}{c|}{\bfseries Smart City}
				\\  \cline{2-9}  
				& {\bfseries 10} & {\bfseries 20} & {\bfseries 50}  & {\bfseries 80} & {\bfseries 10} & {\bfseries 20} & {\bfseries 50}  & {\bfseries 80}  \\
				\hline
				10 & 87  & 89   & 91 & 94 & 78  & 83 & 98 & 100    \\  \hline			
				20 & 96  & 89   & 91  & 94  & 83  & 83 & 98  & 100  \\  \hline				
				50 & 100  & 100   & 96  & 94   & 99  & 99 & 98  & 100   \\  \hline					
				80 & 100  & 100   & 100  & 100 & 100  & 100 & 100  & 100  \\  \hline								
			\end{tabular}
		}			
		\label{tbl:ratiopatterns_Appro_Exact}
	\end{minipage}
	\vspace{-0.1in}
\end{table*}
\begin{table*}[!t]	
	\caption{Summary of Interesting Patterns}
	\vspace{-0.15in}
	\centering
	\resizebox{\textwidth}{!}{
		\begin{tabular}{ |p{16.5cm}|c|c| }
			\hline  \thead{Patterns} & \thead{Supp. (\%)} & \thead{Conf. (\%)}\\
			\hline  
			(P1) ([05:58, 08:24] First Floor Lights) $\succcurlyeq$ ([05:58, 06:59] Upstairs Bathroom Lights) $\succcurlyeq$ ([05:59, 06:06] Microwave) &  20 & 30  \\
			\hline
			(P2) ([06:00, 07:01] Upstairs Bathroom Lights) $\succcurlyeq$ ([06:40,  06:46] Upstairs Bathroom Plugs) &  30 & 55   \\
			\hline 
			(P3) ([18:00, 18:30] Lights Dining Room) $\rightarrow$ ([18:31, 20:16] Children Room Plugs) $\between$ ([19:00, 22:31] Lights Living Room) & 20  &  20 \\
			\hline
			(P4) ([15:59, 16:05] Hallway Lights) $\rightarrow$ ([17:58, 18:29] Kitchen Lights $\succcurlyeq$ ([18:00, 18:18] Plug In Kitchen) $\succcurlyeq$ ([18:08, 18:15] Microwave) &  20 & 25 \\
			\hline
			(P5) ([06:02, 06:19] Kitchen Lights) $\rightarrow$ ([06:05, 06:12] Microwave) $\between$ ([06:09, 06:11] Kettle) & 20 & 35 \\
			\hline 
			(P6) ([18:10,18:15] Kitchen App) $\rightarrow$ ([18:15,19:00] Lights Plugs) $\succcurlyeq$ ([18:20,18:25] Microwave) $\rightarrow$ ([18:25,18:55] Cooktop)   &  25 & 50
			\\ \hline 
			(P7) ([16:45, 17:30] Washer) $\rightarrow$ ([17:40,18:55] Dryer) $\rightarrow$ ([19:05, 20:10] Dining Room Lights)  $\succcurlyeq$ ([19:10, 19:30] Cooktop)
			&  10 & 30
			\\ \hline 
			(P8) ([06:10, 07:00] Kitchen Lights) $\succcurlyeq$ ([06:10, 06:15] Kettle) $\rightarrow$ ([06:30, 06:40] Toaster) $\rightarrow$ ([06:45, 06:48] Microwave)   &  25 & 40
			\\ \hline 
			(P9) ([18:00, 18:25] Kitchen Lights) $\succcurlyeq$ ([18:00, 18:05] Kettle) $\rightarrow$ ([18:05, 18:10] Microwave) $\rightarrow$ ([19:35, 20:50] Washer)   &  20 & 40
			\\ \specialrule{1.5pt}{1pt}{1pt}
			(P10) Heavy Rain $\succcurlyeq$ Unclear Visibility $\succcurlyeq$ Overcast Cloudiness $\rightarrow$ High Motorist Injury   &  5 & 30
			\\ \hline 
			(P11) Extremely Unclear Visibility $\succcurlyeq$ High Snow $\succcurlyeq$ High Motorist Injury   &  3 & 45 \\
			\hline
			(P12) Very Strong Wind $\rightarrow$ High Motorist Injury & 5 & 40 \\
			\hline
			(P13) Frost Temperature $\rightarrow$ Medium Cyclist Injury & 5& 20 \\
			\hline
			(P14) Strong Wind $\rightarrow$ High Pedestrian Killed & 4 & 30\\
			\hline
			(P15) Strong Wind $\rightarrow$ High Motorist Killed  & 4 & 10
			\\ \specialrule{1.5pt}{1pt}{1pt}
			(P16) [2.12 seconds] Negation   $\succcurlyeq$ [0.61 seconds] Left Head Tilt-side   $\succcurlyeq$ [0.27 seconds] Lowered Eye-brows  & 5 & 10\\
			\hline 
			(P17) [1.53 seconds] Wh-question $\succcurlyeq$ [0.36 seconds] Lowered Eye-brows   $\rightarrow$ [0.05 seconds] Blinking Eye-aperture  & 10 & 15\\
			\hline 
			(P18) [1.69 seconds] Wh-question  $\succcurlyeq$ [0.35 seconds] Right Head Tilt-side  $\succcurlyeq$ [0.27 seconds] Lowered Eye-brows  & 5 & 5\\
			\hline 
			(P19) [1.92 seconds] Wh-question   $\succcurlyeq$ [0.82 seconds] Squint Eye-aperture  $\rightarrow$ [0.13 seconds] Forward Body Lean  & 1 & 5\\
			 \hline 
		\end{tabular} 
	}
	\label{tbl:interestingPatterns}
\end{table*}
\begin{table*}[!t]
	\centering
	\vspace{-0.1in}
	\begin{minipage}{.49\linewidth}
		\caption{Runtime Comparison (seconds)}
		\vspace{-0.15in}
		\resizebox{\columnwidth}{!}{
			\begin{tabular}{|c|c|c|c|c|c|c|c|}
				\hline 
				\multirow{3}{*}{Supp. (\%)} & \multirow{3}{*}{Methods}   & \multicolumn{6}{c|}{\bfseries Conf. (\%)}
				\\  \cline{3-8}  
				& & \multicolumn{3}{c|}{\bfseries NIST}  & \multicolumn{3}{c|}{\bfseries Smart City}  
				\\  \cline{3-8}  
				& & {\bfseries 20} & {\bfseries 50}  & {\bfseries 80} & {\bfseries 20} & {\bfseries 50}    & {\bfseries 80} \\
				\hline
				\multirow{8}{*}{20}  & H-DFS  & 73864.39  & 8967.15   & 1538.49  & 2516.64  & 223.47 &  10.27    \\  \cline{2-8}  
				& \centering IEMiner  & 69440.62  & 7965.41  & 622.79  & 1419.51  & 130.80   & 8.59    \\  \cline{2-8}  
				&\centering TPMiner  & 31445.99  & 7702.02  & 533.95  & 418.25  & 118.89   & 6.66    \\  \cline{2-8}  
				& \centering Z-Miner   & 19063.24  & 2409.22  & 160.19  & 194.86 & 33.60  & 4.85   \\  \cline{2-8}  
				&\centering E-HTPGM  & 3968.19 & 672.45   & 109.08  & 86.36  & 16.89   &  2.85  \\  \cline{2-8} 
				&\centering A-HTPGM & {\bfseries 1174.28}  & {\bfseries 262.56}   &  {\bfseries 55.48}    & {\bfseries 37.54}   & {\bfseries 8.46}   &  {\bfseries 0.70}   \\  \hline  
				
				\multirow{8}{*}{50}  & H-DFS  & 6268.88  & 5170.72   & 1296.01  & 453.47&	88.32&		9.82    \\  \cline{2-8}  
				& \centering IEMiner  & 5497.78  & 4581.10  & 564.48  & 300.80 &	73.81&	7.81    \\  \cline{2-8}  
				&\centering TPMiner  & 3483.02  & 2976.37  & 512.23  & 	118.89&	37.54&	6.14  \\  \cline{2-8}  
				& \centering Z-Miner   & 2971.26  & 2061.75  & 149.81  & 92.22 & 21.05  & 1.70    \\  \cline{2-8}  				
				&\centering E-HTPGM  & 573.50 &	365.30 &	80.19  & 23.84 &	8.76 &	0.82    \\  \cline{2-8}    
				&\centering A-HTPGM & {\bfseries 309.37}&	{\bfseries 207.46}&	{\bfseries 47.86}  & {\bfseries 3.71} & {\bfseries 1.69}	 & {\bfseries 0.68}	   \\  \hline
				
				\multirow{8}{*}{80}  & H-DFS  & 1057.21  & 867.73   & 761.61  &  13.27&	8.39&		4.41     \\  \cline{2-8}  
				& \centering IEMiner  & 954.99  & 460.93  & 355.19 & 9.59 &	5.47&	4.37      \\  \cline{2-8}  
				&\centering TPMiner  & 899.25  & 412.01  & 306.91  &  6.66 &	3.44&	3.37   \\  \cline{2-8} 
				& \centering Z-Miner   & 241.87  & 170.64  & 139.74  & 3.19 & 1.23  & 1.19    \\  \cline{2-8}     
				&\centering E-HTPGM  & 143.66 &	93.55 &	63.51 & 1.47 &	0.58 &	0.47    \\  \cline{2-8}   
				&\centering A-HTPGM &  {\bfseries 63.71}	&	{\bfseries 51.35} & {\bfseries 41.26}	 & {\bfseries  0.51}& {\bfseries 0.35}	 & {\bfseries 0.21}	 \\  \hline  
			\end{tabular}	
		}
		\label{tbl:runtimeBaselines}
	\end{minipage}%
	\begin{minipage}{.505\linewidth}
		\caption{Memory Usage Comparison (MB)}
		\vspace{-0.15in}
		\resizebox{\columnwidth}{!}{
			\begin{tabular}{|c|c|c|c|c|c|c|c|c|c|}
				\hline 
				\multirow{3}{*}{Supp. (\%)} & \multirow{3}{*}{Methods}   & \multicolumn{6}{c|}{\bfseries Conf. (\%)}
				\\  \cline{3-8}  
				& & \multicolumn{3}{c|}{\bfseries NIST}  & \multicolumn{3}{c|}{\bfseries Smart City}  
				\\  \cline{3-8}  
				& & {\bfseries 20} & {\bfseries 50}  & {\bfseries 80} & {\bfseries 20} & {\bfseries 50}   & {\bfseries 80} \\
				\hline
				\multirow{8}{*}{20}  & H-DFS  & 11976.25  & 4382.12   & 1143.17  & 1293.28&	470.49	&107.89     \\  \cline{2-8}  
				& \centering IEMiner  & 7241.96  & 1613.96  & 705.51 & 1197.74&	460.52	&65.92    \\  \cline{2-8}  
				&\centering TPMiner  & 6558.48  & 1216.96  & 700.75 & 1002.82&	254.26	& 61.23    \\  \cline{2-8}  
				& \centering Z-Miner   & 91875.84  & 17642.01  & 5241.76  & 1690.75 & 602.08  & 149.77    \\  \cline{2-8} 				   
				&\centering E-HTPGM  & 1748.93 &	732.39 &	571.48  & 510.30 &	140.76 &	40.48    \\  \cline{2-8}   
				&\centering A-HTPGM & {\bfseries 875.29}& {\bfseries 674.44}	 & {\bfseries 562.77}	 & {\bfseries 161.63} & {\bfseries 85.95}	& {\bfseries 32.56}	    \\  \hline  
				
				\multirow{8}{*}{50}  & H-DFS  &  3744.73  & 3173.70   & 940.48 & 1040.56&	412.14	&	92.81    \\  \cline{2-8}  
				& \centering IEMiner  & 1455.14  & 1155.31  & 663.52 &  870.64&	 353.18 &	60.87    \\  \cline{2-8}  
				&\centering TPMiner  &  1109.89  & 909.38  & 600.73  & 660.66&	150.68&	58.98  \\  \cline{2-8}  
				& \centering Z-Miner   & 16278.14  & 10277.83  & 2153.03  & 1195.59 & 505.16  & 117.64    \\  \cline{2-8} 				
				&\centering  E-HTPGM  &  621.77 &	  424.36 &		 345.94 &  139.50 &	 119.08 &	 34.69   \\  \cline{2-8}   
				&\centering  A-HTPGM  &{\bfseries  319.59} &	{\bfseries  227.06}&	{\bfseries  186.70}	  & {\bfseries  83.55}  & {\bfseries  62.16}	&{\bfseries  29.26}	 \\ \hline
				
				\multirow{8}{*}{80}  & H-DFS  & 877.13 &	726.56 &	641.43  &  249.78&	139.59&		63.65    \\  \cline{2-8}  
				& \centering IEMiner  &  657.46  & 609.25  & 549.25 &  149.45	&119.83&	59.59    \\  \cline{2-8}  
				&\centering TPMiner  &  575.98  & 512.86  & 475.22 &  119.59&	69.91&		58.63    \\  \cline{2-8}  
				& \centering  Z-Miner   &  1934.23 &  1735.01 &  1613.09  &  263.27 &  153.16  &  93.23    \\  \cline{2-8}  				  
				&\centering  E-HTPGM  &   313.99	&  261.78	&	 153.26 &   52.93	& 36.96 &	 29.89    \\  \cline{2-8}    
				&\centering  A-HTPGM  &  {\bfseries  257.32}	& {\bfseries  187.29}& {\bfseries  106.87}	 & {\bfseries  35.75} & {\bfseries  31.74}	& {\bfseries  25.28}    \\ \hline  
			\end{tabular}
		}
		\label{tbl:memoryBaselines}
	\end{minipage}
	\vspace{-0.1in}
\end{table*}

\begin{table*}[!t]
	\centering
	\begin{minipage}{.65\linewidth}
		\caption{Pruned Time Series and Events from A-HTPGM}
		\vspace{-0.15in}
		\resizebox{\columnwidth}{1.2cm}{
			\begin{tabular}{|c|c|c|c|c|c|c|c|c|c|c|c|c|}
				\hline   
				\multirow{4}{*}{\# Attr.} & \multicolumn{6}{c|}{\bfseries NIST} & \multicolumn{6}{c|}{\bfseries Smart City}
				\\  \cline{2-13}  
				& \multicolumn{3}{c|}{\bfseries \# Pruned Time Series}  & \multicolumn{3}{c|}{\bfseries \# Pruned Events} & \multicolumn{3}{c|}{\bfseries \# Pruned Time Series}  & \multicolumn{3}{c|}{\bfseries \# Pruned Events}    
				\\  \cline{2-13}  
				& {\bfseries 20-20} & {\bfseries 20-50} & {\bfseries 20-80}  & {\bfseries 20-20} & {\bfseries 20-50} & {\bfseries 20-80} & {\bfseries 20-20} & {\bfseries 20-50} & {\bfseries 20-80} & {\bfseries 20-20} & {\bfseries 20-50} & {\bfseries 20-80} \\
				\hline
				200 & 23 & 55&	83&	46&	110&	166&	11&	27&	43&	27&	87&	135 \\  \hline		
				
				400 & 37&	101&	157&	74&	202&	314&	17&	49&	81&	57&	197&	309 \\  \hline	
				
				600 & 45&	141&	225&	90&	282&	450&	32&	80&	128&	96&	316&	492   \\  \hline		
				
				800 & 54&	182&	294&	108&	364&	588&	41&	105&	169&	129&	429&	669  \\  \hline	
				
				1000 & 83&	243&	383&	166&	486&	766&	51&	131&	211&	163&	543&	847  \\  \hline									
			\end{tabular}
		}
		\label{tbl:prunedAttributesEvents}
	\end{minipage}
	\begin{minipage}{.34\linewidth}
		\caption{Building $\mathcal{D}_{\text{SYB}}$ and $\mathcal{D}_{\text{SEQ}}$}
		\vspace{-0.15in}
		\resizebox{\linewidth}{1.2cm}{
				\begin{tabular}{ |c|c|c|c|c| }
				\hline 
				\multirow{3}{*}{Dataset} & \multicolumn{2}{c|}{\bfseries $\mathcal{D}_{\text{SYB}}$} & \multicolumn{2}{c|}{\bfseries $\mathcal{D}_{\text{SEQ}}$}\\  
				\cline{2-5}
				& \multirow{2}{*}{\bfseries Time (sec)} & \multirow{2}{*}{\bfseries Storage (MB)} & \multirow{2}{*}{\bfseries Time (sec)} & \multirow{2}{*}{\bfseries Storage (MB)} \\ 
				& &  &  &  \\ 
				\hline 
				\centering NIST& 24.92 & 10.3 & 21.60 & 4.2  \\
				\hline 
				\centering UKDALE& 19.88 & 24.1 & 8.95 & 11.4 \\
				\hline
				\centering DataPort& 11.32 & 17.7 & 20.62 & 2.9 \\
				\hline
				\centering Smart City& 17.41 & 21.9 & 13.76 & 7.8 \\
				\hline
				\centering ASL& 14.47 & 5.8 & 10.05 & 1.5 \\
				\hline
			\end{tabular}
		}			
		\label{tbl:cost_building_DSEQ}
	\end{minipage}
	\vspace{-0.12in}
\end{table*}

As shown in Table \ref{tbl:runtimeBaselines}, A-HTPGM achieves the best runtime among all methods, and E-HTPGM has better runtime than the baselines. On the tested datasets, the range and average speedups of A-HTPGM compared to other methods are: $[1.21$-$4.82]$ and $2.31$ (E-HTPGM), $[2.52$-$25.86]$ and $7.85$ (Z-Miner), $[7.43$-$69.68]$ and $21.65$ (TPMiner), $[8.61$-$188.16]$ and $40.75$ (IEMiner), and $[14.50$-$332.98]$ and $61.36$ (H-DFS). The speedups of E-HTPGM compared to the baselines are: $[1.47$-$5.64]$ and $3.19$ on average (Z-Miner), $[3.59$-$30.97]$ and $9.08$ on avg. (TPMiner), $[4.63$-$78.41]$ and $15.86$ on avg. (IEMiner), and $[5.54$-$118.21]$ and $23.37$ on avg. (H-DFS).
Note that the time to compute MI and $\mu$ for the NIST and the smart city datasets in Table \ref{tbl:runtimeBaselines} are $28.01$ and $20.82$ seconds, respectively.  

Moreover, A-HTPGM is most efficient, i.e., achieves highest speedup and memory saving, when the support threshold is low, e.g., $\sigma=20\%$. This is because typical datasets often contain many patterns with very low support and confidence. Thus, using A-HTPGM to prune uncorrelated series early helps save computational time and resources. However, the speedup comes at the cost of a small loss in accuracy (discussed in Sections \ref{sec:scalability} and \ref{sec:accuracy}).

In terms of memory consumption, as shown in Table \ref{tbl:memoryBaselines}, A-HTPGM is the most efficient method, while E-HTPGM is more efficient than the baselines. The range and the average memory consumption of A-HTPGM compared to other methods are: $[1.1$-$3.2]$ and $1.6$ (E-HTPGM), $[3.7$-$105.1]$ and $19.1$ (Z-Miner), $[1.3$-$7.9]$ and $3.4$ (TPMiner), $[1.4$-$10.4]$ and $4.5$ (IEMiner), and $[2.1$-$13.9]$ and $6.7$ (H-DFS). The memory usage of E-HTPGM compared to the baselines are: $[2.9$-$52.5]$ and $11.4$ on avg. (Z-Miner), $[1.2$-$4.7]$ and $2.1$ on average (TPMiner), $[1.3$-$6.2]$ and $2.7$ on avg. (IEMiner), and $[1.9$-$7.5]$ and $4.1$ on avg. (H-DFS).

Finally, in Table \ref{tbl:cost_building_DSEQ}, we provide the pre-processing times to convert the raw time series to $\mathcal{D}_{\text{SYB}}$, and $\mathcal{D}_{\text{SYB}}$ to $\mathcal{D}_{\text{SEQ}}$. We also report the sizes of $\mathcal{D}_{\text{SYB}}$ and $\mathcal{D}_{\text{SEQ}}$ stored on disk. We see that while the storage costs for $\mathcal{D}_{\text{SYB}}$ and $\mathcal{D}_{\text{SEQ}}$ are small, the pre-processing times are $10$-$25$ seconds. This is a one-time cost which can be reused for many mining runs, making it negligible in all non-trivial cases.

\vspace{-0.1in}\subsubsection{Scalability evaluation on synthetic datasets}\label{sec:scalability}

As discussed in Section \ref{sec:FTPMfTSMining}, the complexity of HTPGM is driven by two main factors: (1) the number of temporal sequences, and (2) the number of time series. The  evaluation on real-world datasets has shown that E-HTPGM and A-HTPGM outperform the baselines significantly in both runtimes and memory usage.  
However, to further assess the scalability, we scale these two factors on synthetic datasets. Specifically, starting from the real-world datasets, we generate $10$ times more sequences, and create up to $1000$ synthetic time series. We evaluate the scalability using two configurations: varying the number of sequences, and varying the number of time series. 

Figs. \ref{fig:scaleSequence_Energy} and \ref{fig:scaleSequence_WeatherCollision} show the runtimes of A-HTPGM, E-HTPGM and the baselines when the number of sequences changes (y-axis is in log scale).  
The range and average speedups of A-HTPGM w.r.t. other methods are: [$1.5$-$3.7$] and $2.5$ (E-HTPGM), [$3.1$-$13.6$] and $8.1$ (Z-Miner), [$5.1$-$31.2$] and $16.8$ (TPMiner), [$6.4$-$45.8$] and $24.9$ (IEMiner), and [$9.4$-$59.1$] and $31.8$ (H-DFS). In particular, A-HTPGM obtains even higher speedup for more sequences. Similarly, the range and average speedups of E-HTPGM are: [$1.6$-$5.3$] and $3.2$ (Z-Miner), [$2.2$-$12.1$] and $6.7$ (TPMiner), [$3.5$-$17.4$] and $10.1$ (IEMiner), and [$4.9$-$22.8$] and $12.9$ (H-DFS).

Figs. \ref{fig:scaleAttribute_Energy} and \ref{fig:scaleAttribute_WeatherCollision} compare the runtimes of A-HTPGM with other methods when changing the number of time series (y-axis is in log scale). It is seen that, A-HTPGM achieves even higher speedup with more time series. 
The range and average speedups of A-HTPGM are: [$2.1$-$4.9$] and $2.9$ (E-HTPGM), [$2.9$-$10.4$] and $6.8$ (Z-Miner), [$3.6$-$21.5$] and $12.8$ (TPMiner), [$4.7$-$30.2$] and $18.1$ (IEMiner), and [$6.1$-$39.6$] and $23.2$ (H-DFS), and of E-HTPGM are: [$1.4$-$4.1$] and $2.4$ (Z-Miner), [$1.7$-$8.1$] and $4.4$ (TPMiner), [$2.3$-$11.3$] and $6.2$ (IEMiner), and [$2.7$-$16.3$] and $8.1$ (H-DFS).

\input{graph/scalabilitySequenceAttribute}
\begin{figure*}[!t]
	\vspace{-0.1in}
	\begin{minipage}[t]{1\columnwidth} 
		\centering
		\begin{subfigure}{0.32\columnwidth}
			\centering
			\resizebox{\linewidth}{!}{
				\begin{tikzpicture}[scale=0.6]
					\begin{axis}[
						compat=newest,
						xlabel={Sequence (\%)},
						ylabel={Runtime (sec)}, 
						label style={font=\Huge},
						ticklabel style = {font=\Huge},
						xmin=20, xmax=100,
						ymin=0, ymax=6000,
						xtick={20,40,60,80,100},
						legend columns=-1,
						legend entries = {NoPrune, Apriori, Trans, All},
						legend style={nodes={scale=0.55,  transform shape}, font=\Large},
						legend to name={legendpruning},
						ymode=log,
						log basis y={10},
						ymajorgrids=true,
						grid style=dashed,
						line width=1.75pt
						]
						\addplot[
						color=blue,
						mark=oplus,
						mark size=4pt,
						] 
						coordinates {
							(20,2331.21)(40,3508.36)(60,4309.30)(80,4743.70)(100,5284.17)
						};
						
						\addplot[
						color=black,
						mark=triangle,
						mark size=4pt,
						] 
						coordinates {
							(20,2030.74)(40,3083.86)(60,4034.55)(80,4614.97)(100,5121.43)
						};
						
						\addplot[
						color=teal,
						mark=square*,
						mark size=4pt,
						]	
						coordinates {
							(20,1978.45)(40,2732.02)(60,3327.80)(80,3894.23)(100,4359.78)
						};
						
						\addplot[
						color=red,
						mark=*,
						mark size=4pt,
						] 
						coordinates {
							(20,1683.58)(40,2222.11)(60,2792.11)(80,3249.28)(100,3648.87)
						};
					\end{axis}
				\end{tikzpicture}
			}
			\caption{Varying \% Seq.}
		\end{subfigure}
		\begin{subfigure}{0.32\columnwidth}
			\centering
			\resizebox{\linewidth}{!}{
				\begin{tikzpicture}[scale=0.6]
					\begin{axis}[
						compat=newest,
						xlabel={Confidence (\%)},
						ylabel={Runtime (sec)}, 
						label style={font=\Huge},
						ticklabel style = {font=\Huge},
						xmin=20, xmax=100,
						ymin=0, ymax=8000,
						xtick={20,40,60,80,100},
						legend columns=-1,
						legend entries = {NoPrune, Apriori, Trans, All},
						legend style={nodes={scale=0.55,  transform shape}, font=\Large},
						legend to name={legendpruning},
						ymode=log,
						log basis y={10},
						ymajorgrids=true,
						grid style=dashed,
						line width=1.75pt
						]
						\addplot[
						color=blue,
						mark=oplus,
						mark size=4pt,
						] 	
						coordinates {
							(20,6540.54)(40,3513.02)(60,1510.59)(80,222.48)(100,19.24)
						};
						
						\addplot[
						color=black,
						mark=triangle,
						mark size=4pt,
						]	
						coordinates {
							(20,4558.87)(40,2083.1)(60,809.06)(80,100.69)(100,4.86)
						};
						
						\addplot[
						color=teal,
						mark=square*,
						mark size=4pt,
						] 
						coordinates {
							(20,3281.28)(40,1052.29)(60,391.48)(80,59.54)(100,11.75)
						};
						
						\addplot[
						color=red,
						mark=*,
						mark size=4pt,
						] 
						coordinates {
							(20,2061.61)(40,628.13)(60,206.79)(80,51.55)(100,0.50)
						};
					\end{axis}
				\end{tikzpicture}
			}
			\caption{Varying Conf.}
		\end{subfigure}
		\begin{subfigure}{0.32\columnwidth}
			\centering
			\resizebox{\linewidth}{!}{
				\begin{tikzpicture}[scale=0.6]
					\begin{axis}[
						compat=newest,
						xlabel={Support (\%)},
						ylabel={Runtime (sec)}, 
						label style={font=\Huge},
						ticklabel style = {font=\Huge},
						xmin=20, xmax=100,
						ymin=0, ymax=8000,
						xtick={20,40,60,80,100},
						legend columns=-1,
						legend entries = {NoPrune, Apriori, Trans, All},
						legend style={nodes={scale=0.55,   transform shape}, font=\Large},
						legend to name={legendpruning},
						ymode=log,
						log basis y={10},
						ymajorgrids=true,
						grid style=dashed,
						line width=1.75pt
						]
						\addplot[
						color=blue,
						mark=oplus,
						mark size=4pt,
						] 	
						coordinates {
							(20,6540.54)(40,2314.73)(60,836.60)(80,192.89)(100,0.81)
						};
						
						\addplot[
						color=black,
						mark=triangle,
						mark size=4pt,
						]	
						coordinates {
							(20,4558.87)(40,1500.53)(60,417.82)(80,100.20)(100,0.44)
						};
						
						\addplot[
						color=teal,
						mark=square*,
						mark size=4pt,
						] 
						coordinates {
							(20,3281.28)(40,958.68)(60,160.78)(80,42.14)(100,0.26)
						};
						
						\addplot[
						color=red,
						mark=*,
						mark size=4pt,
						] 
						coordinates {
							(20,2061.61)(40,614.51)(60,107.44)(80,31.47)(100,0.17)
						};
					\end{axis}
				\end{tikzpicture}
			}
			\caption{Varying Supp.}
		\end{subfigure}
		\vspace{-0.1in}
		\ref{legendpruning}
		\vspace{-0.05in}
		\caption{Runtimes of E-HTPGM on NIST}
		\label{fig:performance}
	\end{minipage}
	\hspace{0.2in}  
	\begin{minipage}[t]{1\columnwidth} 
		\centering
		\begin{subfigure}{0.32\columnwidth}
			\centering
			\resizebox{\linewidth}{!}{
				\begin{tikzpicture}[scale=0.6]
					\begin{axis}[
						compat=newest,
						xlabel={Sequence (\%)},
						ylabel={Runtime (sec)}, 
						label style={font=\Huge},
						ticklabel style = {font=\Huge},
						xmin=20, xmax=100,
						ymin=0, ymax=60000,
						xtick={20,40,60,80,100},
						legend columns=-1,
						legend entries = {NoPrune, Apriori, Trans, All},
						legend style={nodes={scale=0.55,   transform shape}, font=\Large},
						legend to name={legendpruning},
						ymode=log,
						log basis y={10},
						ymajorgrids=true,
						grid style=dashed,
						line width=1.75pt
						]
						\addplot[
						color=blue,
						mark=oplus,
						mark size=4pt,
						] 
						coordinates {
							(20,10264.14)(40,16755.73)(60,30520.38)(80,37807.50)(100,52075.58)
						};
						
						\addplot[
						color=black,
						mark=triangle,
						mark size=4pt,
						] 
						coordinates {
							(20,6861.25)(40,11797.50)(60,19908.41)(80,26141.33)(100,36699.55)
						};
						
						\addplot[
						color=teal,
						mark=square*,
						mark size=4pt,
						]	
						coordinates {
							(20,1665.45)(40,2842.28)(60,4580.18)(80,6082.17)(100,7581.47)
						};
						
						\addplot[
						color=red,
						mark=*,
						mark size=4pt,
						] 
						coordinates {
							(20,1161.39)(40,1842.56)(60,3029.91)(80,4095.47)(100,5188.48)
						};
					\end{axis}
				\end{tikzpicture}
			}
			\caption{Varying \% Seq.}
		\end{subfigure}
		\begin{subfigure}{0.32\columnwidth}
			\centering
			\resizebox{\linewidth}{!}{
				\begin{tikzpicture}[scale=0.6]
					\begin{axis}[
						compat=newest,
						xlabel={Confidence (\%)},
						ylabel={Runtime (sec)}, 
						label style={font=\Huge},
						ticklabel style = {font=\Huge},
						xmin=20, xmax=100,
						ymin=0, ymax=55000,
						xtick={20,40,60,80,100},
						legend columns=-1,
						legend entries = {NoPrune, Apriori, Trans, All},
						legend style={nodes={scale=0.55,   transform shape}, font=\Large},
						legend to name={legendpruning},
						ymode=log,
						log basis y={10},
						ymajorgrids=true,
						grid style=dashed,
						line width=1.75pt
						]
						\addplot[
						color=blue,
						mark=oplus,
						mark size=4pt,
						] 	
						coordinates {
							(20,51583.07)(40,23293.53)(60,10845.18)(80,4350.60)(100,18.78)
						};
						
						\addplot[
						color=black,
						mark=triangle,
						mark size=4pt,
						]	
						coordinates {
							(20,34967.54)(40,14122.69)(60,6435.51)(80,2265.75)(100,6.52)
						};
						
						\addplot[
						color=teal,
						mark=square*,
						mark size=4pt,
						] 
						coordinates {
							(20,7635.89)(40,4484.24)(60,2708.75)(80,906.67)(100,10.58)
						};
						
						\addplot[
						color=red,
						mark=*,
						mark size=4pt,
						] 
						coordinates {
							(20,5200.04)(40,3093.80)(60,1648.89)(80,704.43)(100,3.31)
						};
					\end{axis}
				\end{tikzpicture}
			}
			\caption{Varying Conf.}
		\end{subfigure}
		\begin{subfigure}{0.32\columnwidth}
			\centering
			\resizebox{\linewidth}{!}{
				\begin{tikzpicture}[scale=0.6]
					\begin{axis}[
						compat=newest,
						xlabel={Support (\%)},
						ylabel={Runtime (sec)}, 
						label style={font=\Huge},
						ticklabel style = {font=\Huge},
						xmin=20, xmax=100,
						ymin=0, ymax=55000,
						xtick={20,40,60,80,100},
						legend columns=-1,
						legend entries = {NoPrune, Apriori, Trans, All},
						legend style={nodes={scale=0.55,   transform shape}, font=\Large},
						legend to name={legendpruning},
						ymode=log,
						log basis y={10},
						ymajorgrids=true,
						grid style=dashed,
						line width=1.75pt
						]
						\addplot[
						color=blue,
						mark=oplus,
						mark size=4pt,
						] 	
						coordinates {
							(20,51583.07)(40,22757.74)(60,10413.09)(80,4224.74)(100,4.15)
						};
						
						\addplot[
						color=black,
						mark=triangle,
						mark size=4pt,
						]	
						coordinates {
							(20,34967.54)(40,15447.70)(60,8089.10)(80,3108.48)(100,4.22)
						};
						
						\addplot[
						color=teal,
						mark=square*,
						mark size=4pt,
						] 
						coordinates {
							(20,7635.89)(40,3295.53)(60,1689.15)(80,731.91)(100,3.15)
						};
						
						\addplot[
						color=red,
						mark=*,
						mark size=4pt,
						] 
						coordinates {
							(20,5200.04)(40,2275.97)(60,1080.71)(80,367.01)(100,2.77)
						};
					\end{axis}
				\end{tikzpicture}
			}
			\caption{Varying Supp.}
		\end{subfigure}
		\vspace{-0.1in}
		\ref{legendpruning}
		\vspace{-0.05in}
		\caption{Runtimes of E-HTPGM on Smart City}
		\label{fig:performance1}
	\end{minipage}      
	\vspace{-0.15in}  
\end{figure*}  
\begin{figure}[!h]
	\centering
	\begin{minipage}{1\columnwidth}
		\hspace{0.042in}
		\centering
		\begin{subfigure}[t]{0.32\columnwidth}
			\centering
			\resizebox{\linewidth}{!}{
				\begin{tikzpicture}[scale=0.6]
				\begin{axis}[
				compat=newest,
				style={very thick},
				xlabel={Confidence (\%)},
				ylabel={Cumulative Probability}, 
				label style={font=\Huge},
				ticklabel style = {font=\Huge},
				xmin=0, xmax=100,
				ymin=0, ymax=1,
				xtick={0,20,40,60,80,100},
				legend columns=1,
				legend entries = {supp=10\%, supp=20\%, supp=30\%, supp=40\%},
				legend style={at={(0.95,0.2)},anchor=east,font=\Huge},
				ymajorgrids=true,
				grid style=dashed,
				line width=1.75pt
				]
				\addplot[
				smooth,
				color=blue,
				solid
				]
				coordinates {
					(5,0.0)(20,0.82)(40,0.94)(60,0.99)(80,1)(100,1)
				};
				\addplot[
				smooth,
				color=red,
				dashed
				]
				coordinates {
					(10,0.0)(20,0.81)(40,0.88)(60,0.96)(80,0.99)(100,1)
				};
				\addplot[
				smooth,
				color=teal,
				densely dashdotted
				]
				coordinates {
					(20,0.0)(30,0.71)(60,0.91)(80,1)(100,1)
				};
				\addplot[
				smooth,
				color=black,
				loosely dotted
				]
				coordinates {
					(30,0)(50,0.63)(60,0.85)(80,1)(100,1)
				};
				
				\end{axis}
				\end{tikzpicture}
			}
			\caption{NIST}
		\end{subfigure}
		\begin{subfigure}[t]{0.32\columnwidth}
			\centering
			\resizebox{\linewidth}{!}{
				\begin{tikzpicture}[scale=0.6]
				\begin{axis}[
				compat=newest,
				style={very thick},
				xlabel={Confidence (\%)},
				ylabel={Cumulative Probability}, 
				label style={font=\Huge},
				ticklabel style = {font=\Huge},
				xmin=0, xmax=100,
				ymin=0, ymax=1,
				xtick={0,20,40,60,80,100},
				legend columns=1,
				legend entries = {supp=10\%, supp=20\%, supp=30\%, supp=40\%},
				legend style={at={(0.95,0.2)},anchor=east,font=\Huge},
				ymajorgrids=true,
				grid style=dashed,
				line width=1.75pt
				]
				\addplot[
				smooth,
				color=blue,
				solid
				]
				coordinates {
					(0,0.0)(20,0.86)(60,0.98)(80,1)(100,1)
				};
				\addplot[
				smooth,
				color=red,
				dashed
				]
				coordinates {
					(5,0.0)(20,0.85)(60,0.93)(80,0.99)(100,1)
				};
				\addplot[
				smooth,
				color=teal,
				densely dashdotted
				]
				coordinates {
					(10,0.0)(30,0.77)(60,0.89)(80,0.97)(100,1)
				};
				\addplot[
				smooth,
				color=black,
				loosely dotted
				]
				coordinates {
					(15,0)(30,0.32)(50,0.51)(60,0.70)(80,0.96)(100,1)
				};
				
				\end{axis}
				\end{tikzpicture}
			}
			\caption{Smart City}
		\end{subfigure}
		\begin{subfigure}[t]{0.32\columnwidth}
		\centering
		\resizebox{\linewidth}{!}{
			\begin{tikzpicture}[scale=0.6]
			\begin{axis}[
			compat=newest,
			style={very thick},
			xlabel={Confidence (\%)},
			ylabel={Cumulative Probability}, 
			label style={font=\Huge},
			ticklabel style = {font=\Huge},
			xmin=0, xmax=6,
			ymin=0, ymax=1,
			xtick={1,2,3,4,5,6},
			xticklabels = {0.5,1,5,10,20,30},
			legend columns=1,
			legend entries = {supp=0.5\%, supp=1\%, supp=5\%, supp=10\%},
			legend cell align={left},
			legend style={at={(0.95,0.2)},anchor=east,font=\Huge},
			ymajorgrids=true,
			grid style=dashed,
			line width=1.75pt
			]
			\addplot[
			smooth,
			color=blue,
			solid
			]
			coordinates {
				(0,0)(1,0.6)(2,0.7)(3,0.92)(4,0.95)(5,0.99)(6,1)
			};
			\addplot[
			smooth,
			color=red,
			dashed
			]
			coordinates {
				(0,0)(1,0.55)(2,0.65)(3,0.88)(4,0.93)(5,0.99)(6,1)
			};
			\addplot[
			smooth,
			color=teal,
			densely dashdotted
			]
			coordinates {
				(0,0)(1,0.2)(2,0.58)(3,0.84)(4,0.90)(5,0.96)(6,1)
			};
			\addplot[
			smooth,
			color=black,
			loosely dotted
			]
			coordinates {
				(0,0)(1,0.1)(2,0.5)(3,0.81)(4,0.86)(5,0.93)(6,1)
			};
			
			\end{axis}
			\end{tikzpicture}
		}
		\caption{ASL}
	\end{subfigure}	
		\vspace{-0.2in}
		\caption{Cumulative probability of pruned patterns}    
		\label{fig:cumProbPrunedAtt}
	\end{minipage}		   
\end{figure}

In Figs. \ref{fig:scaleAttribute_Energy} and \ref{fig:scaleAttribute_WeatherCollision}, to illustrate the computation time of MI and $\mu$, we add an additional bar chart for A-HTPGM. Each bar represents the runtime of A-HTPGM with two separate components: the time to compute MI and $\mu$ (top red), and the mining time (bottom blue). However, note that for each dataset, we only need to compute MI and $\mu$ once (the computed values are used across the mining process with different support and confidence thresholds). Thus, the times to compute MI and $\mu$, for example, in Figs. \ref{fig:scaleAttribute_Energy_sub1}, \ref{fig:scaleAttribute_Energy_sub2}, and \ref{fig:scaleAttribute_Energy_sub3}, are added only for comparison and are not all actually used. 

Moreover, most baselines fail for the larger configurations in the scalability study, e.g., Z-Miner on the NIST dataset when $\sigma$$=$$\delta$$=$$20\%$ (Fig. \ref{fig:scaleSequence_Energy_20_20}), and Z-Miner, TPMiner, IEMiner and H-DFS when the number of time series grows to $1000$ (Fig. \ref{fig:scaleAttribute_Energy_sub1}). 
The scalability test shows that A-HTPGM and E-HTPGM can scale well on big datasets, both vertically (many sequences) and horizontally (many time series), unlike the baselines.

Furthermore, the number of time series and events pruned by A-HTPGM in the scalability test are provided in Table \ref{tbl:prunedAttributesEvents}. Here, we can see that high confidence threshold leads to more time series (events) to be pruned. This is because confidence has a direct relationship with MI, therefore, high confidence results in higher $\mu$, and thus, more pruned time series.

\vspace{-0.1in}\subsubsection{Evaluation of the pruning techniques in E-HTPGM}\label{sec_exact}
We compare different versions of E-HTPGM to understand how effective the pruning techniques are:  (1) NoPrune: E-HTPGM with no pruning, (2) Apriori: E-HTPGM with Apriori-based pruning (Lemmas \ref{lem2}, \ref{lem3}), (3) Trans: E-HTPGM with transitivity-based pruning (Lemmas \ref{lem:transitivity}, \ref{lem:filter}, \ref{lem5}, \ref{lem6}), and (4) All: E-HTPGM applied both pruning techniques. 

We use $3$ different configurations that vary: the number of sequences, the confidence, and the support. Figs. \ref{fig:performance}, \ref{fig:performance1} 
show the results (the y-axis is in log scale). It can be seen that (All)-E-HTPGM achieves the best performance among all versions. Its speedup w.r.t. (NoPrune)-E-HTPGM ranges from $5$ up to $60$ 
depending on the configurations, showing that the proposed prunings are very effective in improving E-HTPGM performance. Furthermore, (Trans)-E-HTPGM delivers larger speedup than (Apriori)-E-HTPGM. The average speedup is from $8$ to $20$  
for (Trans)-E-HTPGM, and from $3$ to $12$ for (Apriori)-E-HTPGM. However, applying both always yields better speedup than applying either of them.

\vspace{-0.09in}\subsubsection{Evaluation of A-HTPGM}\label{sec:accuracy}
We proceed to evaluate the accuracy of A-HTPGM and the quality of patterns pruned by A-HTPGM. 

To evaluate the accuracy, we compare the patterns extracted by A-HTPGM and E-HTPGM. Table \ref{tbl:ratiopatterns_Appro_Exact} shows the accuracies of A-HTPGM for different supports and confidences. It is seen that, A-HTPGM obtains high accuracy ($\ge 71\%$) when $\sigma$ and $\delta$ are low, e.g., $\sigma=\delta=10\%$, and very high accuracy ($\ge 95\%$) when $\sigma$ and $\delta$ are high, e.g., $\sigma=\delta=50\%$.  
Next, we analyze the quality of patterns pruned by A-HTPGM. These patterns are extracted from the uncorrelated time series. Fig. \ref{fig:cumProbPrunedAtt} shows the cumulative distribution of the confidences of the pruned patterns. It is seen that most of these patterns have low confidences, and can thus safely be pruned. For NIST and Smart City datasets, $80\%$ of pruned patterns have confidences less than $20\%$ when the support is $10\%$ and $20\%$, and $70\%$ of pruned patterns have confidences less than $30\%$ when the support is $30\%$. For the ASL dataset, $80\%$ of pruned patterns have confidences less than $5\%$.  

\textbf{Other experiments:} 
We analyze the effects of the tolerance buffer $\epsilon$, and the overlapping duration t$_{\text{ov}}$ to the quality of extracted patterns. The analysis can be seen in the full paper \cite{ho2020efficient}.

%\vspace{-0.08in}
\section{Conclusion and Future Work}\label{sec:conclusion}%\vspace{-0.05in}
This paper presents our comprehensive Frequent Temporal Pattern Mining from Time Series (FTPMfTS) solution that offers: (1) an end-to-end FTPMfTS process to mine frequent temporal patterns from time series, (2) an efficient and exact Hierarchical Temporal Pattern Graph Mining (E-HTPGM) algorithm that employs efficient data structures and multiple pruning techniques to achieve fast mining, and (3) an approximate A-HTPGM that uses mutual information to prune unpromising time series, allows HTPGM to scale on big datasets. Extensive experiments conducted on real world and synthetic datasets show that both A-HTPGM and E-HTPGM outperform the baselines, consume less memory, and scale well to big datasets. Compared to the baselines, the approximate A-HTPGM delivers an order of magnitude speedup on large synthetic datasets and up to $2$ orders of magnitude speedup on real-world datasets. %achieves up to $2$ orders of magnitude speedup on real-world datasets, and up to an order of magnitude speedup on large scale synthetic datasets. 
In future work, we plan to extend HTPGM to prune at the event level to further improve its performance.

\begin{appendices}
	\appendix
	\section{Detailed Proofs of Complexities, Lemmas and Theorems}
\subsection{Mutual exclusive property of temporal relations}\label{app:mutualexclusive}
\textbf{Property 1. }\textit{(Mutual exclusive) Consider the set of temporal relations $\Re=$ \{Follows, Contains, Overlaps\}. Let $E_i$ and $E_j$ be two temporal events, and $e_i$ occurring during $[t_{s_i}, t_{e_i}]$, $e_j$ occurring during $[t_{s_j}, t_{e_j}]$ be their corresponding event instances, and $\epsilon$ be the tolerance buffer. The relations in $\Re$ are mutually exclusive on $E_i$ and $E_j$.} 
\begin{proof}
	\textbf{$\ast$ Case 1:} Assume the relation \textbf{Follows($E_{i_{\triangleright e_i}}$, $E_{j_{\triangleright e_j}}$)} holds between $E_i$ and $E_j$. Thus, we have: 
	\begin{equation}
	t_{e_i} \pm \epsilon \le t_{s_j}
	\label{eq:mutualcase11}
	\end{equation}
	and:
	\begin{equation}
	t_{s_j} < t_{e_j} \Rightarrow t_{e_i} \pm \epsilon < t_{e_j}
	\label{eq:mutualcase12}
	\end{equation}
	
	Hence, Contains($E_{i_{\triangleright e_i}}$, $E_{j_{\triangleright e_j}}$) cannot exist between $E_i$ and $E_j$, since Contains($E_{i_{\triangleright e_i}}$, $E_{j_{\triangleright e_j}}$) holds iff ${(t_{s_i} \le t_{s_j})} \wedge$ $(t_{e_i} \pm \epsilon \ge t_{e_j})$ (contradict Eq. \eqref{eq:mutualcase12}). 
	
	Similarly, Overlaps($E_{i_{\triangleright e_i}}$, $E_{j_{\triangleright e_j}}$) cannot exist between $E_i$ and $E_j$ since Overlaps($E_{i_{\triangleright e_i}}$, $E_{j_{\triangleright e_j}}$) holds iff ${(t_{s_i} < t_{s_j})} \wedge$ $(t_{e_i} \pm \epsilon < t_{e_j})$ $\wedge$ $(t_{e_i}-t_{s_j} \ge d_o \pm \epsilon)$ (contradict Eq. \eqref{eq:mutualcase11}).
	
	In conclusion, if Follows($E_{i_{\triangleright e_i}}$, $E_{j_{\triangleright e_j}}$) holds between $E_i$ and $E_j$, then the two remaining relations cannot exist between $E_i$ and $E_j$.
	
	\textbf{$\ast$ Case 2:} Assume the relation \textbf{Contains($E_{i_{\triangleright e_i}}$, $E_{j_{\triangleright e_j}}$)} holds between $E_i$ and $E_j$. Thus, we have: 
	\begin{equation}
	t_{s_i} \le t_{s_j}
	\label{eq:mutualcase21}
	\end{equation}
	\begin{equation}
		t_{e_i} \pm \epsilon \ge t_{e_j}
	\label{eq:mutualcase22}
	\end{equation}
	
	Hence, Follows($E_{i_{\triangleright e_i}}$, $E_{j_{\triangleright e_j}}$) cannot exist between $E_i$ and $E_j$ since Follows($E_{i_{\triangleright e_i}}$, $E_{j_{\triangleright e_j}}$) holds iff $t_{e_i} \pm \epsilon < t_{e_j}$ (contradict Eq. \eqref{eq:mutualcase22}).
	
	Similarly, Overlaps($E_{i_{\triangleright e_i}}$, $E_{j_{\triangleright e_j}}$) cannot exist between $E_i$ and $E_j$, since Overlaps($E_{i_{\triangleright e_i}}$, $E_{j_{\triangleright e_j}}$) holds iff ${(t_{s_i} < t_{s_j})} \wedge$ $(t_{e_i} \pm \epsilon < t_{e_j})$ $\wedge$ $(t_{e_i}-t_{s_j} \ge d_o \pm \epsilon)$ (contradict Eq. \eqref{eq:mutualcase22}). 
	
	In conclusion, if Contains($E_{i_{\triangleright e_i}}$, $E_{j_{\triangleright e_j}}$) holds between $E_i$ and $E_j$, then the two remaining relations cannot exist between $E_i$ and $E_j$.
	
	\textbf{$\ast$ Case 3:} Assume the relation \textbf{Overlaps($E_{i_{\triangleright e_i}}$, $E_{j_{\triangleright e_j}}$)} holds between $E_i$ and $E_j$. Thus, we have: 
	\begin{equation}
	t_{s_i} < t_{s_j}
	\label{eq:mutualcase31}
	\end{equation}
	\begin{equation}
	t_{e_i} \pm \epsilon < t_{e_j}
		\label{eq:mutualcase32}
	\end{equation}
	\begin{equation}
	t_{e_i}-t_{s_j} \ge d_o \pm \epsilon \Rightarrow t_{s_j} \le t_{e_i} - d_o \pm \epsilon
		\label{eq:mutualcase33}
	\end{equation}
	
	Hence, Follows($E_{i_{\triangleright e_i}}$, $E_{j_{\triangleright e_j}}$) cannot exist between $E_i$ and $E_j$, since Follows($E_{i_{\triangleright e_i}}$, $E_{j_{\triangleright e_j}}$) holds iff $t_{e_i} \pm \epsilon < t_{s_j}$ (contradict Eq. \eqref{eq:mutualcase33}).
	
	Similarly, Contains($E_{i_{\triangleright e_i}}$, $E_{j_{\triangleright e_j}}$) cannot exist between $E_i$ and $E_j$, since Contains($E_{i_{\triangleright e_i}}$, $E_{j_{\triangleright e_j}}$) holds iff $t_{e_i} \pm \epsilon \ge t_{e_j}$ (contradict Eq. \eqref{eq:mutualcase32}).
	
	In conclusion, if Overlaps($E_{i_{\triangleright e_i}}$, $E_{j_{\triangleright e_j}}$) holds between $E_i$ and $E_j$, then the two remaining relations cannot exist between $E_i$ and $E_j$.
\end{proof}

\subsection{Mining frequent single event}\label{app:proofcomplexity1event}
\textit{Complexity:} 
The complexity of finding frequent single events is $O(m \cdot$$\mid$$\mathcal{D}_{\text{SEQ}}$$\mid)$, where $m$ is the number of distinct events.
\begin{proof}
	Computing the support for each event $E_i$ takes $O(\mid$ $\mathcal{D}_{\text{SEQ}}$$\mid)$ (to count the set bits of the \textit{bitmap} $b_{E_i}$ of length $\mid$$\mathcal{D}_{\text{SEQ}}$$\mid$). Thus, counting the support for $m$ events takes $O(m \cdot$$\mid$$\mathcal{D}_{\text{SEQ}}$$\mid)$.
\end{proof}

\subsection{Lemma \ref{lem1}} \label{app:prooflem1}
\textbf{Lemma \ref{lem1}. }\textit{Let $m$ be the number of events in $\mathcal{D}_{\text{SEQ}}$, and $h$ be the longest length of a temporal pattern. The total number of temporal patterns in HPG from L$_1$ to L$_h$ is $O(m^h3^{h^2})$.}

Lemma \ref{lem1} shows that the exponential search space of HTPGM is driving by: the number of distinct events ($m$) which is correlated to the number of time series in the datasets, the max pattern length ($h$), and the number of temporal relations ($3$). A dataset of just a few hundred events can create a very large search space with billions of candidate patterns. 

\begin{proof}
At L$_1$, the number of nodes is: $N_1=m \sim O(m)$. At L$_2$, the number of permutations of $m$ distinct events taken $2$ at a time is: $P(m,2)$. However, since the same event can form a pair of events with itself, e.g., (SOn,SOn), the total number of nodes at L$_2$ is: $N_2=P(m,2)+m$ $\sim O(m^2)$. Each node in $N_2$ can form $3$ different temporal relations, and thus, the total number of 2-event patterns in L$_2$ is: $N_2\times 3^1$ $\sim O(m^23^1)$. Similarly, the number of 3-event nodes at L$_3$ is: $N_3= P(m,3)+P(m,2)+m$ $\sim O(m^3)$, and the number of 3-event patterns is: $N_3 \times 3^3$ $\sim O(m^33^3)$. At level L$_h$, the number of nodes is $O(m^h)$, while the number of h-event patterns is $O(m^h \times 3^{\frac{1}{2}h(h-1)})$ $\sim O(m^h3^{h^2})$. Therefore, the total number of temporal patterns from L$_1$ to L$_h$ in HPG is $O(m)+O(m^23^1)+O(m^33^3)+...+O(m^h3^{h^2}) \sim O(m^h3^{h^2})$.
\end{proof}

\subsection{Lemma \ref{lem2}} \label{app:prooflem2}
\textbf{Lemma \ref{lem2}. }\textit{	Let $P$ be a 2-event pattern formed by an event pair $(E_i, E_j)$. Then, $\textit{supp}(P) \le \textit{supp}(E_i,E_j)$.}

From Lemma \ref{lem2}, the support of a pattern is at most the support of its events. Thus, infrequent event pairs cannot form frequent patterns and thereby, can be safely pruned.

\begin{proof}
	Derived directly from Defs. 3.11, 3.12, 3.13 and 3.14. 
\end{proof}

\subsection{Lemma \ref{lem3}} \label{app:prooflem3}
\textbf{Lemma \ref{lem3}. }\textit{Let $(E_i, E_j)$ be a pair of events occurs in a 2-event pattern $P$.  Then \textit{conf}($P$) $\le$ \textit{conf}($E_i,E_j$).}

From Lemma \ref{lem3}, the confidence of a pattern $P$ is always at most the confidence of its events. Thus, a low-confidence event pair cannot form any high-confidence patterns and therefore, can be safely pruned.

\begin{proof}
	Can derived directly from Defs. 3.15 and 3.16. 
\end{proof}

\subsection{Mining frequent 2-event pattern} \label{app:proofcomplexity2event}
\textit{Complexity:} Let $m$ be the number of frequent single events in L$_1$, and $i$ be the average number of event instances of each frequent event. The complexity of frequent 2-event pattern mining is $O(m^2 i^2\mid$$\mathcal{D}_{\text{SEQ}}$$\mid$$^2)$.  

\begin{proof}
	The Cartesian product of $m$ events in L$_1$ generates $m^2$ event pairs. To compute the support of $m^2$ event pairs, we count the set bits of the \textit{bitmap} that takes $O(m^2$$\mid$$\mathcal{D}_{\text{SEQ}}$$\mid)$.
	
	For each node in L$_2$, we need to compute the support and confidence of their temporal relations, which takes $O(i^2$$\mid$$\mathcal{D}_{\text{SEQ}}$$\mid^2$$)$. We have potentially $m^2$ nodes. And thus, it takes $O(m^2 i^2$$\mid$$\mathcal{D}_{\text{SEQ}}$$\mid^2$$)$.
	
	The overall complexity is:  $O(m^2$$\mid$$\mathcal{D}_{\text{SEQ}}$$\mid$ $+$ $m^2 i^2$$\mid$$\mathcal{D}_{\text{SEQ}}$$\mid^2$$) \sim \\O(m^2 i^2$$\mid$$\mathcal{D}_{\text{SEQ}}$$\mid^2$$)$.
\end{proof}

\subsection{Lemma \ref{lem:transitivity}} \label{app:prooflemtransitivity}
\textbf{Lemma \ref{lem:transitivity}.}
\textit{Let $S=<e_1$,..., $e_{n-1}>$ be a temporal sequence that supports an (n-1)-event pattern $P=<(r_{12}$, $E_{1_{\triangleright e_1}}$, $E_{2_{\triangleright e_2}})$,..., $(r_{(n-2)(n-1)}$, $E_{{n-2}_{\triangleright e_{n-2}}}$, $E_{{n-1}_{\triangleright e_{n-1}}})>$. Let $e_n$ be a new event instance added to $S$ to create the temporal sequence $S^{'}$$=$$<e_1, ..., e_{n}>$.} 
	
\textit{The set of temporal relations $\Re$ is transitive on $S^{'}$: $\forall e_i \in S^{'}$, $i < n$, $\exists r \in \Re$ s.t. $r(E_{i_{\triangleright e_i}}$,$E_{n_{\triangleright e_n}})$ holds.}

Lemma \ref{lem:transitivity} says that given a temporal sequence $S$, a new event instance added to $S$ will always form at least one temporal relation with existing instances in $S$. This is due to the temporal transitivity property, as the time interval of new event instance will have a temporal order with the time intervals of existing instances.

\begin{proof}
	Since $S^{'}=<e_1, ..., e_{n}>$ is a temporal sequence, the event instances in $S^{'}$ are chronologically ordered by their start times. Then, $\forall e_i \in S^{'}, i \neq n$: $t_{s_i} \le t_{s_n}$. We have: 
	\begin{itemize}
		\item If $t_{e_i} \pm \epsilon \le t_{s_n}$, then $E_{i_{\triangleright e_i}}$ $\rightarrow E_{n_{\triangleright e_n}}$.
		\item If ${(t_{s_i} \le t_{s_n})} \wedge$ $(t_{e_i} \pm \epsilon \ge t_{e_n})$, then $E_{i_{\triangleright e_i}} \succcurlyeq E_{n_{\triangleright e_n}}$.
		\item If ${(t_{s_i} < t_{s_n})} \wedge$ $(t_{e_i} \pm \epsilon < t_{e_n})$ $\wedge$ $(t_{e_i}-t_{s_n} \ge d_o \pm \epsilon)$ where $d_o$ is the minimal overlapping duration, then $E_{i_{\triangleright e_i}} \between E_{n_{\triangleright e_n}}$. 
	\end{itemize}
\end{proof}

\subsection{Lemma \ref{lem:filter}} \label{app:prooflemfilter}
\textbf{Lemma \ref{lem:filter}.}
\textit{Let $N_{k-1}=(E_1,...,E_{k-1})$ be a frequent (\textit{k-1})-event combination, and $E_k$ be a frequent single event. The combination $N_k= N_{k-1} \cup E_k$ can form frequent k-event temporal patterns if $\forall E_i \in N_{k-1}$, $\exists r \in \Re$ s.t. $r(E_i,E_k)$ is a frequent temporal relation.}
	
From Lemma \ref{lem:filter}, 
	only single events in L$_1$ that occur in L$_{k-1}$ should be used to create k-event combinations, as those that do not occur in L$_{k-1}$ cannot form frequent temporal patterns and thus, should not be considered.
	
\begin{proof}
	Let $p_k$ be any k-event pattern formed by $N_k$. Then $p_k$ is a list of $\frac{1}{2}k(k-1)$ triples $(E_i,r_{ij},E_j)$ where each represents a relation $r(E_i,E_j)$ between two events. In order for $p_k$ to be frequent, each of the relations in $p_k$ must be frequent (Defs. 3.11 and 3.15, and Lemma \ref{lem:transitivity}). However, since $\nexists E_i \in N_{k-1}$ such that $r(E_i,E_k)$ is frequent, $p_k$ is not frequent.
\end{proof}

\subsection{Lemma \ref{lem5}} \label{app:prooflem5}
\textbf{Lemma \ref{lem5}.}
	\textit{Let $P$ and $P^{'}$ be two temporal patterns. If $P^{'} \subseteq P$, then \textit{conf}($P^{'}$) $ \geq$ \textit{conf}($P$).}
	
Lemma \ref{lem5} says that the confidence of a pattern $P$ is always at most the confidence of its sub-patterns. Thus, a temporal pattern at level L$_{k-1}$ that is infrequent and/or low-confidence cannot be part of frequent and high-confidence patterns at L$_k$.

\begin{proof}
	Can be derived directly from Def. 3.16.
\end{proof}

\subsection{Lemma \ref{lem6}} \label{app:prooflem6}
\textbf{Lemma \ref{lem6}.}
	\textit{Let $P$ and $P^{'}$ be two temporal patterns. If $P^{'} \subseteq P$ and $\frac{\textit{supp}(P^{'})}{\max_{1 \leq k \leq \mid P \mid}\{\textit{supp}(E_k)\}}_{E_k \in P} \leq \delta$, then \textit{conf($P$)} $\leq \delta$.}

 From Lemma \ref{lem6}, a temporal pattern $P$ cannot be high-confidence if any of its sub-patterns are low-confidence.
 
\begin{proof} We have:
	\begin{align}
	\small
	\text{\it{conf(P)}} &= \frac{\textit{supp}(P)}{\max_{1 \leq k \leq \mid P \mid}\{\textit{supp}(E_k) \}} \nonumber \\ &\leq \frac{\textit{supp}(P^{'})}{\max_{1 \leq k \leq \mid P \mid}\{\textit{supp}(E_k) \}} \leq \delta \nonumber
	\end{align}
\end{proof}

\subsection{Mining frequent k-event pattern}\label{app:proofcomplexitykevent}
\textit{Complexity:} Let $r$ be the average number of frequent (k-1)-event patterns in L$_{k-1}$. The complexity of frequent k-event pattern mining is $O($$\mid$$\textit{1Freq}$$\mid$ $\cdot$ $\mid$$L_{k-1}$$\mid$ $\cdot$ $r$ $\cdot$ $k^2$$\cdot$$\mid$$\mathcal{D}_{\text{SEQ}}$$\mid$$)$.

\begin{proof}
	
	For each frequent (k-1)-event pattern of a node in L$_k$, we need to compute the support and confidence of $\frac{1}{2}(k-1)(k-2)$ triples, which takes $O(\frac{1}{2}(k-1)(k-2)$$\mid$$\mathcal{D}_{\text{SEQ}}$$\mid$$) \sim O(k^2$$\mid$$\mathcal{D}_{\text{SEQ}}$$\mid$$)$.
	
	We have $\mid$\textit{1Freq}$\mid$ $\times$ $\mid$$L_{k-1}$$\mid$ nodes in L$_k$, each has $r$ frequent (k-1)-event patterns. Thus, the total complexity is: $O($$\mid$$\textit{1Freq}$$\mid$ $\cdot$ $\mid$$L_{k-1}$$\mid$ $\cdot$ $r$ $\cdot$ $k^2$$\cdot$$\mid$$\mathcal{D}_{\text{SEQ}}$$\mid$$)$.
\end{proof}

\subsection{Lemma \ref{lem:supportconnection1}}\label{app:prooflemsupportconnection1}
\textbf{Lemma \ref{lem:supportconnection1}.} 
	\textit{Let $\textit{supp}(X_1,Y_1)_{\mathcal{D}_{\text{SYB}}}$ and $\textit{supp}(X_1,Y_1)_{\mathcal{D}_{\text{SEQ}}}$ be the supports of $(X_1,Y_1)$ in $\mathcal{D}_{\text{SYB}}$ and $\mathcal{D}_{\text{SEQ}}$, respectively. We have the following relation:  $\textit{supp}(X_1,Y_1)_{\mathcal{D}_{\text{SYB}}} \leq \textit{supp}(X_1,Y_1)_{\mathcal{D}_{\text{SEQ}}}$.}
	
From Lemma \ref{lem:supportconnection1}, if an event pair is frequent in $\mathcal{D}_{\text{SYB}}$, it is also frequent in $\mathcal{D}_{\text{SEQ}}$. 

\begin{proof}	
	Recall that when converting $\mathcal{D}_{\text{SYB}}$ to $\mathcal{D}_{\text{SEQ}}$, we divide the symbolic time series in $\mathcal{D}_{\text{SYB}}$ into equal length temporal sequences. Let $n$ be the length of each symbolic time series in $\mathcal{D}_{\text{SYB}}$, and $m$ be the length of each temporal sequence. 
	The number of temporal sequences obtained in $\mathcal{D}_{\text{SEQ}}$ is: $\lceil \frac{n}{m} \rceil$.
	
	The support of $(X_1,Y_1)$ in $D_{\text{SYB}}$ is computed as: 
	\begin{equation}
	supp(X_1,Y_1)_{D_{\text{SYB}}} = \frac{\sum_{i=1}^{\lceil \frac{n}{m} \rceil} \sum_{j=1}^{m} s_{ij}} {n}
	\label{eq:supportSYBSEQ1}
	\end{equation}
	where
	\[s_{ij} = \begin{cases}
	\small
	1, & \parbox[l]{1\columnwidth}{\text{if $(X_1,Y_1)$ occurs in row j of the sequence $s_i$ in  $\mathcal{D}_{\text{SYB}}$}}\\
	0,              &\text{otherwise}
	\end{cases}\]
	Intuitively, Eq. \ref{eq:supportSYBSEQ1} computes the relative support of $(X_1,Y_1)$ in $\mathcal{D}_{\text{SYB}}$ by counting the number of times $(X_1,Y_1)$ occurs in $\mathcal{D}_{\text{SYB}}$, and then dividing to the size of $\mathcal{D}_{\text{SYB}}$.  
	
	On the other hand, the support of $(X_1,Y_1)$ in $\mathcal{D}_{\text{SEQ}}$ is computed by counting the number of sequences in $\mathcal{D}_{\text{SEQ}}$ that $(X_1,Y_1)$ occurs. Note that if $(X_1,Y_1)$ occurs more than one in the same sequence, we will count only $1$ for that sequence. 
	Thus, we have: 
	\begin{equation}
	\textit{supp}(X_1,Y_1)_{\mathcal{D}_{\text{SEQ}}} = \frac{\sum_{i=1}^{\lceil \frac{n}{m} \rceil} g_i} {n/m} = \frac{m \cdot \sum_{i=1}^{\lceil \frac{n}{m} \rceil} g_i} {n} 
	\label{eq:supportSYBSEQ2}
	\end{equation}
	where 
	\[g_i = \begin{cases}
	1,& \text{if $(X_1,Y_1)$ occurs in the sequence $g_i$ in $\mathcal{D}_{\text{SEQ}}$}\\
	0,              & \text{otherwise}
	\end{cases}\]
	\\
	Compare Eqs. \ref{eq:supportSYBSEQ1} and \ref{eq:supportSYBSEQ2}, we have:
	\begin{equation}
	\sum_{i=1}^{\lceil \frac{n}{m} \rceil} \sum_{j=1}^{m} s_{ij} \le m \cdot \sum_{i=1}^{\lceil \frac{n}{m} \rceil} g_i
	\end{equation}
	Hence: 
	\begin{equation}
	\textit{supp}(X_1,Y_1)_{\mathcal{D}_{\text{SYB}}} \leq \textit{supp}(X_1,Y_1)_{\mathcal{D}_{\text{SEQ}}}
	\end{equation}
	
\end{proof}

\subsection{Theorem \ref{theorem:bound}}\label{app:prooftheorembound}
\textbf{Theorem \ref{theorem:bound}.}   (Lower bound of the confidence)
	 \textit{Let $\sigma$ and $\mu$ be the minimum support and mutual information thresholds, respectively. Assume that $(X_1,Y_1)$ is frequent in $ \mathcal{D}_{\text{SEQ}}$, i.e., $\textit{supp}(X_1,Y_1)_{\mathcal{D}_{\text{SEQ}}} \ge \sigma$. If the NMI\hspace{0.04in} $\widetilde{I}(\mathcal{X}_S$;$\mathcal{Y}_S) \ge \mu$, then the confidence of $(X_1,Y_1)$ in $\mathcal{D}_{\text{SEQ}}$ has a lower bound: \vspace{-0.1in}
\begin{align} \vspace{-0.1in}
\small
\textit{conf}(X_1,Y_1)_{\mathcal{D}_{\text{SEQ}}} \ge  \sigma \cdot \lambda_1^{\frac{1-\mu}{\sigma} } \cdot \left(\frac{n_x - 1}{1 - \sigma}\right)^{\frac{\lambda_2}{\sigma}}    
\end{align}
where: $n_x$ is the number of symbols in $\Sigma_X$, $\lambda_1$ is the minimum support of $X_i \in \mathcal{X}_S, \forall i$, and $\lambda_2$ is the support of $(X_i, Y_j) \in (\mathcal{X}_S, \mathcal{Y}_S)$ such that $\frac{p(X_i,Y_j)}{p(Y_j)}$ is minimal, $\forall (i \neq 1$ $\&$ $j \neq 1)$.} 
 
\begin{proof} 

From Eq. \eqref{eq:NMI}, we have:
\begin{equation}
\small
\widetilde{I}(\mathcal{X}_S;\mathcal{Y}_S)= 1 - \frac{H(\mathcal{X}_S \vert \mathcal{Y}_S)}{H(\mathcal{X}_S)} \ge \mu
\end{equation} 
Hence:
\begin{equation}
\small
\frac{H(\mathcal{X}_S \vert \mathcal{Y}_S)}{H(\mathcal{X}_S)} \le  1-\mu
\label{eq:30}
\end{equation} 
First, we derive a lower bound for $\frac{H(\mathcal{X}_S \vert \mathcal{Y}_S)}{H(\mathcal{X}_S)}$. We have:
\begin{align}
\small
\frac{H(\mathcal{X}_S \vert \mathcal{Y}_S)}{H(\mathcal{X}_S)} &= \frac{p(X_1, Y_1) \cdot \log p(X_1 \vert Y_1) }   {\sum_{i} p(X_i) \cdot \log p(X_i)} \nonumber \\ &+ \frac{\sum_{i \neq 1 \& j \neq 1} p(X_i, Y_j) \cdot \log \frac{p(X_i , Y_j)} {p(Y_j)}}{\sum_{i} p(X_i) \cdot \log p(X_i)}
\label{eq:31}
\end{align}		
We first consider the numerator in Eq. \eqref{eq:31}. Suppose that:
\begin{align}
\small
\frac{p(X_m,Y_n)}{p(Y_n)} = \min \lbrace\frac{p(X_i,Y_j)}{p(Y_j)}\rbrace, \forall (i \neq 1 \& j \neq 1)
\end{align}
Then, applying the min-max inequality theorem \cite{beckenbach1961introduction}, we have:
\begingroup
\begin{align}
\small
\frac{p(X_m,Y_n)}{p(Y_n)} \leq \frac{\sum_{i \neq 1 \& j \neq 1} p(X_i,Y_j)}{\sum_{i \neq 1 \& j \neq 1} p(Y_j)} = \frac{1-p(X_1,Y_1)}{n_x-p(Y_1)}
\end{align}
\begin{align}
\small
\Rightarrow \log \frac{p(X_m,Y_n)}{p(Y_n)} \leq \log \frac{1-p(X_1,Y_1)}{n_x-p(Y_1)} 
\end{align}
\begin{align}
\small
\Rightarrow p(X_m,Y_n) \cdot \log \frac{p(X_m,Y_n))}{p(Y_n)} &\leq p(X_m,Y_n) \cdot \log \frac{1-p(X_1,Y_1)}{n_x-p(Y_1)} 
\label{eq:minmax1}
\end{align}
\endgroup
Now, consider the second term of the numerator in Eq. \eqref{eq:31}. Since we have $\log \frac{p(X_i , Y_j)} {p(Y_j)} < 0$, hence:
\begin{align}
\small
\sum_{i \neq 1 \& j \neq 1} p(X_i, Y_j) \cdot \log \frac{p(X_i , Y_j)} {p(Y_j)} &\leq p(X_m,Y_n) \cdot \log \frac{p(X_m,Y_n)}{p(Y_n)} 
\label{eq:3new}
\end{align}
From Eqs \eqref{eq:minmax1}, \eqref{eq:3new}, it follows that:
\begin{align}
\small
\sum_{i \neq 1 \& j \neq 1} p(X_i, Y_j) \cdot \log \frac{p(X_i , Y_j)} {p(Y_j)} &\leq p(X_m,Y_n) \cdot \log \frac{1-p(X_1,Y_1)}{n_x-p(Y_1)} \nonumber \\
&= \lambda_2 \cdot \log \frac{1-p(X_1,Y_1)}{n_x-p(Y_1)}
\label{eq:3new1}
\end{align}
where $\lambda_2=p(X_m,Y_n)$.
\\
Next, we consider the denominator in Eq. \eqref{eq:31}. Suppose that:
\begin{align}
\small
p(X_k) = \min \lbrace p(X_i)\rbrace, \forall i 
\end{align}
Then we have:
\begin{align}
\small
p(X_i) &\ge p(X_k), \forall i \nonumber \\
\Rightarrow \log  p(X_i) &\ge \log p(X_k) \nonumber \\
\Rightarrow p(X_i)\log  p(X_i) &\ge p(X_i)\log p(X_k) \nonumber \\
\Rightarrow \sum_{i} p(X_i)\log  p(X_i) &\ge \sum_{i} p(X_i)\log p(X_k) \nonumber \\
&= \log p(X_k) \sum_{i} p(X_i) \nonumber\\
&= \log p(X_k) \nonumber\\
&= \log \lambda_1
\label{eq:minmax2}
\end{align}
where $\lambda_1 = p(X_k)$.
\\
Replace Eq. \eqref{eq:3new1} and \eqref{eq:minmax2} into Eq. \eqref{eq:31}, we get:		

\begin{align}
\small
\frac{H(\mathcal{X}_S \vert \mathcal{Y}_S)}{H(\mathcal{X}_S)} \geq  \frac{p(X_1, Y_1) \cdot \log p(X_1 \vert Y_1) + \lambda_2 \cdot \log \frac{1-p(X_1,Y_1)}{n_x-p(Y_1)}}{\log \lambda_1} 
\label{eq:33}
\end{align}	
Next, we derive the confidence lower bound of $(X_1,Y_1)$ in $\mathcal{D}_{\text{SYB}}$. Assuming that $\textit{supp}(X_1,Y_1)_{\mathcal{D}_{\text{SEQ}}} \geq\textit{supp}(X_1,Y_1)_{\mathcal{D}_{\text{SYB}}} \geq \sigma$. We consider two cases.

\textbf{$\ast$ Case 1:} Assuming that $\textit{supp}(Y_{1})_{\mathcal{D}_{\text{SYB}}} \ge \textit{supp}(X_{1})_{\mathcal{D}_{\text{SYB}}}$. Thus, the confidence of $(X_1,Y_1)$ in $\mathcal{D}_{\text{SYB}}$ is computed as
\begin{align}
\small
\textit{conf}(X_1,Y_1)_{\mathcal{D}_{\text{SYB}}} = \frac{\textit{supp}(X_1,Y_1)_{\mathcal{D}_{\text{SYB}}}}{\textit{supp}(Y_{1})_{\mathcal{D}_{\text{SYB}}}} = \frac{p(X_1,Y_1)}{p(Y_1)}
\end{align}	
\\	
Since we have: $\log \frac{p(X_1,Y_1)}{p(Y_1)} < 0$, and $\sigma \le p(X_1,Y_1)$, we can deduce: 
\begin{equation}
\small
p(X_1,Y_1) \cdot \log \frac{p(X_1,Y_1)}{p(Y_1)} \leq \sigma \cdot \log \frac{p(X_1,Y_1)}{p(Y_1)}
\label{eq:case1_21}
\end{equation}
\\
We also have: $1-p(X_1,Y_1) \leq 1-\sigma$, and $ n_x - 1 \leq n_x - p(Y_1)$, hence:
\begin{align}
\small
\frac{1-p(X_1,Y_1)}{n_x- p(Y_1)} &\leq \frac{1-\sigma}{ n_x - 1} \nonumber \\
\Rightarrow \lambda_2 \cdot \log \frac{1-p(X_1,Y_1)}{n_x- p(Y_1)} &\leq \lambda_2 \cdot \log \frac{1-\sigma}{ n_x - 1}
\label{eq:case1_22}
\end{align}
\\
From Eqs. \eqref{eq:case1_21}, \eqref{eq:case1_22}, we have: 
\begin{align}
\small
p(X_1, Y_1) \cdot \log \frac{p(X_1,Y_1)}{p(Y_1)} + \lambda_2 \cdot \log \frac{1-p(X_1,Y_1)}{n_x-p(Y_1)} \leq \nonumber 
 &\\ \sigma \cdot \log \frac{p(X_1,Y_1)}{p(Y_1)} + \lambda_2 \cdot \log \frac{1-\sigma}{n_x - 1} \le 0
\label{eq:case1_2}
\end{align}
Replace Eq. \eqref{eq:case1_2} into the numerator of Eq. \eqref{eq:33}, we get:
\begin{align}
\small
\frac{H(\mathcal{X}_S \vert \mathcal{Y}_S)}{H(\mathcal{X}_S)} \geq \frac{\sigma \cdot \log \frac{p(X_1,Y_1)}{p(Y_1)} + \lambda_2 \cdot \log \frac{1-\sigma}{n_x - 1}} {\log \lambda_1} 
\label{eq:case1_3}
\end{align}			
\\	
From Eqs. \eqref{eq:30} and \eqref{eq:case1_3}, it follows that:
	\begin{align}
	\small
	(1-\mu) &\geq \frac{\sigma \cdot \log \frac{p(X_1,Y_1)}{p(Y_1)} + \lambda_2 \cdot \log \frac{1-\sigma}{n_x - 1}} {\log \lambda_1} 
	\label{eq:case1_4}
	\end{align}
Hence:
	\begin{align}
	\small
	\textit{conf}(X_1,Y_1)_{\mathcal{D}_{\text{SYB}}} = \frac{p(X_1,Y_1)}{p(Y_1)} \geq  \lambda_1^{\frac{1-\mu}{\sigma} } \cdot \left(\frac{n_x - 1}{1-\sigma}\right)^\frac{\lambda_2}{\sigma}
	\label{eq:case1_5}
	\end{align}
	
\textbf{$\ast$ Case 2:} Assuming that $\textit{supp}(Y_{1})_{\mathcal{D}_{\text{SYB}}} < \textit{supp}(X_{1})_{\mathcal{D}_{\text{SYB}}}$. Thus, the confidence of $(X_1,Y_1)$ in $\mathcal{D}_{\text{SYB}}$ is computed as 
\begin{equation}
\small
\textit{conf}(X_1,Y_1)_{\mathcal{D}_{\text{SYB}}} = \frac{\textit{supp}(X_1,Y_1)_{\mathcal{D}_{\text{SYB}}}}{\textit{supp}(X_{1})_{\mathcal{D}_{\text{SYB}}}} = \frac{p(X_1,Y_1)}{p(X_1)}
\label{eq:case2_1}
\end{equation}
From Eq. \eqref{eq:33}, we have:
	\begingroup
	\begin{align} 
	\small
	\frac{H(\mathcal{X}_S \vert \mathcal{Y}_S)}{H(\mathcal{X}_S)} &\geq  \frac{p(X_1, Y_1) \cdot \log \left( \frac{p(X_1,Y_1)}{p(X_1)} \frac{p(X_1)}{p(Y_1)} \right) + \lambda_2 \cdot \log \frac{1-p(X_1,Y_1)}{n_x-p(Y_1)}} {\log \lambda_1} 	\label{eq:case2_2} \\
	&\geq \frac{\sigma \cdot \log \left( \frac{p(X_1,Y_1)}{p(X_1)} \cdot \frac{1}{\sigma} \right)+ \lambda_2 \cdot \log \frac{1-\sigma}{n_x - 1}} {\log \lambda_1} 
	\label{eq:case2_3}
	\end{align}
	\endgroup
From Eqs. \eqref{eq:30}, \eqref{eq:case2_3}, it follows that:
	\begin{align}
	\small
	 (1-\mu) \geq \frac{\sigma \cdot \log \left( \frac{p(X_1,Y_1)}{p(X_1)} \cdot \frac{1}{\sigma} \right)+ \lambda_2 \cdot \log \frac{1-\sigma}{n_x - 1}} {\log \lambda_1}
	 \label{eq:case2_4}
	\end{align}
Hence:
	\begin{align}
	\small
	\textit{conf}(X_1,Y_1)_{\mathcal{D}_{\text{SYB}}} = \frac{p(X_1,Y_1)}{p(X_1)} \geq \lambda_1^{\frac{1-\mu}{\sigma} } \cdot \left(\frac{n_x - 1}{1-\sigma} \right)^\frac{\lambda_2}{\sigma} \cdot \sigma
	\label{eq:case2_5}
	\end{align}
	\\
	In Eq. \eqref{eq:case2_5}, we have $\sigma < 1$. Thus, from Eqs.  \eqref{eq:case1_5}, \eqref{eq:case2_5}, the confidence lower bound of $(X_1,Y_1)$ in $\mathcal{D}_{\text{SYB}}$ in both cases is:
	\begingroup
	\begin{align}
	\small
	\textit{conf}(X_1,Y_1)_{\mathcal{D}_{\text{SYB}}}  &\geq  \lambda_1^{\frac{1-\mu}{\sigma} } \cdot \left(\frac{n_x - 1}{1-\sigma}\right)^\frac{\lambda_2}{\sigma}
	\label{eq:15}
	\end{align}
	\endgroup
	\\
	Next, we derive the confidence of $(X_1,Y_1)$ in the temporal sequence database $\mathcal{D}_{\text{SEQ}}$. From Lemma \ref{lem:supportconnection1}, we have:
	\begin{align}
	\small
	\textit{supp}(X_1)_{\mathcal{D}_{\text{SYB}}} \le \textit{supp}(X_1)_{\mathcal{D}_{\text{SEQ}}} %\le \sigma_{\max}
	\label{eq:confSeq1}
	\end{align} 
	\begin{align}
	\small
	\textit{supp}(Y_1)_{\mathcal{D}_{\text{SYB}}} \le \textit{supp}(Y_1)_{\mathcal{D}_{\text{SEQ}}} %\le \sigma_{\max}
	\label{eq:confSeq2}
	\end{align} 
	\begin{align}
	\small
	\textit{supp}(X_1,Y_1)_{\mathcal{D}_{\text{SYB}}} \le \textit{supp}(X_1,Y_1)_{\mathcal{D}_{\text{SEQ}}} %\le \sigma_{\max}
	\label{eq:confSeq3}
	\end{align} 
	\\
	Without loss of generality, we assume $\textit{supp}(X_1)_{\mathcal{D}_{\text{SEQ}}}$ $\ge$ $\textit{supp}(Y_1)_{\mathcal{D}_{\text{SEQ}}}$. Hence, the confidence of $(X_1,Y_1)$ in $\mathcal{D}_{\text{SEQ}}$ is computed as
	\begingroup
	\begin{align}
	\small
	\textit{conf}(X_1,Y_1)_{\mathcal{D}_{\text{SEQ}}} &=\frac{\textit{supp}(X_1,Y_1)_{\mathcal{D}_{\text{SEQ}}}}{\textit{supp}(X_1)_{\mathcal{D}_{\text{SEQ}}}} \nonumber \\ &\geq \frac{\textit{supp}(X_1,Y_1)_{\mathcal{D}_{\text{SYB}}}}{\textit{supp}(X_1)_{\mathcal{D}_{\text{SEQ}}}} \nonumber \\ &=
	\frac{\textit{supp}(X_1,Y_1)_{\mathcal{D}_{\text{SYB}}}}{\textit{supp}(X_1)_{\mathcal{D}_{\text{SYB}}}} \cdot \frac{\textit{supp}(X_1)_{\mathcal{D}_{\text{SYB}}}}{\textit{supp}(X_1)_{\mathcal{D}_{\text{SEQ}}}} \nonumber \\ &=
	\textit{conf}(X_1,Y_1)_{\mathcal{D}_{\text{SYB}}} \cdot \frac{\textit{supp}(X_1)_{\mathcal{D}_{\text{SYB}}}}{\textit{supp}(X_1)_{\mathcal{D}_{\text{SEQ}}}}
	\label{eq:confSeq4}
	\end{align}	
	\endgroup
	\\
Since we have $\sigma \le \textit{supp}(X_1)_{\mathcal{D}_{\text{SYB}}}$ and $\textit{supp}(X_1)_{\mathcal{D}_{\text{SEQ}}} \le 1$, it follows that:
\begin{align}
\small
\frac{\textit{supp}(X_1)_{\mathcal{D}_{\text{SYB}}}}{\textit{supp}(X_1)_{\mathcal{D}_{\text{SEQ}}}} \ge \sigma
\label{eq:confSeq4_1}
\end{align}

Finally, from Eqs. \eqref{eq:15}, \eqref{eq:confSeq4} and \eqref{eq:confSeq4_1}, we can derive the confidence lower bound of $(X_1,Y_1)$ in $\mathcal{D}_{\text{SEQ}}$:
\begin{align}
	\small
	\textit{conf}(X_1,Y_1)_{\mathcal{D}_{\text{SEQ}}}  &\geq \sigma \cdot \lambda_1^{\frac{1-\mu}{\sigma} } \cdot \left(\frac{n_x - 1}{1-\sigma}\right)^\frac{\lambda_2}{\sigma} 
\end{align}

\end{proof}

\section{Additional Experimental Results}
\subsection{Baselines comparison}
Tables \ref{tbl:runtimeBaselinesAppendix} and \ref{tbl:memoryBaselinesAppendix} show the comparison results between A-HTPGM, E-HTPGM and the baselines on the UKDALE, DataPort, and ASL datasets. A-HTPGM achieves the best performance (both runtime and memory usage) among all methods, and E-HTPGM has better performance than the baselines. 
The range and average speedups of E-HTPGM compared to the baselines are: $[1.1-5.4]$ and $2.2$ (Z-Miner), $[3.1-12.7]$ and $6.1$ (TPMiner), $[3.7-30.1]$ and $10.8$ (IEMiner), and $[4.3-35.1]$ and $12.8$ (H-DFS). 
Instead, the speedups of A-HTPGM compared to E-HTPGM and the baselines respectively are: $[1.5-8.8]$ and $4.1$ (E-HTPGM), $[2.3-14.2]$ and $7.9$ (Z-Miner), $[11.7-38.1]$ and $21.2$ (TPMiner), $[15.3-116.3]$ and $40.6$ (IEMiner), and $[17.2-135.4]$ and $48.2$ (H-DFS). 
Note that the time to compute MI and $\mu$ for the UKDALE, DataPort, and ASL datasets in Table \ref{tbl:runtimeBaselinesAppendix} are $17.82$, $8.37$, $11.27$ seconds, respectively. 

In average, on the tested datasets, E-HTPGM consumes ${\sim}3.8$ times less memory than the baselines due to the applied pruning techniques, while A-HTPGM uses ${\sim}4.7$ times less memory than E-HTPGM and the baselines by pruning uncorrelated series. 

\begin{table*}[!h]
	\centering
	\caption{Runtime Comparison (seconds)}
	\begin{minipage}{.59\linewidth}
		\resizebox{\columnwidth}{!}{
			\begin{tabular}{|c|c|c|c|c|c|c|c|}
				\hline 
				\multirow{3}{*}{Supp. (\%)} & \multirow{3}{*}{Methods}   & \multicolumn{6}{c|}{\bfseries Conf. (\%)}
				\\  \cline{3-8}  
				& & \multicolumn{3}{c|}{\bfseries UKDALE} & \multicolumn{3}{c|}{\bfseries DataPort}   
				\\  \cline{3-8}  
				& & {\bfseries 20} & {\bfseries 50}  & {\bfseries 80} & {\bfseries 20} & {\bfseries 50}  & {\bfseries 80} \\
				\hline
				\multirow{5}{*}{20}  & H-DFS  & 20130.09  & 3207.59   & 909.24  & 1802.76  & 1256.29 & 225.45   \\  \cline{2-8}  
				& \centering IEMiner  & 15308.61  & 2841.64   & 816.51  & 1359.95  & 1006.34   & 208.62      \\  \cline{2-8}  
				&\centering TPMiner  & 9045.28  & 2317.69   & 601.73  & 633.64  & 360.19   & 101.96    \\  \cline{2-8}  
				& \centering Z-Miner   & 2867.95  & 869.58 & 123.87  & 214.82 & 110.35  & 85.04    \\  \cline{2-8} 
				&\centering E-HTPGM  & 1958.01  & 653.08  & 92.33  & 94.29 & 73.16   & 29.58    \\  \cline{2-8} 
				&\centering A-HTPGM & {\bfseries 571.14}  & {\bfseries 173.51}  & {\bfseries 41.93} & {\bfseries 25.35}  & {\bfseries 12.42}   &  {\bfseries 8.23}  \\  \hline  
				
				\multirow{5}{*}{50}  & H-DFS  & 6208.26  & 1831.54   & 559.25  & 526.24  & 391.59 & 110.47   \\  \cline{2-8}  
				& \centering IEMiner  & 5481.08  & 1570.18   & 426.13  & 460.37  & 321.78   & 93.64    \\  \cline{2-8}    
				&\centering TPMiner  & 3251.94  & 1007.26   & 381.07  & 267.64  & 164.83   & 84.81      \\  \cline{2-8}  
				& \centering Z-Miner   & 963.36  & 853.56 & 118.70  & 106.59 & 72.84  & 50.17    \\  \cline{2-8} 
				&\centering E-HTPGM  & 269.68  & 159.25  & 71.78 & 73.27 & 52.17   & 25.28    \\  \cline{2-8} 
				&\centering A-HTPGM & {\bfseries 110.87}  & {\bfseries 83.11}  & {\bfseries 26.66} & {\bfseries 9.37}  & {\bfseries 6.35}   &  {\bfseries 4.16}  \\  \hline 
				
				\multirow{5}{*}{80}  & H-DFS  & 915.37  & 509.04   & 328.15  & 42.19  & 21.68 & 12.09   \\  \cline{2-8}  
				& \centering IEMiner  & 852.81  & 454.26   & 294.38  & 35.09  & 16.95   & 10.36      \\  \cline{2-8}  
				&\centering TPMiner  & 597.34  & 400.15   & 251.34  & 27.61  & 13.82   & 9.08    \\  \cline{2-8}  
				& \centering Z-Miner   & 123.28  & 115.76 & 104.41  & 13.07 & 7.51  & 4.17    \\  \cline{2-8} 
				&\centering E-HTPGM  & 82.78  & 63.34  & 41.92  & 3.12 & 2.08   & 1.13    \\  \cline{2-8} 
				&\centering A-HTPGM & {\bfseries 32.33}  & {\bfseries 21.20}  & {\bfseries 19.12} & {\bfseries 1.08}  & {\bfseries 0.62}   &  {\bfseries 0.41}  \\  \hline 
			\end{tabular}	
		}
	\end{minipage}
	\begin{minipage}{.382\linewidth}
		\resizebox{\columnwidth}{!}{
			\begin{tabular}{|c|c|c|c|c|}
				\hline 
				\multirow{3}{*}{Supp. (\%)} & \multirow{3}{*}{Methods}   & \multicolumn{3}{c|}{\bfseries Conf. (\%)}
				\\  \cline{3-5}  
				& & \multicolumn{3}{c|}{\bfseries ASL}  
				\\  \cline{3-5}  
				& & {\bfseries 0.5} & {\bfseries 1}  & {\bfseries 10} \\
				\hline
				\multirow{5}{*}{0.5}  & H-DFS  & 2028.32  & 1523.57   & 354.61     \\  \cline{2-5}  
				& \centering IEMiner  & 1567.21  & 1307.83  & 301.53     \\  \cline{2-5}  
				&\centering TPMiner  & 674.25  & 428.46  & 113.04    \\  \cline{2-5}  
				& \centering Z-Miner   & 239.24  & 140.92 & 36.35    \\  \cline{2-5} 
				&\centering E-HTPGM  & 126.28 & 91.28   & 24.23   \\  \cline{2-5} 
				&\centering A-HTPGM & {\bfseries 57.15}  & {\bfseries 11.25}   &  {\bfseries 9.09}   \\  \hline  
				
				\multirow{5}{*}{1}  & H-DFS  & 667.28 &	460.59 &	136.71    \\  \cline{2-5}  
				& \centering IEMiner  & 547.63 &	384.03 &	116.27    \\  \cline{2-5}  
				&\centering TPMiner  & 209.53 &	136.19 &	82.14  \\  \cline{2-5}  
				& \centering Z-Miner   & 68.95  & 58.01 & 26.13    \\  \cline{2-5} 
				&\centering E-HTPGM  & 41.25 & 35.93   & 13.79   \\  \cline{2-5} 
				&\centering A-HTPGM & {\bfseries 5.91}  & {\bfseries 4.07}   &  {\bfseries 3.03}   \\  \hline 
				
				\multirow{5}{*}{10}  & H-DFS  & 45.61 &	25.59 &	5.59   \\  \cline{2-5}  
				& \centering IEMiner  & 40.27 &	21.91 &	5.07  \\  \cline{2-5}  
				&\centering TPMiner  & 15.64 &	9.24 &	3.51  \\  \cline{2-5}  
				& \centering Z-Miner   & 2.07  & 1.58 & 1.07    \\  \cline{2-5} 
				&\centering E-HTPGM  & 1.95 & 0.73   & 0.52   \\  \cline{2-5} 
				&\centering A-HTPGM & {\bfseries 0.89}  & {\bfseries 0.48}   &  {\bfseries 0.21}   \\  \hline 
			\end{tabular}	
		}
	\end{minipage}
	\label{tbl:runtimeBaselinesAppendix}
\end{table*}

\begin{table*}[!h]
	\centering
	\caption{Memory Usage Comparison (MB)}
	\begin{minipage}{.59\linewidth}
		\resizebox{\columnwidth}{!}{
			\begin{tabular}{|c|c|c|c|c|c|c|c|}
				\hline 
				\multirow{3}{*}{Supp. (\%)} & \multirow{3}{*}{Methods}   & \multicolumn{6}{c|}{\bfseries Conf. (\%)}
				\\  \cline{3-8}  
				& & \multicolumn{3}{c|}{\bfseries UKDALE} & \multicolumn{3}{c|}{\bfseries DataPort}   
				\\  \cline{3-8}  
				& & {\bfseries 20} & {\bfseries 50}  & {\bfseries 80} & {\bfseries 20} & {\bfseries 50}  & {\bfseries 80} \\
				\hline
				\multirow{5}{*}{20}  & H-DFS  & 13164.37  & 3258.19   & 1834.82  & 957.41  & 405.86 & 118.37   \\  \cline{2-8}  
				& \centering IEMiner  & 11307.08  & 3014.82   & 1740.35  & 915.73  & 397.61   & 105.94      \\  \cline{2-8}  
				&\centering TPMiner  & 7521.59  & 2641.58   & 1354.26  & 842.62  & 243.94   & 93.68    \\  \cline{2-8}  
				& \centering Z-Miner   & 21068.68  & 5660.88 & 2172.18  & 1512.52 & 1137.61  & 198.02   \\  \cline{2-8} 
				&\centering E-HTPGM  & 1018.93  & 899.92   & 488.93  & 399.42  & 123.62   & 50.03    \\  \cline{2-8} 
				&\centering A-HTPGM & {\bfseries 893.09}  & {\bfseries 681.97}  & {\bfseries 391.58} & {\bfseries 245.20}  & {\bfseries 74.41}   &  {\bfseries 42.77}  \\  \hline   	
				\multirow{5}{*}{50}  & H-DFS  & 8643.31  & 2257.36   & 1294.35  & 720.09  & 322.26 & 106.46   \\  \cline{2-8}  
				& \centering IEMiner  & 7652.61  & 2010.53   & 1034.31  & 685.94  & 300.57   & 98.63      \\  \cline{2-8}  
				&\centering TPMiner  & 3134.85  & 1623.25   & 921.08  & 469.33  & 219.63   & 91.54    \\  \cline{2-8}  
				& \centering Z-Miner   & 9599.25  & 3593.45 & 1437.48  & 918.71 & 594.86  & 130.22   \\  \cline{2-8} 
				&\centering E-HTPGM  & 894.02  & 720.07   & 284.89  & 175.55  & 88.11   & 40.69    \\  \cline{2-8} 
				&\centering A-HTPGM & {\bfseries 554.86}  & {\bfseries 391.40}  & {\bfseries 219.34} & {\bfseries 129.19}  & {\bfseries 58.12}   &  {\bfseries 31.99}  \\  \hline  
				
				\multirow{5}{*}{80}  & H-DFS  & 1943.05  & 1103.62   & 827.34  & 301.04  & 143.67 & 58.94   \\  \cline{2-8}  
				& \centering IEMiner  & 1826.37  & 1004.31   & 807.92  & 297.62  & 120.83   & 52.63      \\  \cline{2-8}  
				&\centering TPMiner  & 1406.52  & 867.38   & 759.37  & 172.93  & 112.54   & 43.82    \\  \cline{2-8}  
				& \centering Z-Miner   & 2267.86  & 1365.96 & 1065.80  & 416.66 & 197.51  & 61.30   \\  \cline{2-8} 
				&\centering E-HTPGM  & 396.65  & 276.50   & 206.24  & 80.06  & 68.22   & 27.14    \\  \cline{2-8} 
				&\centering A-HTPGM & {\bfseries 287.12}  & {\bfseries 186.16}  & {\bfseries 150.97} & {\bfseries 57.66}  & {\bfseries 35.82}   &  {\bfseries 19.56}  \\  \hline  
			\end{tabular}	
		}
	\end{minipage}
	\begin{minipage}{.376\linewidth}
		\resizebox{\columnwidth}{!}{
			\begin{tabular}{|c|c|c|c|c|c|c|}
				\hline 
				\multirow{3}{*}{Supp. (\%)} & \multirow{3}{*}{Methods}   & \multicolumn{3}{c|}{\bfseries Conf. (\%)}
				\\  \cline{3-5}  
				& & \multicolumn{3}{c|}{\bfseries ASL}
				\\  \cline{3-5}  
				& & {\bfseries 0.5} & {\bfseries 1}  & {\bfseries 10} \\
				\hline
				\multirow{5}{*}{0.5}  & H-DFS  & 1007.24 &	416.48 &	126.37     \\  \cline{2-5}  
				& \centering IEMiner  & 946.81 &	409.21 & 120.54    \\  \cline{2-5}  
				&\centering TPMiner  & 914.52 &	250.46 &	103.59    \\  \cline{2-5}  
				& \centering Z-Miner   & 1116.61  & 736.76 & 201.16    \\  \cline{2-5}
				&\centering E-HTPGM  & 169.17  & 138.88 & 80.04    \\  \cline{2-5} 
				&\centering A-HTPGM & {\bfseries 123.75}& {\bfseries 115.24}	 & {\bfseries 61.66}	    \\  \hline  
				
				\multirow{5}{*}{1}  & H-DFS  &  765.24 &	372.18 &	118.37   \\  \cline{2-5}  
				& \centering IEMiner  & 713.48 &	351.22 &	110.73    \\  \cline{2-5}  
				&\centering TPMiner  &  421.67 &	233.47 &	97.35  \\  \cline{2-5}  
				& \centering Z-Miner   & 811.62  & 492.59 & 197.88    \\  \cline{2-5}
				&\centering E-HTPGM  & 153.69  & 125.07 & 48.78    \\  \cline{2-5} 
				&\centering A-HTPGM & {\bfseries 114.05}& {\bfseries 84.46}	 & {\bfseries 32.26}	    \\  \hline 
				
				\multirow{5}{*}{10}  & H-DFS  & 324.67 &	110.27 &	69.47     \\  \cline{2-5}  
				& \centering IEMiner  &  318.32 &	107.33 &	60.45    \\  \cline{2-5}  
				&\centering TPMiner  &  191.08	& 100.81 &	 55.34    \\  \cline{2-5} 
				& \centering Z-Miner   & 343.04  & 142.99 & 134.41    \\  \cline{2-5}
				&\centering E-HTPGM  & 57.04  & 51.28 & 43.85    \\  \cline{2-5} 
				&\centering A-HTPGM & {\bfseries 52.78}& {\bfseries 38.52}	 & {\bfseries 23.91}	    \\  \hline  
			\end{tabular}
		}
	\end{minipage}
	\label{tbl:memoryBaselinesAppendix}
\end{table*}

\begin{table*}[h!]
	\centering
	\begin{minipage}{.47\linewidth}
		\caption{The Accuracy and the Number of Extracted Patterns from A-HTPGM on UKDALE}
		\resizebox{\columnwidth}{!}{
			\begin{tabular}{|c|c|c|c|c|c|c|c|c|}
				\hline 
				\multirow{4}{*}{Supp. (\%)} & \multicolumn{8}{c|}{\bfseries Conf. (\%)} \\  
				\cline{2-9}  
				& \multicolumn{8}{c|}{\bfseries UKDALE}
				\\  \cline{2-9}  
				& \multicolumn{4}{c|}{\bfseries Accuracy (\%)}  & \multicolumn{4}{c|}{\bfseries \# Patterns} 
				\\  \cline{2-9}
				& {\bfseries 10} & {\bfseries 20} & {\bfseries 50}  & {\bfseries 80} & {\bfseries 10} & {\bfseries 20} & {\bfseries 50}    & {\bfseries 80} \\
				\hline
				10 & 70  & 75   & 90 & 100 & 1647209 & 1058644 & 101327 & 20341  
				\\  \hline			
				20 & 73  & 78   & 95 & 100 & 924612 & 842061 & 72649 & 15934   
				\\  \hline				
				50 & 89  & 93   & 98 & 100 & 69367 & 62478 & 58326 & 13517   
				\\  \hline					
				80 & 100  & 100   & 100 & 100 & 12584 & 11358 & 10659 & 10659 
				\\  \hline					
			\end{tabular}
		}
		\label{tbl:ratiopatterns_Appro_Exact_Appendix1}
	\end{minipage}
	\hspace{0.4in}
	\begin{minipage}{.44\linewidth}
		\caption{The Accuracy and the Number of Extracted Patterns from A-HTPGM on DataPort}
		\resizebox{\columnwidth}{!}{
			\begin{tabular}{|c|c|c|c|c|c|c|c|c|}
				\hline 
				\multirow{4}{*}{Supp. (\%)} & \multicolumn{8}{c|}{\bfseries Conf. (\%)} \\  
				\cline{2-9}  
				& \multicolumn{8}{c|}{\bfseries DataPort}
				\\  \cline{2-9}  
				& \multicolumn{4}{c|}{\bfseries Accuracy (\%)}  & \multicolumn{4}{c|}{\bfseries \# Patterns} 
				\\  \cline{2-9}
				& {\bfseries 10} & {\bfseries 20} & {\bfseries 50}  & {\bfseries 80} & {\bfseries 10} & {\bfseries 20} & {\bfseries 50}    & {\bfseries 80} \\
				\hline
				10 & 75  & 80   & 94 & 100 & 653225 & 226241 & 120654 & 9851  
				\\  \hline			
				20 & 80  & 83   & 95 & 100 & 204587 & 184217 & 45462 & 8126   
				\\  \hline				
				50 & 94  & 95   & 100 & 100 & 117562 & 100259 & 42096 & 7194   
				\\  \hline					
				80 & 98  & 100   & 100 & 100 & 9051 & 8462 & 6428 & 5203 
				\\  \hline							
			\end{tabular}
		}
		\label{tbl:ratiopatterns_Appro_Exact_Appendix2}
	\end{minipage}
\end{table*}

\subsection{Evaluation of the pruning techniques in E-HTPGM}
In this section, we report the evaluation results of E-HTPGM on UKDALE, DataPort and ASL datasets. We use $3$ different configurations that vary: the number of sequences, the confidence, and the support. Figs. \ref{fig:performance_AppendixUkdale}, \ref{fig:performance_AppendixDataport}, \ref{fig:performance_AppendixASL} show the results (the y-axis is in log scale). It can be seen that All-E-HTPGM achieves the best performance among all versions. Its speedup w.r.t. NoPrune-E-HTPGM ranges from $2.3$ up to $9.8$ depending on the configurations, showing that the proposed prunings are very effective in improving E-HTPGM performance. Furthermore, Trans-E-HTPGM brings larger speedup than Apriori-E-HTPGM. The speedup range is from $1.5$ to $5.2$ for Trans-E-HTPGM, and from $1.3$ to $3.6$ for Apriori-E-HTPGM. However, applying both always yields better speedup than applying either of them.

\begin{figure}[!h]
	\centering
	\begin{minipage}[t]{1\columnwidth} 
		\centering
		\begin{subfigure}{0.32\columnwidth}
			\centering
			\resizebox{\linewidth}{!}{
				\begin{tikzpicture}[scale=0.6]
				\begin{axis}[
				compat=newest,
				xlabel={Sequence (\%)},
				ylabel={Runtime (sec)}, 
				label style={font=\Huge},
				ticklabel style = {font=\Huge},
				xmin=20, xmax=100,
				ymin=0, ymax=12000,
				xtick={20,40,60,80,100},
				legend columns=-1,
				legend entries = {NoPrune, Apriori, Trans, All},
				legend style={nodes={scale=0.55, transform shape}},
				legend to name={legendpruning},
				ymode=log,
				log basis y={10},
				ymajorgrids=true,
				grid style=dashed,
				line width=1.75pt
				]
				\addplot[
				color=blue,
				mark=oplus,
				mark size=4pt,
				] 
				coordinates {
					(20,1087.76)(40,3223.14)(60,5564.81)(80,8832.87)(100,10341.15)	
				};
				
				\addplot[
				color=black,
				mark=triangle,
				mark size=4pt,
				] 
				coordinates {
					(20,575.98)(40,1642.54)(60,3358.46)(80,4949.09)(100,6120.89)	
				};
				
				\addplot[
				color=teal,
				mark=square*,
				mark size=4pt,
				]	
				coordinates {
					(20,422.85)(40,942.92)(60,2318.43)(80,3489.78)(100,4561.09)	
				};
				
				\addplot[
				color=red,
				mark=*,
				mark size=4pt,
				] 
				coordinates {
					(20,217.96)(40,691.65)(60,1175.55)(80,2312.14)(100,3114.04)	
				};
				\end{axis}
				\end{tikzpicture}
			}
			\caption{Varying \% Seq.}
		\end{subfigure}
		\begin{subfigure}{0.32\columnwidth}
			\centering
			\resizebox{\linewidth}{!}{
				\begin{tikzpicture}[scale=0.6]
				\begin{axis}[
				compat=newest,
				xlabel={Confidence (\%)},
				ylabel={Runtime (sec)}, 
				label style={font=\Huge},
				ticklabel style = {font=\Huge},
				xmin=20, xmax=100,
				ymin=0, ymax=12000,
				xtick={20,40,60,80,100},
				legend columns=-1,
				legend entries = {NoPrune, Apriori, Trans, All},
				legend style={nodes={scale=0.55, transform shape}},
				legend to name={legendpruning},
				ymode=log,
				log basis y={10},
				ymajorgrids=true,
				grid style=dashed,
				line width=1.75pt
				]
				\addplot[
				color=blue,
				mark=oplus,
				mark size=4pt,
				] 	
				coordinates {
					(20,10341.15)(40,8023.79)(60,3561.45)(80,1088.49)(100,124.31)	
				};
				
				\addplot[
				color=black,
				mark=triangle,
				mark size=4pt,
				]	
				coordinates {
					(20,6120.89)(40,4084.12)(60,1702.88)(80,298.85)(100,55.69)		
				};
				
				\addplot[
				color=teal,
				mark=square*,
				mark size=4pt,
				] 
				coordinates {
					(20,4561.09)(40,2308.55)(60,1290.45)(80,398.91)(100,84.16)	
				};
				
				\addplot[
				color=red,
				mark=*,
				mark size=4pt,
				] 
				coordinates {
					(20,3114.04)(40,1442.85)(60,735.67)(80,152.25)(100,26.21)	
				};
				\end{axis}
				\end{tikzpicture}
			}
			\caption{Varying Conf.}
		\end{subfigure}
		\begin{subfigure}{0.32\columnwidth}
			\centering
			\resizebox{\linewidth}{!}{
				\begin{tikzpicture}[scale=0.6]
				\begin{axis}[
				compat=newest,
				xlabel={Support (\%)},
				ylabel={Runtime (sec)}, 
				label style={font=\Huge},
				ticklabel style = {font=\Huge},
				xmin=20, xmax=100,
				ymin=0, ymax=12000,
				xtick={20,40,60,80,100},
				legend columns=-1,
				legend entries = {NoPrune, Apriori, Trans, All},
				legend style={nodes={scale=0.55, transform shape}},
				legend to name={legendpruning},
				ymode=log,
				log basis y={10},
				ymajorgrids=true,
				grid style=dashed,
				line width=1.75pt
				]
				\addplot[
				color=blue,
				mark=oplus,
				mark size=4pt,
				] 	
				coordinates {
					(20,10341.15)(40,6882.85)(60,2688.39)(80,805.58)(100,95.11)
				};
				
				\addplot[
				color=black,
				mark=triangle,
				mark size=4pt,
				]	
				coordinates {
					(20,6120.89)(40,4160.72)(60,1454.70)(80,520.88)(100,67.05)
				};
				
				\addplot[
				color=teal,
				mark=square*,
				mark size=4pt,
				] 
				coordinates {
					(20,4561.09)(40,2010.21)(60,575.47)(80,156.43)(100,26.21)
				};
				
				\addplot[
				color=red,
				mark=*,
				mark size=4pt,
				] 
				coordinates {
					(20,3114.04)(40,1154.69)(60,400.82)(80,124.90)(100,21.79)
				};
				\end{axis}
				\end{tikzpicture}
			}
			\caption{Varying Supp.}
		\end{subfigure}
		\vspace{-0.05in}
		\ref{legendpruning}
		\vspace{-0.05in}
		\caption{Runtimes of E-HTPGM on UKDALE}
		\label{fig:performance_AppendixUkdale}
	\end{minipage} 
\end{figure}

\begin{figure}[!h]
	\centering
	\vspace{-0.2in}
	\begin{minipage}[t]{1\columnwidth} 
		\centering
		\begin{subfigure}{0.32\columnwidth}
			\centering
			\resizebox{\linewidth}{!}{
				\begin{tikzpicture}[scale=0.6]
				\begin{axis}[
				compat=newest,
				xlabel={Sequence (\%)},
				ylabel={Runtime (sec)}, 
				label style={font=\Huge},
				ticklabel style = {font=\Huge},
				xmin=20, xmax=100,
				ymin=0, ymax=1000,
				xtick={20,40,60,80,100},
				legend columns=-1,
				legend entries = {NoPrune, Apriori, Trans, All},
				legend style={nodes={scale=0.55, transform shape}},
				legend to name={legendpruning},
				ymode=log,
				log basis y={10},
				ymajorgrids=true,
				grid style=dashed,
				line width=1.75pt
				]
				\addplot[
				color=blue,
				mark=oplus,
				mark size=4pt,
				] 
				coordinates {
					(20,185.60)(40,309.59)(60,538.41)(80,734.16)(100,925.78)
				};
				
				\addplot[
				color=black,
				mark=triangle,
				mark size=4pt,
				] 
				coordinates {
					(20,133.06)(40,230.83)(60,326.91)(80,461.31)(100,679.10)
				};
				
				\addplot[
				color=teal,
				mark=square*,
				mark size=4pt,
				]	
				coordinates {
					(20,101.93)(40,189.91)(60,279.85)(80,393.78)(100,504.53)
				};
				
				\addplot[
				color=red,
				mark=*,
				mark size=4pt,
				] 
				coordinates {
					(20,76.06)(40,137.77)(60,198.22)(80,276.41)(100,368.58)
				};
				\end{axis}
				\end{tikzpicture}
			}
			\caption{Varying \% Seq.}
		\end{subfigure}
		\begin{subfigure}{0.32\columnwidth}
			\centering
			\resizebox{\linewidth}{!}{
				\begin{tikzpicture}[scale=0.6]
				\begin{axis}[
				compat=newest,
				xlabel={Confidence (\%)},
				ylabel={Runtime (sec)}, 
				label style={font=\Huge},
				ticklabel style = {font=\Huge},
				xmin=20, xmax=100,
				ymin=0, ymax=1000,
				xtick={20,40,60,80,100},
				legend columns=-1,
				legend entries = {NoPrune, Apriori, Trans, All},
				legend style={nodes={scale=0.55, transform shape}},
				legend to name={legendpruning},
				ymode=log,
				log basis y={10},
				ymajorgrids=true,
				grid style=dashed,
				line width=1.75pt
				]
				\addplot[
				color=blue,
				mark=oplus,
				mark size=4pt,
				] 	
				coordinates {
					(20,925.78)(40,450.29)(60,212.53)(80,60.14)(100,7.33)
				};
				
				\addplot[
				color=black,
				mark=triangle,
				mark size=4pt,
				]	
				coordinates {
					(20,679.10)(40,309.10)(60,133.04)(80,20.65)(100,2.98)
				};
				
				\addplot[
				color=teal,
				mark=square*,
				mark size=4pt,
				] 
				coordinates {
					(20,454.53)(40,200.51)(60,84.89)(80,34.80)(100,4.48)
				};
				
				\addplot[
				color=red,
				mark=*,
				mark size=4pt,
				] 
				coordinates {
					(20,308.58)(40,128.10)(60,67.26)(80,15.00)(100,1.57)
				};
				\end{axis}
				\end{tikzpicture}
			}
			\caption{Varying Conf.}
		\end{subfigure}
		\begin{subfigure}{0.32\columnwidth}
			\centering
			\resizebox{\linewidth}{!}{
				\begin{tikzpicture}[scale=0.6]
				\begin{axis}[
				compat=newest,
				xlabel={Support (\%)},
				ylabel={Runtime (sec)}, 
				label style={font=\Huge},
				ticklabel style = {font=\Huge},
				xmin=20, xmax=100,
				ymin=0, ymax=1000,
				xtick={20,40,60,80,100},
				legend columns=-1,
				legend entries = {NoPrune, Apriori, Trans, All},
				legend style={nodes={scale=0.55, transform shape}},
				legend to name={legendpruning},
				ymode=log,
				log basis y={10},
				ymajorgrids=true,
				grid style=dashed,
				line width=1.75pt
				]
				\addplot[
				color=blue,
				mark=oplus,
				mark size=4pt,
				] 	
				coordinates {
					(20,925.78)(40,397.85)(60,195.62)(80,52.21)(100,5.39)
				};
				
				\addplot[
				color=black,
				mark=triangle,
				mark size=4pt,
				]	
				coordinates {
					(20,679.10)(40,189.07)(60,68.86)(80,17.71)(100,2.11)
				};
				
				\addplot[
				color=teal,
				mark=square*,
				mark size=4pt,
				] 
				coordinates {
					(20,454.53)(40,251.21)(60,108.05)(80,28.22)(100,2.95)
				};
				
				\addplot[
				color=red,
				mark=*,
				mark size=4pt,
				] 
				coordinates {
					(20,308.58)(40,134.16)(60,46.02)(80,11.52)(100,1.37)
				};
				\end{axis}
				\end{tikzpicture}
			}
			\caption{Varying Supp.}
		\end{subfigure}
		\vspace{-0.05in}
		\ref{legendpruning}
		\vspace{-0.05in}
		\caption{Runtimes of E-HTPGM on DataPort}
		\label{fig:performance_AppendixDataport}
	\end{minipage} 
\end{figure}

\begin{figure}[!h]
	\centering
	\vspace{-0.2in}
	\begin{minipage}[t]{1\columnwidth} 
		\centering
		\begin{subfigure}{0.32\columnwidth}
			\centering
			\resizebox{\linewidth}{!}{
				\begin{tikzpicture}[scale=0.6]
				\begin{axis}[
				compat=newest,
				xlabel={Sequence (\%)},
				ylabel={Runtime (sec)}, 
				label style={font=\Huge},
				ticklabel style = {font=\Huge},
				xmin=20, xmax=100,
				ymin=0, ymax=700,
				xtick={20,40,60,80,100},
				legend columns=-1,
				legend entries = {NoPrune, Apriori, Trans, All},
				legend style={nodes={scale=0.55, transform shape}},
				legend to name={legendpruning},
				ymode=log,
				log basis y={10},
				ymajorgrids=true,
				grid style=dashed,
				line width=1.75pt
				]
				\addplot[
				color=blue,
				mark=oplus,
				mark size=4pt,
				] 
				coordinates {
					(20,184.84)(40,238.85)(60,267.11)(80,340.00)(100,531.57)
				};
				
				\addplot[
				color=black,
				mark=triangle,
				mark size=4pt,
				] 
				coordinates {
					(20,128.60)(40,147.38)(60,182.85)(80,218.62)(100,259.12)
				};
				
				\addplot[
				color=teal,
				mark=square*,
				mark size=4pt,
				]	
				coordinates {
					(20,90.85)(40,104.49)(60,113.26)(80,149.08)(100,175.57)
				};
				
				\addplot[
				color=red,
				mark=*,
				mark size=4pt,
				] 
				coordinates {
					(20,58.04)(40,79.47)(60,86.77)(80,102.35)(100,132.91)
				};
				\end{axis}
				\end{tikzpicture}
			}
			\caption{Varying \% Seq.}
		\end{subfigure}
		\begin{subfigure}{0.32\columnwidth}
			\centering
			\resizebox{\linewidth}{!}{
				\begin{tikzpicture}[scale=0.6]
				\begin{axis}[
				compat=newest,
				xlabel={Confidence (\%)},
				ylabel={Runtime (sec)}, 
				label style={font=\Huge},
				ticklabel style = {font=\Huge},
				xmin=1, xmax=5,
				ymin=0, ymax=700,
				xtick={1,2,3,4,5},
				xticklabels = {0.5,1,5,10,30},
				legend columns=-1,
				legend entries = {NoPrune, Apriori, Trans, All},
				legend style={nodes={scale=0.55, transform shape}},
				legend to name={legendpruning},
				ymode=log,
				log basis y={10},
				ymajorgrids=true,
				grid style=dashed,
				line width=1.75pt
				]
				\addplot[
				color=blue,
				mark=oplus,
				mark size=4pt,
				] 	
				coordinates {
					(1,531.57)(2,202.56)(3,76.61)(4,50.26)(5,18.32)	
				};
				
				\addplot[
				color=black,
				mark=triangle,
				mark size=4pt,
				]	
				coordinates {
					(1,259.12)(2,91.72)(3,28.21)(4,20.51)(5,11.61)	
				};
				
				\addplot[
				color=teal,
				mark=square*,
				mark size=4pt,
				] 
				coordinates {
					(1,175.57)(2,55.77)(3,15.77)(4,10.25)(5,5.85)	
				};
				
				\addplot[
				color=red,
				mark=*,
				mark size=4pt,
				] 
				coordinates {
					(1,132.91)(2,24.75)(3,8.52)(4,5.10)(5,3.34)	
				};
				\end{axis}
				\end{tikzpicture}
			}
			\caption{Varying Conf.}
		\end{subfigure}
		\begin{subfigure}{0.32\columnwidth}
			\centering
			\resizebox{\linewidth}{!}{
				\begin{tikzpicture}[scale=0.6]
				\begin{axis}[
				compat=newest,
				xlabel={Support (\%)},
				ylabel={Runtime (sec)}, 
				label style={font=\Huge},
				ticklabel style = {font=\Huge},
				xmin=1, xmax=5,
				ymin=0, ymax=700,
				xtick={1,2,3,4,5},
				xticklabels = {0.5,1,5,10,30},
				legend columns=-1,
				legend entries = {NoPrune, Apriori, Trans, All},
				legend style={nodes={scale=0.55, transform shape}},
				legend to name={legendpruning},
				ymode=log,
				log basis y={10},
				ymajorgrids=true,
				grid style=dashed,
				line width=1.75pt
				]
				\addplot[
				color=blue,
				mark=oplus,
				mark size=4pt,
				] 	
				coordinates {
					(1,531.57)(2,178.48)(3,69.38)(4,53.77)(5,16.51)
				};
				
				\addplot[
				color=black,
				mark=triangle,
				mark size=4pt,
				]	
				coordinates {
					(1,259.12)(2,96.03)(3,35.69)(4,23.42)(5,7.31)
				};
				
				\addplot[
				color=teal,
				mark=square*,
				mark size=4pt,
				] 
				coordinates {
					(1,175.57)(2,64.84)(3,21.64)(4,16.65)(5,11.26)
				};
				
				\addplot[
				color=red,
				mark=*,
				mark size=4pt,
				] 
				coordinates {
					(1,132.91)(2,20.94)(3,9.88)(4,6.26)(5,4.09)
				};
				\end{axis}
				\end{tikzpicture}
			}
			\caption{Varying Supp.}
		\end{subfigure}
		\vspace{-0.05in}
		\ref{legendpruning}
		\vspace{-0.05in}
		\caption{Runtimes of E-HTPGM on ASL}
		\label{fig:performance_AppendixASL}
	\end{minipage} 
\end{figure}

\subsection{Evaluation of A-HTPGM} \label{sec:accuracy_Appendix}
Table \ref{tbl:ratiopatterns_Appro_Exact_Appendix1}, \ref{tbl:ratiopatterns_Appro_Exact_Appendix2}, and \ref{tbl:ratiopatterns_Appro_Exact_Appendix3} show the accuracies and the numbers of extracted patterns from A-HTPGM for different support and confidence thresholds. It can be seen that, A-HTPGM obtains high accuracy ($\ge 70\%$) when $\sigma$ and $\delta$ are low, e.g, $\sigma = \delta = 10\%$, and very high accuracy ($\ge 90\%$) when $\sigma$ and $\delta$ are high, e.g, $\sigma = \delta =50\%$.

\begin{table}[!t]
	\centering
	\caption{The Accuracy and the Number of Extracted Patterns from A-HTPGM on ASL}
	\vspace{-0.1in}
	\resizebox{\columnwidth}{1.3cm}{
		\begin{tabular}{|c|c|c|c|c|c|c|c|c|}
			\hline 
			\multirow{4}{*}{Supp. (\%)} & \multicolumn{8}{c|}{\bfseries Conf. (\%)} \\  
			\cline{2-9}  
			& \multicolumn{8}{c|}{\bfseries ASL}
			\\  \cline{2-9}  
			& \multicolumn{4}{c|}{\bfseries Accuracy (\%)}  & \multicolumn{4}{c|}{\bfseries \# Patterns} 
			\\  \cline{2-9}
			& {\bfseries 0.5} & {\bfseries 1} & {\bfseries 5}  & {\bfseries 10} & {\bfseries 0.5} & {\bfseries 1} & {\bfseries 5}    & {\bfseries 10} \\
			\hline
			0.5 & 82  & 85   & 95 & 98 & 86261 & 56747 & 3413 & 1960  
			\\  \hline			
			1 & 87  & 87   & 95  & 100 & 24862 & 16561 & 3005 & 1902  
			\\  \hline				
			5 & 91  & 97   & 100  & 100 & 5810 & 5612 & 1612 & 1484  
			\\  \hline					
			10 & 94  & 98   & 100  & 100 & 1396 & 1106 & 1008 & 901 
			\\  \hline					
		\end{tabular}
	}
	\label{tbl:ratiopatterns_Appro_Exact_Appendix3}
\end{table}

Next, we analyze the quality of patterns pruned by A-HTPGM. These patterns are extracted from the uncorrelated time series. 
Fig. \ref{fig:cumProbPrunedAtt_Appendix} shows the cumulative distribution of the confidences for the pruned patterns. It is seen that most of these patterns have low confidences. For UKDALE and DataPort, $85\%$ of patterns have confidences less than $20\%$ when the support is $10\%$ and $20\%$, and $75\%$ of patterns have confidences less than $30\%$ when the support is $30\%$. For ASL, $80\%$ of patterns have confidences less than $5\%$. Such patterns are likely not interesting to explore, and thus, are pruned by A-HTPGM. 

\begin{figure}[!h]
	\centering
		\begin{subfigure}[t]{0.32\columnwidth}
			\centering
			\resizebox{\linewidth}{!}{
				\begin{tikzpicture}[scale=0.6]
				\begin{axis}[
				compat=newest,
				style={very thick},
				xlabel={Confidence (\%)},
				ylabel={Cumulative Probability}, 
				label style={font=\Huge},
				ticklabel style = {font=\Huge},
				xmin=0, xmax=100,
				ymin=0, ymax=1,
				xtick={0,20,40,60,80,100},
				legend columns=1,
				legend entries = {supp=10\%, supp=20\%, supp=30\%, supp=40\%},
				legend style={at={(0.95,0.2)},anchor=east,font=\Large},
				ymajorgrids=true,
				grid style=dashed,
				line width=1.75pt
				]
				\addplot[
				smooth,
				color=blue,
				solid
				]
				coordinates {
					(0,0)(2,0.04)(20,0.89)(30,0.97)(40,0.99)(60,0.99)(80,0.99)(100,1)
				};
				\addplot[
				smooth,
				color=red,
				dashed
				]
				coordinates {
					(0,0)(10,0.02)(20,0.81)(40,0.90)(50,0.96)(60,0.98)(80,0.99)(100,1)
				};
				\addplot[
				smooth,
				color=teal,
				densely dashdotted
				]
				coordinates {
					(15,0)(20,0.1)(30,0.71)(40,0.81)(50,0.90)(60,0.93)(70,0.96)(80,0.98)(100,1)
				};
				\addplot[
				smooth,
				color=black,
				loosely dotted
				]
				coordinates {
					(20,0)(30,0.1)(50,0.67)(60,0.78)(70,0.89)(80,0.94)(90,0.98)(100,1)
				};
				
				\end{axis}
				\end{tikzpicture}
			}
			\caption{UKDALE}
		\end{subfigure}
		\begin{subfigure}[t]{0.32\columnwidth}
			\centering
			\resizebox{\linewidth}{!}{
				\begin{tikzpicture}[scale=0.6]
				\begin{axis}[
				compat=newest,
				style={very thick},
				xlabel={Confidence (\%)},
				ylabel={Cumulative Probability}, 
				label style={font=\Huge},
				ticklabel style = {font=\Huge},
				xmin=0, xmax=100,
				ymin=0, ymax=1,
				xtick={0,20,40,60,80,100},
				legend columns=1,
				legend entries = {supp=10\%, supp=20\%, supp=30\%, supp=40\%},
				legend style={at={(0.95,0.2)},anchor=east,font=\Large},			
				ymajorgrids=true,
				grid style=dashed,
				line width=1.75pt
				]
				\addplot[
				smooth,
				color=blue,
				solid
				]
				coordinates {
					(0,0)(2,0.04)(15,0.75)(20,0.89)(30,0.93)(40,0.96)(60,0.99)(80,0.99)(100,1)
				};
				\addplot[
				smooth,
				color=red,
				dashed
				]
				coordinates {
					(0,0)(5,0.05)(15,0.50)(20,0.85)(40,0.92)(50,0.95)(60,0.98)(80,0.99)(100,1)
				};
				\addplot[
				smooth,
				color=teal,
				densely dashdotted
				]
				coordinates {
					(0,0)(10,0.03)(15,0.27)(20,0.4)(30,0.73)(40,0.85)(50,0.92)(60,0.95)(70,0.96)(80,0.98)(100,1)
				};
				\addplot[
				smooth,
				color=black,
				loosely dotted
				]
				coordinates {
					(0,0)(15,0.02)(20,0.2)(30,0.4)(50,0.62)(60,0.78)(70,0.85)(80,0.96)(90,0.98)(100,1)
				};
				
				\end{axis}
				\end{tikzpicture}
			}
			\caption{DataPort}
		\end{subfigure}
		\begin{subfigure}[t]{0.32\columnwidth}
			\centering
			\resizebox{\linewidth}{!}{
				\begin{tikzpicture}[scale=0.6]
				\begin{axis}[
				compat=newest,
				style={very thick},
				xlabel={Confidence (\%)},
				ylabel={Cumulative Probability}, 
				label style={font=\Huge},
				ticklabel style = {font=\Huge},
				xmin=0, xmax=6,
				ymin=0, ymax=1,
				xtick={1,2,3,4,5,6},
				xticklabels = {0.5,1,5,10,20,30},
				legend columns=1,
				legend entries = {supp=0.5\%, supp=1\%, supp=5\%, supp=10\%},
				legend cell align={left},
				legend style={at={(0.95,0.2)},anchor=east,font=\Huge},	
				ymajorgrids=true,
				grid style=dashed,
				line width=1.75pt
				]
				\addplot[
				smooth,
				color=blue,
				solid
				]
				coordinates {
					(0,0)(1,0.6)(2,0.7)(3,0.92)(4,0.95)(5,0.99)(6,1)
				};
				\addplot[
				smooth,
				color=red,
				dashed
				]
				coordinates {
					(0,0)(1,0.55)(2,0.65)(3,0.88)(4,0.93)(5,0.99)(6,1)
				};
				\addplot[
				smooth,
				color=teal,
				densely dashdotted
				]
				coordinates {
					(0,0)(1,0.2)(2,0.58)(3,0.84)(4,0.90)(5,0.96)(6,1)
				};
				\addplot[
				smooth,
				color=black,
				loosely dotted
				]
				coordinates {
					(0,0)(1,0.1)(2,0.5)(3,0.81)(4,0.86)(5,0.93)(6,1)
				};
				
				\end{axis}
				\end{tikzpicture}
			}
			\caption{ASL}
		\end{subfigure}	
		\vspace{0.05in}
		\caption{Cumulative probability of pruned patterns}
		\label{fig:cumProbPrunedAtt_Appendix}  
\end{figure}

\subsection{Evaluation of the tolerance buffer $\epsilon$}
We evaluate the impact of the buffer $\epsilon$ on extracted patterns. Tables \ref{tbl:epsilon_Appendix1} and \ref{tbl:epsilon_Appendix2} report the number of extracted patterns for different $\epsilon$ values, and the corresponding percentages of pattern loss compared to $\epsilon=0$. For NIST and DataPort datasets, the percentage of pattern loss is the lowest with $\epsilon=1$ minute. For UKDALE, the percentage of pattern loss is the lowest with $\epsilon=1$ and $2$ minutes. And for Smart City, the percentage of pattern loss is very low, and the losses among different $\epsilon$ values are not very much different since there is low noise level in the dataset. For ASL, the percentage of pattern loss is the lowest when $\epsilon=30$ frames.

\begin{figure}[!h]
	\centering
	\vspace{-0.15in}
		\begin{subfigure}[t]{0.32\columnwidth}
			\centering
			\resizebox{\linewidth}{!}{
				\begin{tikzpicture}[scale=0.6]
				\begin{axis}[
				compat=newest,
				style={very thick},
				xlabel={Confidence (\%)},
				ylabel={Cumulative Probability}, 
				label style={font=\Huge},
				ticklabel style = {font=\Huge},
				xmin=20, xmax=100,
				ymin=0, ymax=1,
				xtick={20,40,60,80,100},
				legend columns=1,
				legend entries = {$\epsilon$=15 minutes, $\epsilon$=10 minutes, $\epsilon$=5 minutes},
				legend style={at={(0.95,0.2)},anchor=east,font=\Large},	
				ymajorgrids=true,
				grid style=dashed,
				line width=1.75pt
				]
				\addplot[
				smooth,
				color=blue,
				solid
				]
				coordinates {
					(20,0.0)(30,0.26)(40,0.35)(50,0.52)(60,0.60)(80,0.67)(100,1)
				};
				\addplot[
				smooth,
				color=teal,
				densely dashdotted
				]
				coordinates {
					(20,0.0)(30,0.18)(40,0.29)(50,0.34)(60,0.45)(80,0.56)(100,1)
				};
				\addplot[
				smooth,
				color=black,
				loosely dotted
				]
				coordinates {
					(20,0)(30,0.21)(40,0.32)(50,0.50)(60,0.55)(70,0.57)(80,0.65)(90,0.85)(100,1)
				};
				
				\end{axis}
				\end{tikzpicture}
			}
			\caption{Smart City}
		\end{subfigure}
		\begin{subfigure}[t]{0.32\columnwidth}
			\centering
			\resizebox{\linewidth}{!}{
				\begin{tikzpicture}[scale=0.6]
				\begin{axis}[
				compat=newest,
				style={very thick},
				xlabel={Confidence (\%)},
				ylabel={Cumulative Probability}, 
				label style={font=\Huge},
				ticklabel style = {font=\Huge},
				xmin=1, xmax=5,
				ymin=0, ymax=1,
				xtick={1,2,3,4,5},
				xticklabels = {1,5,10,20,30},
				legend columns=1,
				legend entries = {$\epsilon$=60 frames, $\epsilon$=45 frames, $\epsilon$=30 frames},
				legend style={at={(0.95,0.2)},anchor=east,font=\Large},	
				ymajorgrids=true,
				grid style=dashed,
				line width=1.75pt
				]
				\addplot[
				smooth,
				color=blue,
				solid
				]
				coordinates {
					(1,0.0)(2,0.41)(3,0.55)(4,0.69)(5,1)
				};
				\addplot[
				smooth,
				color=teal,
				densely dashdotted
				]
				coordinates {
					(1,0.0)(2,0.31)(3,0.51)(4,0.62)(5,1)
				};
				\addplot[
				smooth,
				color=black,
				loosely dotted
				]
				coordinates {
					(1,0.0)(2,0.21)(3,0.35)(4,0.50)(5,1)
				};
				
				\end{axis}
				\end{tikzpicture}
			}
			\caption{ASL}
		\end{subfigure}		
		\caption{Cumulative probability of pattern loss}
		\label{fig:cumProbDiffPattern_Appendix2}
\end{figure}
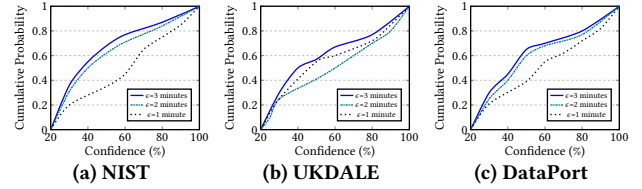
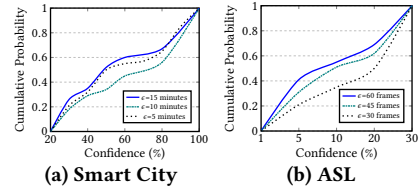

Next, we analyze the quality of extracted patterns using different $\epsilon$ values. Specifically, for each specific $\epsilon$ value, we filter the patterns that are not in the result set of $\epsilon=0$, and calculate the cumulative distribution of the confidences for them. Figs. \ref{fig:cumProbDiffPattern_Appendix1} and \ref{fig:cumProbDiffPattern_Appendix2} show the corresponding results on different datasets. For NIST and DataPort, $65\%$ of patterns have confidences greater than $50\%$ when $\epsilon$ is $1$ minute, and $40\%$ of patterns have confidences greater than $50\%$ when $\epsilon$ is $2$ and $3$ minutes. For UKDALE, $60\%$ of patterns have confidences greater than $50\%$ when $\epsilon$ is $2$ minutes, and $45\%$ of patterns have confidences greater than $50\%$ when $\epsilon$ is $1$ and $3$ minutes. For Smart City, $62\%$ of patterns have confidences greater than $50\%$ when $\epsilon$ is $10$ minutes, and $49\%$ of patterns have confidences greater than $50\%$ when $\epsilon$ is $5$ and $15$ minutes. For ASL, $70\%$ of patterns have confidences greater than $10\%$ when epsilon is $30$ frames, and $50\%$ of patterns have confidences greater than $50\%$ when epsilon is $45$ and $60$ frames. 

\begin{table}[!t]
	\centering
	\caption{Number of Extracted Patterns and Percentages of Pattern Loss on NIST, UKDALE, and DataPort}
	\vspace{-0.1in}
	\resizebox{\columnwidth}{!}{
		\begin{tabular}{|c|c|c|c|c|c|c|}
			\hline 
			\multirow{2}{*}{$\epsilon$ value} &\multicolumn{2}{c|}{\bfseries NIST} &\multicolumn{2}{c|}{\bfseries UKDALE} &\multicolumn{2}{c|}{\bfseries DataPort} \\  
			\cline{2-7}  
			& {\bfseries \# Patterns} & {\bfseries Patterns (\%)} & {\bfseries \# Patterns} & {\bfseries Patterns (\%)} & {\bfseries \# Patterns} & {\bfseries Patterns (\%)} \\
			\hline
			1 minute & 1938081 & 0.42 & 1076811 & 0.31 & 220324 & 0.73 
			\\  \hline							
			2 minutes & 1864630  & 4.19   & 1076248 & 0.36 & 210348 & 5.22
			\\  \hline							
			3 minutes & 1812857  & 6.85   & 1056532 & 2.18 & 202153 & 8.91 
			\\  \hline							
		\end{tabular}
	}
	\label{tbl:epsilon_Appendix1}
\end{table}

\begin{table}[!t]
	\centering
	\caption{Number of Extracted Patterns and Percentages of Pattern Loss on Smart City and ASL}
	\vspace{-0.1in}
	\begin{minipage}{0.48\linewidth}
		\resizebox{\columnwidth}{!}{
			\begin{tabular}{|c|c|c|}
				\hline 
				\multirow{2}{*}{$\epsilon$ value} &\multicolumn{2}{c|}{\bfseries Smart City} \\  
				\cline{2-3}  
				& {\bfseries \# Patterns} & {\bfseries Patterns (\%)} \\
				\hline
				5 minutes & 1264207 & 0.06 
				\\  \hline							
				10 minutes & 1263902  & 0.08   
				\\  \hline							
				15 minutes & 1263787  & 0.09 
				\\  \hline							
			\end{tabular}
		}
	\end{minipage}
	\begin{minipage}{0.48\linewidth}
		\resizebox{\columnwidth}{!}{
			\begin{tabular}{|c|c|c|}
				\hline 
				\multirow{2}{*}{$\epsilon$ value} &\multicolumn{2}{c|}{\bfseries ASL} \\  
				\cline{2-3}  
				& {\bfseries \# Patterns} & {\bfseries Patterns (\%)} \\
				\hline
				30 frames & 104627 & 0.54 
				\\  \hline							
				45 frames & 101683  & 3.34   
				\\  \hline							
				60 frames & 96475  & 8.29 
				\\  \hline							
			\end{tabular}
		}
	\end{minipage}
	\label{tbl:epsilon_Appendix2}
\end{table}

\subsection{Evaluation of the Splitting Strategy using Overlapping Sequences}
We evaluate the impact of overlapping duration used in the sequences splitting strategy on extracted patterns. Table \ref{tbl:numberPatterns_Overlapping_Appendix} reports the number of extracted patterns with different overlapping durations. For NIST, UKDALE, and Smart City, the number of extracted patterns increases when overlapping duration increases, and the number of extracted patterns becomes stable when the overlapping duration is equal to $2$ hours or more. The same trend also applies for DataPort and ASL, where the number of extracted patterns becomes constant when overlapping duration is greater than $1$ hour and $150$ frames, respectively, for the two datasets. 

\begin{table}[!t]
	\centering
		\caption{Number of Extracted Patterns with Different Overlapping Durations}
		\begin{minipage}{.7\linewidth}
			\vspace{-0.1in}
			\resizebox{\columnwidth}{!}{
			\begin{tabular}{|p{1.5cm}|c|c|c|c|}
				\hline 
				{Overlapping Duration}  & {\bfseries NIST} & {\bfseries UKDALE} & {\bfseries DataPort} & {\bfseries Smart City} \\  \hline 
				0 hour & 1946265 & 1080165 & 221948 & 1264971
				  \\  \hline							
				  1 hour  & 1947491  & 1084565  & 222052 & 1265193
				  \\  \hline							
				  2 hours  & 1947564  & 1085719   & 222052 & 1265232
				  \\  \hline
				  3 hours  & 1947564  & 1085719  & 222052 & 1265232
				  \\  \hline						
			\end{tabular}
			}
		\end{minipage}	
		\begin{minipage}{.26\linewidth}
			\vspace{-0.1in}
			\resizebox{\columnwidth}{!}{
			\begin{tabular}{|p{1.5cm}|c|}
				\hline 
				{Overlapping Duration}  & {\bfseries ASL} \\  \hline 
				0 frame & 105196
				  \\  \hline							
				  150 frames  & 105302 
				  \\  \hline							
				  300 frames  & 105302 
				  \\  \hline
				  450 frames  & 105302 
				  \\  \hline						
			\end{tabular}
			}
		\end{minipage}	
		\label{tbl:numberPatterns_Overlapping_Appendix}
\end{table}

\begin{figure}[!h]
	\centering
	\begin{minipage}{1\linewidth}
		\begin{subfigure}[t]{0.32\columnwidth}
			\centering
			\resizebox{\linewidth}{!}{
				\begin{tikzpicture}[scale=0.6]
				\begin{axis}[
				compat=newest,
				style={very thick},
				xlabel={Confidence (\%)},
				ylabel={Cumulative Probability}, 
				label style={font=\Huge},
				ticklabel style = {font=\Huge},
				xmin=20, xmax=100,
				ymin=0, ymax=1,
				xtick={20,40,60,80,100},
				legend columns=1,
				legend entries = {$\epsilon$=3 minutes, $\epsilon$=2 minutes, $\epsilon$=1 minute},
				legend style={at={(0.95,0.2)},anchor=east,font=\Large},
				ymajorgrids=true,
				grid style=dashed,
				line width=1.75pt
				]
				\addplot[
				smooth,
				color=blue,
				solid
				]
				coordinates {
					(20,0.0)(30,0.36)(40,0.55)(50,0.68)(60,0.77)(80,0.87)(100,1)
				};
				\addplot[
				smooth,
				color=teal,
				densely dashdotted
				]
				coordinates {
					(20,0.0)(30,0.31)(40,0.50)(50,0.62)(60,0.71)(80,0.84)(100,1)
				};
				\addplot[
				smooth,
				color=black,
				loosely dotted
				]
				coordinates {
					(20,0)(30,0.20)(50,0.35)(60,0.45)(70,0.65)(80,0.75)(90,0.85)(100,1)
				};
				
				\end{axis}
				\end{tikzpicture}
			}
			\caption{NIST}
		\end{subfigure}
		\begin{subfigure}[t]{0.32\columnwidth}
			\centering
			\resizebox{\linewidth}{!}{
				\begin{tikzpicture}[scale=0.6]
				\begin{axis}[
				compat=newest,
				style={very thick},
				xlabel={Confidence (\%)},
				ylabel={Cumulative Probability}, 
				label style={font=\Huge},
				ticklabel style = {font=\Huge},
				xmin=20, xmax=100,
				ymin=0, ymax=1,
				xtick={20,40,60,80,100},
				legend columns=1,
				legend entries = {$\epsilon$=3 minutes, $\epsilon$=2 minutes, $\epsilon$=1 minute},
				legend style={at={(0.95,0.2)},anchor=east,font=\Large},
				ymajorgrids=true,
				grid style=dashed,
				line width=1.75pt
				]
				\addplot[
				smooth,
				color=blue,
				solid
				]
				coordinates {
					(20,0.0)(30,0.30)(40,0.50)(50,0.57)(60,0.67)(80,0.77)(100,1)
				};
				\addplot[
				smooth,
				color=teal,
				densely dashdotted
				]
				coordinates {
					(20,0)(25,0.10)(30,0.25)(50,0.41)(60,0.50)(70,0.60)(80,0.70)(90,0.80)(100,1)
				};
				\addplot[
				smooth,
				color=black,
				loosely dotted
				]
				coordinates {
					(20,0)(30,0.25)(50,0.55)(60,0.60)(80,0.72)(90,0.85)(100,1)
				};
				
				\end{axis}
				\end{tikzpicture}
			}
			\caption{UKDALE}
		\end{subfigure}
		\begin{subfigure}[t]{0.32\columnwidth}
			\centering
			\resizebox{\linewidth}{!}{
				\begin{tikzpicture}[scale=0.6]
				\begin{axis}[
				compat=newest,
				style={very thick},
				xlabel={Confidence (\%)},
				ylabel={Cumulative Probability}, 
				label style={font=\Huge},
				ticklabel style = {font=\Huge},
				xmin=20, xmax=100,
				ymin=0, ymax=1,
				xtick={20,40,60,80,100},
				legend columns=1,
				legend entries = {$\epsilon$=3 minutes, $\epsilon$=2 minutes, $\epsilon$=1 minute},
				legend style={at={(0.95,0.2)},anchor=east,font=\Large},
				ymajorgrids=true,
				grid style=dashed,
				line width=1.75pt
				]
				\addplot[
				smooth,
				color=blue,
				solid
				]
				coordinates {
					(20,0.0)(30,0.30)(40,0.45)(50,0.65)(60,0.70)(80,0.80)(100,1)
				};
				\addplot[
				smooth,
				color=teal,
				densely dashdotted
				]
				coordinates {
					(20,0.0)(30,0.25)(40,0.40)(50,0.60)(60,0.68)(80,0.77)(100,1)
				};
				\addplot[
				smooth,
				color=black,
				loosely dotted
				]
				coordinates {
					(20,0)(30,0.21)(50,0.39)(60,0.55)(70,0.62)(80,0.72)(90,0.82)(100,1)
				};
				
				\end{axis}
				\end{tikzpicture}
			}
			\caption{DataPort}
		\end{subfigure}	
		\caption{Cumulative probability of pattern loss}
		\label{fig:cumProbDiffPattern_Appendix1}
	\end{minipage}		   
\end{figure}

\begin{table*}[!h]
	\centering
	\caption{Runtime Comparison using Seven Relations Model (seconds)}
	\vspace{-0.15in}
	\begin{minipage}{.55\linewidth}
		\resizebox{\columnwidth}{!}{
			\begin{tabular}{|c|c|c|c|c|c|c|c|}
				\hline 
				\multirow{3}{*}{Supp. (\%)} & \multirow{3}{*}{Methods}   & \multicolumn{6}{c|}{\bfseries Conf. (\%)}
				\\  \cline{3-8}  
				& & \multicolumn{3}{c|}{\bfseries NIST}  & \multicolumn{3}{c|}{\bfseries Smart City}  
				\\  \cline{3-8}  
				& & {\bfseries 20} & {\bfseries 50}  & {\bfseries 80} & {\bfseries 20} & {\bfseries 50}  & {\bfseries 80}\\
				\hline
				\multirow{2}{*}{20}  & Z-Miner  & 7996.05  & 689.89 & 172.23 & 141.78  & 12.97 & 2.10   \\  \cline{2-8}  
				&\centering E-HTPGM  & 2135.12  & 193.45   & 67.97 & 27.89  & 5.42 & 1.07 \\    \hline
				
				\multirow{2}{*}{50}  & Z-Miner  & 564.41  & 559.17   & 154.61 & 11.25  & 10.84 & 1.90  \\  \cline{2-8}  
				&\centering E-HTPGM  & 149.96  & 145.53   & 52.22 & 2.82  & 2.06 & 0.88 \\    \hline
				
				\multirow{2}{*}{80}  & Z-Miner  & 163.46  & 154.62   & 153.51 & 1.39  & 1.33 & 1.33  \\  \cline{2-8}  
				&\centering E-HTPGM  & 52.60  & 51.70   & 50.01 & 0.51  & 0.50 & 0.48   \\    \hline
			\end{tabular}	
		}
	\end{minipage}
	\begin{minipage}{.35\linewidth}
		\resizebox{\columnwidth}{!}{
			\begin{tabular}{|c|c|c|c|c|}
				\hline 
				\multirow{3}{*}{Supp. (\%)} & \multirow{3}{*}{Methods}   & \multicolumn{3}{c|}{\bfseries Conf. (\%)}
				\\  \cline{3-5}  
				& & \multicolumn{3}{c|}{\bfseries ASL}  
				\\  \cline{3-5}  
				& & {\bfseries 0.5} & {\bfseries 1}  & {\bfseries 10}\\
				\hline
				\multirow{2}{*}{0.5}  & Z-Miner  & 194.02 & 107.62 & 19.41   \\  \cline{2-5}  
				&\centering E-HTPGM  & 120.75  & 86.73   & 5.79 \\    \hline
				
				\multirow{2}{*}{1}  & Z-Miner  & 83.05 & 75.02 & 12.39   \\  \cline{2-5}  
				&\centering E-HTPGM  & 39.72 & 34.27 & 5.42  \\    \hline
				
				\multirow{2}{*}{10}  & Z-Miner  & 3.09 & 2.98 & 2.94   \\  \cline{2-5}  
				&\centering E-HTPGM  & 1.78 & 1.66 & 1.64    \\    \hline
			\end{tabular}	
		}
	\end{minipage}
	\label{tbl:runtimeBaselines7relations}
\end{table*}

\begin{table*}[!h]
	\centering
	\caption{Memory Usage Comparison using Seven Relations Model (MB)}
	\vspace{-0.15in}
	\begin{minipage}{.55\linewidth}
		\resizebox{\columnwidth}{!}{
			\begin{tabular}{|c|c|c|c|c|c|c|c|}
				\hline 
				\multirow{3}{*}{Supp. (\%)} & \multirow{3}{*}{Methods}   & \multicolumn{6}{c|}{\bfseries Conf. (\%)}
				\\  \cline{3-8}  
				& & \multicolumn{3}{c|}{\bfseries NIST}  & \multicolumn{3}{c|}{\bfseries Smart City}  
				\\  \cline{3-8}  
				& & {\bfseries 20} & {\bfseries 50}  & {\bfseries 80} & {\bfseries 20} & {\bfseries 50}  & {\bfseries 80}\\
				\hline
				\multirow{2}{*}{20}  & Z-Miner  & 74988.73 & 6469.43 &  1970.82 & 571.89  & 101.11 & 50.34   \\  \cline{2-8}  
				&\centering E-HTPGM  & 13530.18  & 1937.75   & 666.31 & 464.39  & 74.87 & 32.58 \\    \hline
				
				\multirow{2}{*}{50}  & Z-Miner  & 6123.63  & 6028.29   & 1920.11 & 94.88  & 94.19 & 47.55  \\  \cline{2-8}  
				&\centering E-HTPGM  & 1882.01  & 1786.34   & 664.64 & 71.39  & 70.99 & 32.61 \\    \hline
				
				\multirow{2}{*}{80}  & Z-Miner  & 1878.24  & 1859.17   & 1849.66 & 34.16  & 33.38 & 33.30  \\  \cline{2-8}  
				&\centering E-HTPGM  & 660.86  & 660.60   & 593.95 & 31.36 & 31.36 & 29.85   \\    \hline
			\end{tabular}	
		}
	\end{minipage}
	\begin{minipage}{.35\linewidth}
		\resizebox{\columnwidth}{!}{
			\begin{tabular}{|c|c|c|c|c|}
				\hline 
				\multirow{3}{*}{Supp. (\%)} & \multirow{3}{*}{Methods}   & \multicolumn{3}{c|}{\bfseries Conf. (\%)}
				\\  \cline{3-5}  
				& & \multicolumn{3}{c|}{\bfseries ASL}  
				\\  \cline{3-5}  
				& & {\bfseries 0.5} & {\bfseries 1}  & {\bfseries 10}\\
				\hline
				\multirow{2}{*}{0.5}  & Z-Miner  & 403.79 & 333.66 & 185.32  \\  \cline{2-5}  
				&\centering E-HTPGM  & 284.08  & 231.88   & 66.85 \\    \hline
				
				\multirow{2}{*}{1}  & Z-Miner  & 349.65 & 254.24 & 182.31  \\  \cline{2-5}  
				&\centering E-HTPGM  & 170.56 & 161.02 & 66.54   \\    \hline
				
				\multirow{2}{*}{10}  & Z-Miner  & 131.43 & 131.40 & 124.95  \\  \cline{2-5}  
				&\centering E-HTPGM  & 53.06 & 52.26 & 51.34   \\    \hline
			\end{tabular}	
		}
	\end{minipage}
	\label{tbl:memoryBaselines7relations}
\end{table*}

\subsection{Evaluation between E-HTPGM and Z-Miner using seven temporal relations}
As shown in Table \ref{tbl:runtimeBaselines7relations}, E-HTPGM has better runtime than Z-Miner. On the tested datasets, the range and average speedups of A-HTPGM compared to Z-Miner is: $[1.30$-$3.84]$ and $2.28$. In terms of memory consumption, as shown in Table \ref{tbl:memoryBaselines7relations}, E-HTPGM is more efficient than Z-Miner. The range and the average memory consumption of E-HTPGM compared to Z-Miner is: $[1.4$-$130.7]$ and $16.2$.
\end{appendices}

\end{document}